\documentclass[11pt, onecolumn]{IEEEtran}
\pdfoutput=1
\usepackage[T1]{fontenc}
\usepackage{amsmath}
\usepackage{cancel}
\usepackage{amssymb,stmaryrd}
\usepackage{graphicx}
\usepackage{psfrag}
\usepackage{latexsym}
\usepackage{dsfont}
\usepackage{cite}
\usepackage{extraipa}
\usepackage{esvect}
\usepackage{mathrsfs}
\usepackage{color}
\usepackage{srcltx}
\usepackage{relsize,mdframed}
\usepackage{eufrak,feyn}
\usepackage{hyperref}
\usepackage{float, array, makecell,multirow}
\newtheorem{theorem}{Theorem}

\newtheorem{lemma}{Lemma}

\newtheorem{definition}{Definition}

\newtheorem{remark}{Remark}
\newtheorem{example}{Example}
\newcommand{\qed}{\nobreak \ifvmode \relax \else
      \ifdim\lastskip<1.5em \hskip-\lastskip
      \hskip0.5em plus0em minus0.5em \fi \nobreak
      \vrule height0.4em width0.5em depth0.1em\fi}
\DeclareMathAlphabet{\mathpzc}{OT1}{pzc}{m}{it}
\newcommand{\mbf}[1]{{\boldsymbol{#1}}}
\def\stackrel#1#2{\mathrel{\mathop{#2}\limits^{#1}}}

\newcommand{\set}[1]{{\mathcal{#1}}}

\newcommand{\mc}[1]{{\mathcal{#1}}}

\newcommand{\vnm}[1]{\big|\hspace{-0.4mm}\big|#1\big|\hspace{-0.4mm}\big|}
\newcommand{\ve}{{\varepsilon}}
\DeclareMathOperator{\Exp}{{\mathbb{E}}} % Expectation
\newcommand{\V}{{\set{V}}} % Source Alphabet
\newcommand{\X}{{\set{X}}} % Source Alphabet
\newcommand{\Y}{{\set{Y}}} % Source Alphabet
\renewcommand{\Pr}{\mathbb{P}}

\newcommand{\h}{\mathsf{h}}

\newcommand{\A}{{\set{A}}} % Auxiliary Random Variable Alphabet
\newcommand{\B}{{\set{B}}} % Auxiliary Random Variable Alphabet
\newcommand{\M}{{\set{M}}} % Index set
\newcommand{\Times}{\mathlarger{{\mathlarger{\mathlarger{\times}}}}}
\newcommand{\bigll}{\Big|\hspace{-0.5mm}\Big|}
\newcommand{\biggll}{\bigg|\hspace{-0.5mm} \bigg|}
\newcommand{\Bigll}{\Bigg|\hspace{-0.5mm}\Bigg|}
\newcommand{\dkl}{\mathsf{D_{{\textsf{\tiny KL}}}}}

\newcommand{\pms}{{\mathsmaller \pm}}
\newcommand{\ps}{{\mathsmaller +}}
\newcommand{\ms}{\mathsmaller -}
\definecolor{lgray}{gray}{0.45}
\newcommand{\mkv}{\leftrightarrow} % Markov Chain Symbol

\allowdisplaybreaks[4]
\title{\huge{Strong Coordination over Multi-hop Line Networks}}
\author{\IEEEauthorblockN{Badri N. Vellambi,   J\"{o}rg Kliewer} \thanks{This work was supported by NSF grants CCF-1440014, CCF-1439465, and CCF-1320304. Parts of this work were published in the 2015 IEEE Information Theory Workshop~\cite{BV-MB-JK-Line1} and the 50th Annual
Conference on Information Sciences and Systems~\cite{BV-MB-JK-Line2}.}

\IEEEauthorblockA{New Jersey Institute of Technology\\
Newark, NJ 07102\\
Email: badri.n.vellambi@ieee.org, jkliewer@njit.edu\\}
\and
\IEEEauthorblockN{\color{white} abc \color{black}}\\
 \IEEEauthorblockA{}
\and
\IEEEauthorblockN{Matthieu R.~Bloch}\\
\IEEEauthorblockA{Georgia Institute of Technology\\
Atlanta, GA 30332\\
Email: matthieu.bloch@ece.gatech.edu}}
\IEEEoverridecommandlockouts
\begin{document}
\maketitle

\begin{abstract}
We analyze the problem of strong coordination over a multi-hop line network in which the node initiating the coordination is a terminal network node. We assume that each node has access to a certain amount of randomness that is local to the node, and that the nodes share some common randomness, which are used together with explicit hop-by-hop communication to achieve strong coordination.  We derive the trade-offs among the required rates of communication on the network links, the rates of local randomness available to network nodes, and the rate of common randomness  to realize strong coordination. We present an achievable coding scheme built using multiple layers of channel resolvability codes, and establish several settings in which this scheme is proven to offer the best possible trade-offs. \end{abstract}

\section{Introduction}
Decentralized control is an essential feature in almost all large-scale networks such as the Internet, surveillance systems, sensor networks, traffic and power grid networks. Control in such networks is achieved in a distributed fashion by coordinating various actions and response signals of interest. Communication between various parts of the network serves an effective means to establish coordination.  There exist two modes of communication to enable the overall goal of coordination.
\begin{itemize}
\item Coordination of a system through \emph{explicit} communication refers to settings in which communication signals extrinsic to the control and coordination of the system are sent from one part of system to another to specifically coordinate/control the system~\cite{PC-HP-TC-CordCap}. In this case, the signals used for communication are not part of the signals to be coordinated. 
\item Coordination of a system through \emph{implicit} communication refers to scenarios in which the signals inherently sent from one part of the system to another in its natural operation are also used to coordinate/control the system~\cite{Grover-Allerton2010, PC-LZ-ITW2011, Ranade-Sahai-ISIT2011}. In this case, some of the communication signals form a subset of the signals to be coordinated. \end{itemize}
The problem of coordination through (either modes of) communication is very closely tied to a slew of information-theoretic problems, including intrinsic randomness, resolvability, and random number generation~\cite{Vembu-Verdu-IT-1995, NagaokaNotes-1996, Steinberg-Verdu-IT-1996, TSH-Book}, channel resolvability, channel simulation and synthesis~\cite{Han-Verdu-1993, Steinberg-Verdu-IT-1994, PC-DistChanSynth, MHY-AG-MRA-ChanSim-ISIT12, TSH-Book}, and distributed random variable generation~\cite{AG-VA-GenDepRVs, PC-GenCorRVs}. Consequently, many ideas for the design of codes for these problems heavily feature in the design of coordination codes. Two notions of coordination have been studied in the literature:
\begin{itemize}
\item \emph{empirical} coordination, where the aim is to closely match the empirical distribution of the actions/signals at network nodes with a prescribed target histogram/probability mass function; and 
\item \emph{strong} coordination, where the aim is the generation of actions at various network nodes that are collectively required to resemble the output of a jointly correlated source. In this setting, by observing jointly the actions of the network nodes, a statistician cannot determine (with significant confidence) as to whether the actions were generated by a jointly correlated source or from a coordination scheme. 
\end{itemize}
A compendious introduction to the fundamental limits and optimal coding strategies for empirical and strong coordination in many canonical networks (e.g., one-hop, broadcast, relay networks) can be found in~\cite{PC-HP-TC-CordCap}. However, the majority of the networks considered therein comprised of two or and three terminals. The limits and means of the empirical coordination of a discrete memoryless source with a receiver connected by a point-to-point noisy channel was explored in~\cite{LeTreust-ChannelWithState-ITW2014, LeTreust-ChannelWithState-ISIT2015, LeTreust-ChannelFeedback-ISIT2015}. The effects of causality of encoding and channel feedback were investigated in~\cite{LeTreust-ChannelFeedback-ISIT2015}, and the benefits of channel state information available acausally at the encoder was explored in~\cite{LeTreust-ChannelWithState-ITW2014, LeTreust-ChannelWithState-ISIT2015}.

Coordination over the more general three-terminal setting in the presence of a relay was considered in~\cite{MB-JK-Line2, FH-MHY-AG-MRA-RelayCoord, BereyhiAref-ISIT2013}. Inner and outer bounds of the required rates of communication for coordination were derived in~\cite{MB-JK-Line2, FH-MHY-AG-MRA-RelayCoord}. Note that~\cite{MB-JK-Line2} focuses on strong coordination and only one-way communication, whereas \cite{FH-MHY-AG-MRA-RelayCoord} focuses on strong coordination and two-way communication with actions required only at the end terminals (and not at the relay). Inner and outer bounds for the required rates of communication for coordination over a noiseless triangular network with relay was studied in~\cite{BereyhiAref-ISIT2013}. The fundamental limits and optimal schemes for empirical coordination with implicit communication over multiple-access channels with state were explored in ~\cite{Larousse-T-IT-2014, Larousse-Wigger-ISIT2015}.

In this work, we quantify the network resources required for achieving \emph{strong} correlation in multi-hop line networks. By network resources, we mean three quantities required for establishing strong coordination: (a) the rates of hop-by-hop communication between network nodes; (b) the rate of  randomness locally available to each node; and (c) the required rate of common randomness shared by all network nodes. This problem is closely related to those considered in~\cite{MB-JK-Line1, MB-JK-Line2, SS-PC-SecCasc}. In~\cite{SS-PC-SecCasc}, the strong coordination rate region for two- and multi-hop line networks is characterized under the secrecy constraint that an eavesdropper does not additionally learn anything about the joint statistics of the actions even when they observe the communication on the network links. This work does not consider this additional secrecy requirement. It presents a general achievability scheme that is proven to be optimal in the following cases:
\begin{itemize}
\item when there is sufficient common randomness shared by all the nodes in the network; 
\item when the intermediate nodes operate in a \emph{functional} regime in which intermediate-node processing is a deterministic function of the incoming messages and the common randomness alone; and
\item when common randomness is absent, and the actions form a Markov chain that is aligned with the network topology. 
\end{itemize}
The remainder of this work is organized as follows. Section~\ref{sec-Notation} presents the notation used in this work. Section~\ref{sec-PD} presents the formal definition of the strong coordination problem, and Section~\ref{Sec-Prelims} presents the main results of this work. Finally, this work is concluded in Section~\ref{sec-conclu} followed by appendices containing proofs of relevant results in Section~\ref{Sec-Prelims} and some ancillary results. 
 
\section{Notation}\label{sec-Notation}
For $m,n\in\mathbb{N}$ with $m<n$, $\llbracket m, n\rrbracket \triangleq\{m, m+1,\ldots,n\}$. Uppercase letters (e.g., $X$, $Y$) denote random variables (RVs), and the respective script versions (e.g., $\X$, $\Y$) denote their alphabets. In this work, all alphabets are assumed to be finite. Lowercase letters denote the realizations of random variables (e.g., $x$, $y$). Superscripts indicate the length of vectors. Single subscripts always indicate the node indices. In case of double subscripts, the first indicates the node index, and the next indicates the component (i.e., time) index. Given a finite set ${S}$, $\mathsf{unif}(S)$ denotes the uniform probability mass function (pmf) on the set.
Given a pmf $p_X$, $\mathsf{supp}(p_X)$ indicates the support of $p_X$, and $T_\ve^n[p_X]$ denotes the set of all $\ve$-letter typical sequences of length $n$~\cite{Kramer-MUIF}. Given two pmfs $p$ and $q$ on the same alphabet $\X$, with $\mathsf{supp}(q)\subseteq \mathsf{supp}(p)$, $\dkl(p|| q) = \sum_{x} p(x) \log \frac{p(x)}{q(x)}$. Given an event $E$, $\mathbb{P}(E)$ denotes the probability of occurrence of the event $E$. The expectation operator is denoted by  $\Exp[\cdot]$.
Lastly, $p_{X_1\cdots X_k}^{\otimes n}$ denotes the pmf of $n$ i.i.d. random $k$-tuples, with each $k$-tuple correlated according to pmf $p_{X_1 \cdots X_k}$. 

\section{Problem Definition} \label{sec-PD}
\begin{figure}[!h]
\centering
 \vspace{-3mm} 
 \includegraphics[width=4in]{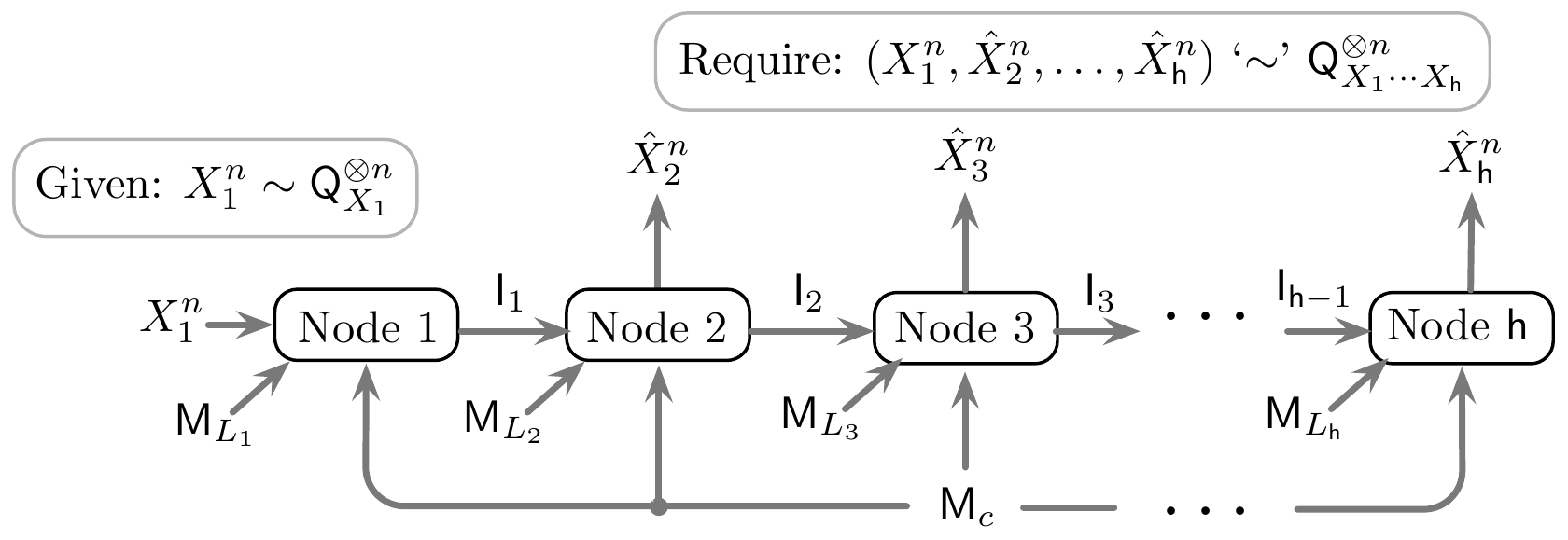}
  \vspace{-7mm}
\caption{The strong coordination problem setup.}
  \vspace{-2mm}
 \label{Fig-1}
\end{figure}

The line coordination problem is a multi-hop extension of the one studied in~\cite{MB-JK-Line1}, and is depicted in Fig.~\ref{Fig-1}. For the sake of completeness, the problem is formally defined here. 

 %---------------------------------------------------------------------------------------------------------------------
%---------------------------------------------------------------------------------------------------------------------
A line network consisting of $\h$ nodes (Nodes $1,\ldots, \h$) and $\h-1$ links  (modeled as noiseless bit pipes) that connect Node $i$ with Node $i+1$, $1\leq i<\h$ is given. Node 1 is specified an action sequence $\{X_{1,i}\}_{i\in\mathbb{N}}$,  an i.i.d process with each $X_{1,i}$ distributed over a finite set $\X_1$ according to pmf $\mathsf{Q}_{X_1}$. Nodes  are assumed to possess local randomness, as well as common randomness shared by all $\h$ nodes to enable \emph{strong coordination} using block codes.  A block code of length $n$ uses $n$ symbols of the specified action (i.e., $X_1^n$), and common and local randomness to generate actions $\hat{X}_i^n$ at Nodes $i$, $i>2$ satisfying the following condition: the joint statistics of actions $(X_1^n, \hat{X}_2^n,\ldots, \hat{X}_\h^n)$ and those of $n$ symbols output by a given discrete memoryless source $\mathsf{Q}_{X_1\cdots X_\h}^{\otimes n}$ are nearly indistinguishable under the variational distance metric.  The overall aim is to characterize the required rates of communication messages and randomness (both common and local) to achieve such strong coordination. The following definitions  are now in order.

\begin{definition} \label{def1}
Given joint pmf $\mathsf{Q}_{X_1\cdots X_\h}$ and $\ve>0$, a strong coordination  $\ve$-code of length $n$ at a rate tuple $(\mathsf{R}_c, \mathsf{R}_1,\ldots, \mathsf{R}_{\h-1}, \rho_1, \ldots, \rho_{\h})\in{\mathbb{R}^{+^{2\h}}}$ is a collection of $\h+1$ independent and uniform random variables $(\mathsf M_c,\mathsf M_{L_1},\ldots,\mathsf M_{L_\h})$, $\h-1$ message-generating functions $\psi_1,\ldots, \psi_{\h-1}$, and $\h-1$ action-generating functions $\phi_2,\ldots,\phi_\h$ such that:
\begin{itemize}
\item Randomness constraints:
\begin{align}
\textrm{[Common]}\,\,& \mathsf{M}_c\sim \mathsf{unif}(\llbracket1,2^{n\mathsf{R}_c}\rrbracket),\\
\textrm{[Local]}\,\,&  \mathsf{M}_{L_i}\sim \mathsf{unif}(\llbracket1,2^{n\rho_i}\rrbracket), \quad i=1,\ldots,\h.
\end{align}
\item Message-generation and action-generation constraints:
\begin{align}
\mathsf I_1&\triangleq \psi_1( \mathsf{M}_{L_1},X_1^n, \mathsf{M}_c) \in \llbracket 1, 2^{n\mathsf{R}_1}\rrbracket,\\
\mathsf I_j&\triangleq \psi_j( \mathsf{M}_{L_j},\mathsf I_{j-1},  \mathsf{M}_c) \in \llbracket 1, 2^{n\mathsf{R}_j}\rrbracket, \,\,\,\, 2\leq j<  \h,\\
\hat{X}_j^n &\triangleq \phi_{j}( \mathsf{M}_{L_j},\mathsf I_{j-1}, \mathsf{M}_c), \;\,\;\,\,\enspace\,\enspace\qquad\;\quad\:\quad\: 2\leq j \leq \h.
\end{align}
\item Strong coordination constraint:
\begin{align}
\vnm{\mathsf{Q}_{X_1}^{\otimes n}{Q}_{\hat{X}_2^n\cdots \hat{X}_\h^n| X_1^n}-\mathsf Q_{X_1\cdots X_\h}^{\otimes n}} \leq \ve,\quad  \label{eqn-PrbReq} \end{align}
where ${Q}_{\hat{X}_2^n\cdots \hat{X}_\h^n| X_1^n}$ is the conditional pmf of the actions generated at Nodes $2,\ldots,\h$ that is induced by the code. $\hfill${\tiny$\blacksquare$}
\end{itemize}
\end{definition}
\begin{definition}
A tuple $\mathbf{R}\triangleq(\mathsf{R}_c, \mathsf{R}_1,..., \mathsf{R}_{\h-1}, \rho_1, ..., \rho_{\h})\in {\mathbb{R}^+}^{2\h}$ is said to be \emph{achievable} for strong coordination of actions according to $\mathsf{Q}_{X_1\cdots X_\h}$ if for any $\ve>0$, there exists a strong coordination  $\ve$-code of some length $n\in\mathbb{N}$ at $\mathbf{R}$. Further, the $2\h$-dimensional strong coordination capacity region is defined as the closure of the set of all achievable rate vectors.  $\hfill${\tiny$\blacksquare$}
\end{definition}
One straightforward observation is that if $H(X_2,\ldots X_\h | X_1)=0$, i.e., $X_2,\ldots, X_\h$ are (deterministic) functions of $X_1$, then there is no need for local or common randomness, and the strong coordination problem becomes purely a communication problem with the following rate region.
\begin{remark}
If $H(X_2,\ldots, X_\h | X_1) = 0$, then $(\mathsf R_c,\mathsf R_1,\ldots, \mathsf R_{\h-1}, \rho_1,\ldots, \rho_h)\in{\mathbb{R}^{+^{2\h}}}$ is achievable iff
\begin{align*}
\quad\qquad\qquad\qquad\qquad\qquad R_\ell &\geq H(X_{\ell+1},\ldots,X_\h),\quad \ell =1,\ldots, \h-1\quad\qquad\qquad\qquad\qquad\qquad\hfill\textrm{\tiny$\blacksquare$}
\end{align*}
\end{remark}
%---------------------------------------------------------------------------------------------------------------------
\begin{figure}[!h]
\centering
 \vspace{-1mm}
 \includegraphics[width=4in]{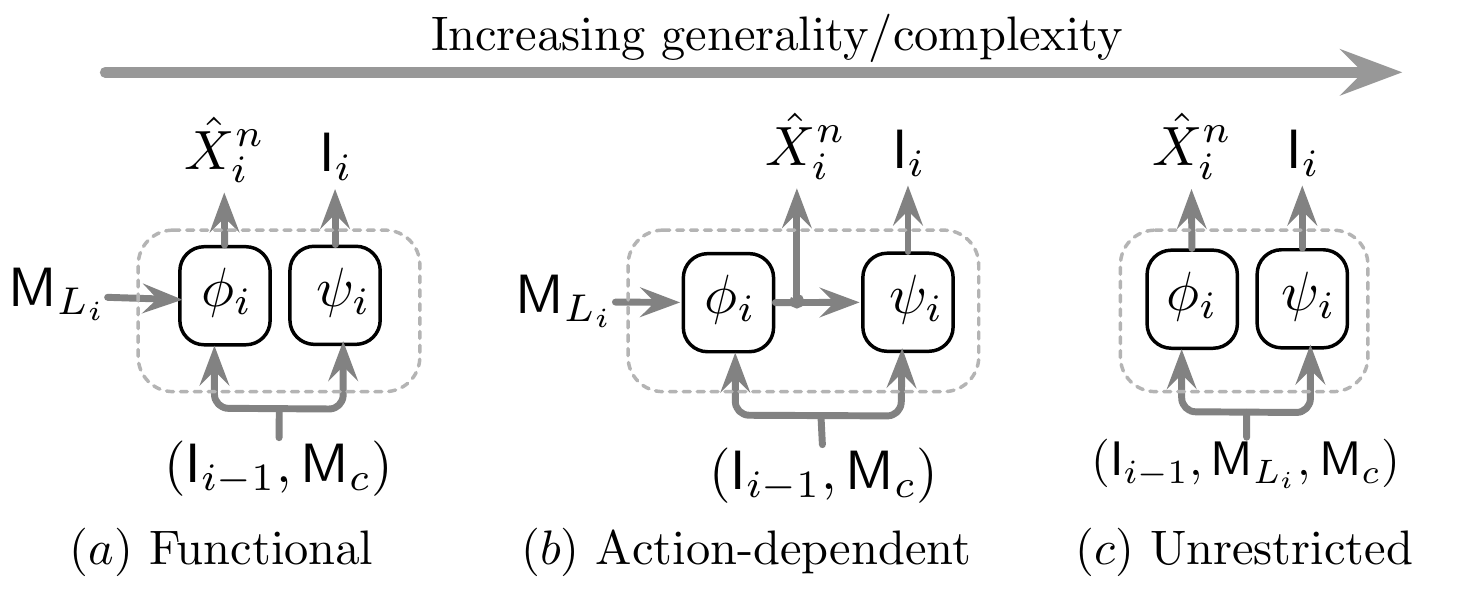}
  \vspace{-3mm}
\caption{Three possible encoder structures.}
 \label{Fig-2}
 \vspace{-3mm}
\end{figure}
%---------------------------------------------------------------------------------------------------------------------
So without loss of generality, we may assume for the rest of this work that the above remark does not apply to the given pmf $\mathsf Q_{X_1,X_2,\ldots, X_\h}$. Before we proceed to the results, we introduce three possible modes of operation for intermediate nodes. These three modes highlighted in Fig.~\ref{Fig-2} vary depending on the message generation at each intermediate node, and are as follows.
\begin{itemize}
\item In the \emph{functional} mode given in Fig.~\ref{Fig-2}\,(a), the outgoing message at each intermediate node is generated from the incoming message and common randomness, i.e., the local randomness at an intermediate node is used only to generate the action corresponding to the node.
\item In the \emph{action-dependent} mode given in Fig.~\ref{Fig-2}\,(b),  the intermediate node uses the incoming message, and local and common randomness to generate the node's action. The outgoing message is then generated using the incoming message, common randomness, and the generated action. Hence, in this mode, local randomness at a node can affect the next-hop message only through the generated action; and finally,
\item In the \emph{unrestricted} mode given in Fig.~\ref{Fig-2}\,(c), both the action and the next-hop message generated at an intermediate node depend on the incoming message, and local and common randomness. 
\end{itemize}

In theory, the set of rate vectors achievable using the unrestricted mode is a superset of those achievable using the action-dependent mode, 
 which is in turn a superset of those achievable by the functional mode. Further, these inclusions are, in general, strict (see the discussion at the end of Section~\ref{Sec-Prelims}). Before we present the achievable coding scheme for strong coordination, we present the following lemmas, which characterize the rate-transfer arguments to the problem at hand. 
\begin{lemma}\label{eqn-URratetransfer}
If  strong coordination is achievable under the unrestricted mode of operation using a common randomness rate $\mathsf R_c$, local randomness rates $(\rho_1,\ldots,\rho_\h)$ and communication rates $(\mathsf R_1,\ldots, \mathsf{R}_{\h-1})$, then:
\begin{itemize}
\item[A.] For any $1\leq \ell \leq \h$ and $\delta \leq \rho_\ell$, strong coordination is also achievable under the unrestrictive mode of operation using a common randomness rate $\mathsf R_c+\delta $, local randomness rates $(\rho_1,\ldots,\rho_{\ell-1}, \rho_\ell -\delta, \rho_{\ell+1},\ldots,\rho_\h)$ and communication rates $(\mathsf R_1,\ldots, \mathsf{R}_{\h-1})$; and 
\item[B.] For any $1<\ell\leq \h$ and $\delta\leq \rho_\ell$, strong coordination is also achievable under the unrestrictive mode of operation using a common randomness rate $\mathsf R_c$, local randomness rates $(\rho_1,\ldots,\rho_{\ell-1}+\delta, \rho_{\ell}-\delta, \rho_{\ell+1},\ldots,\rho_\h)$ and communication rates $(\mathsf R_1,\ldots, \mathsf R_{\ell-2}, \mathsf{R}_{\ell-1}+\delta, \mathsf{R}_\ell,\ldots, \mathsf{R}_{\h-1})$.
  \end{itemize}
 \end{lemma}
 \begin{IEEEproof}
 The first rate-transfer argument follows from the fact that local randomness rate at any node can be transferred onto common randomness rate, because a part of common randomness can always be used by precisely one node in the network to boost its local randomness. The second rate transfer follows from that fact that unused/excess local randomness at a node can be transmitted to the next-hop node to boost its local randomness when intermediate nodes operate in the unrestricted mode.
 \end{IEEEproof}
 
 \begin{lemma}\label{eqn-ADratetransfer}
If  strong coordination is achievable in the action-dependent (or functional) mode of operation using a common randomness rate $\mathsf R_c$, local randomness rates $(\rho_1,\ldots,\rho_\h)$ and communication rates $(\mathsf R_1,\ldots, \mathsf{R}_{\h-1})$, then:
\begin{itemize}
\item[A.] For any $1\leq \ell \leq \h$ and $\delta \leq \rho_\ell$, strong coordination is also achievable under the action-dependent (or functional) mode of operation using a common randomness rate $\mathsf R_c+\delta $, local randomness rates $(\rho_1,\ldots,\rho_{\ell-1}, \rho_\ell -\delta, \rho_{\ell+1},\ldots,\rho_\h)$ and communication rates $(\mathsf R_1,\ldots, \mathsf{R}_{\h-1})$; and 
\item[B.] For any $1<\ell\leq \h$ and $\delta\leq \rho_\ell$, strong coordination is also achievable under the action-dependent (or functional) mode of operation using a common randomness rate $\mathsf R_c$, local randomness rates $(\rho_1+\delta,\ldots,\rho_{\ell-1}, \rho_{\ell}-\delta, \rho_{\ell+1},\ldots,\rho_\h)$ and communication rates $(\mathsf R_1+\delta,\mathsf R_1+\delta, \ldots, \mathsf{R}_{\ell-1}+\delta, \mathsf{R}_\ell,\ldots, \mathsf{R}_{\h-1})$.
  \end{itemize}
 \end{lemma}
 \begin{IEEEproof}
 The first rate-transfer argument follows from the same argument as that for the first rate-transfer argument in Lemma~\ref{eqn-URratetransfer}. The second rate transfer follows from that fact that unused/excess local randomness at the first node can be transmitted to any other node to boost its local randomness. Note that intermediate nodes cannot forward unused/excess local randomness to nodes down the line when operating in the functional or action-dependent mode.
 \end{IEEEproof}

\section{ Results }\label{Sec-Prelims}
We first begin with the inner bound and then present specific settings for which we derive matching outer bounds. Throughout this work, we do not explicitly determine the cardinalities of the auxiliary RV alphabets. The cardinalities can bounded using Carath\'{e}odory's theorem~\cite{YKH-Book}.

\subsection{Inner Bound: An Achievable Scheme} \label{subsec-GenInnerBnd}
\begin{figure}[!h]
\centering
\vspace{-3mm}
 \includegraphics[width=6in]{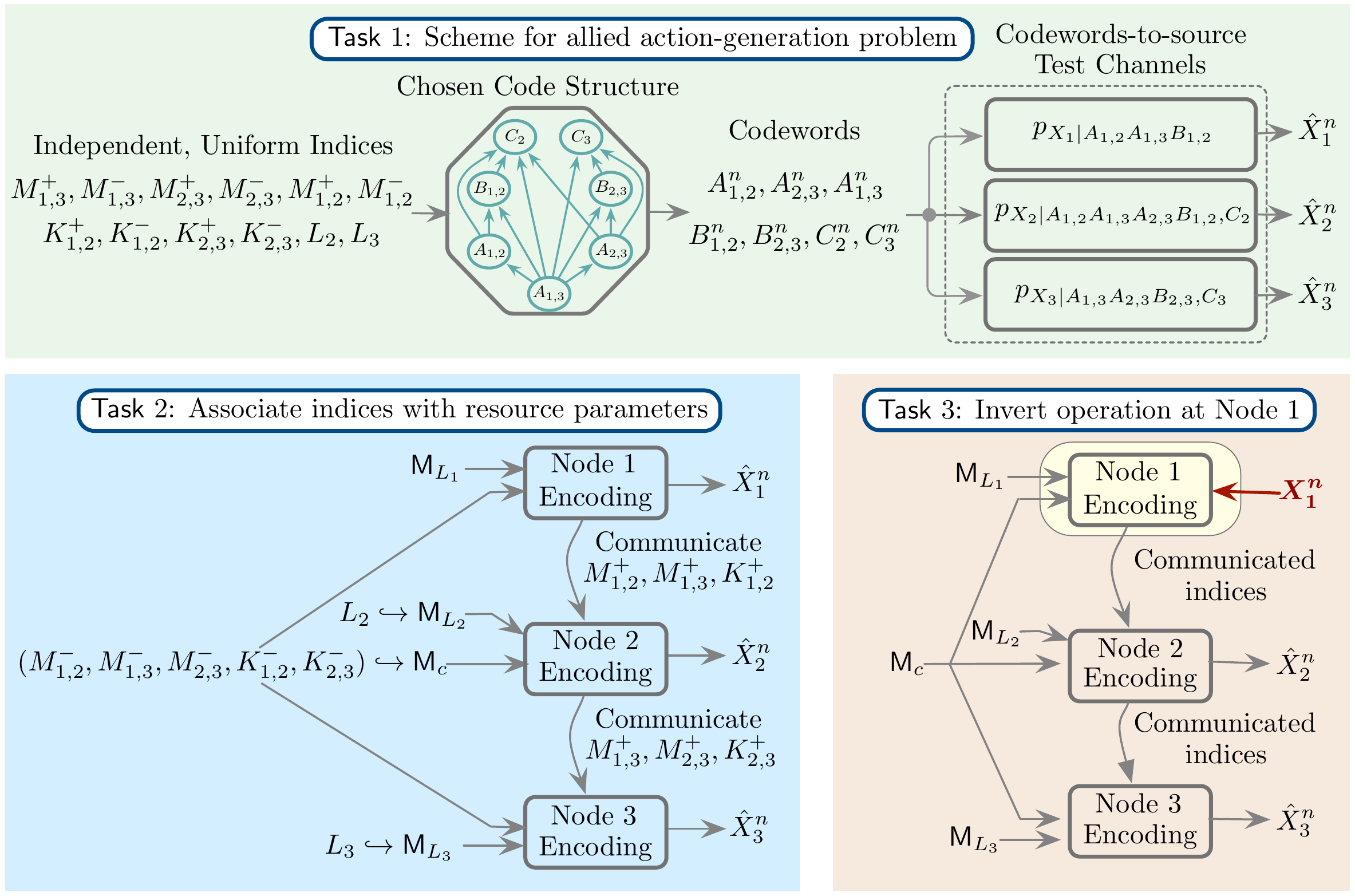}
 \vspace{-2mm}
\caption{Transforming a solution to the problem of generating $\h$ sources to one for the strong coordination problem.}
 \label{Fig-3}
\end{figure}

The approach for the design of strong coordination codes combines ideas from channel resolvability codes~\cite{TSH-Book, PC-GenCorRVs, PC-HP-TC-CordCap} and  channel synthesis~\cite{PC-DistChanSynth}.  In order to design a strong coordination code, we look at an allied problem of generating $\h$ actions $\hat{X}_1^n,\ldots,\hat{X}_\h^n$ from uniform and independent random variables (a.k.a. indices) such that the joint pmf of the generated actions ${Q}_{\hat{X}_1^n\cdots \hat{X}_\h^n}$ satisfies:
\begin{align}
\vnm{{Q}_{\hat{X}_1^n\cdots \hat{X}_\h^n}-\mathsf Q_{X_1\cdots X_\h}^{\otimes n}}_1 \leq \ve.
\end{align}
The approach in this work is the design of the strong coordination code using channel resolvability codes, and has three major tasks as described below.
\begin{itemize}
\item \underline{\textsf{Task} 1}: The first task is to devise a scheme to generate the $\h$ sources, which is termed as the \emph{allied action-generation problem}. To do so, first, a suitable structure of auxiliary RVs is chosen, and a codebook structure based on the chosen auxiliary RV structure is constructed. Independent and uniformly distributed indices are used to select the codewords from the codebook, and appropriate \emph{test channels} are used to generate the $\h$ actions satisfying the above strong coordination requirement. Note that the auxiliary RV and codebook structure, and the corresponding test channels must be such that actions are generated in a distributed fashion. 
\item \underline{\textsf{Task} 2}: The next task is to assign subsets of indices as common randomness, local randomness at each node, and messages to be communicated between nodes.
\item \underline{\textsf{Task} 3}: The last task is to then invert the operation at Node 1, which transforms the operation of generating the action at Node 1 to generating the messages intended of communication from the specified action. 
\end{itemize}
An illustration of the three steps for the three-node setting is given in Fig.~\ref{Fig-3}. Note that much of the detail presented therein such as the exact structure and form of the auxiliary RVs, test channels and the assignments to the network resources (communication, local randomness and common randomness rates) will be presented in due course. Note that in the figure $X \hookrightarrow Y$ indicates that the RV $X$ is associated as a part of the RV $Y$, and hence a part of $Y$ is used to realize $X$. As of now, the figure is only intended to indicate the overall procedure. However, we will repeatedly refer back to this figure (and the tasks) as we develop various technical aspects of the strong coordination scheme. Let us now proceed with the details of this scheme.

%---------------------------------------------------------------------------------------------------------------------
%---------------------------------------------------------------------------------------------------------------------

\subsubsection{Choice of Auxiliary Random Variables (\textsf{Task} 1)}
  We use $\binom{n}{2}+2\h-2$ auxiliary RVs in a specific way to generate the $\h$ actions in the allied problem. For an illustration of the auxiliary RV structure for $\h=3$ for the allied action-generation problem, the reader is directed to Fig.~\ref{Fig-4}.The details of this auxiliary RV structure is as follows:
%---------------------------------------------------------------------------------------------------------------------
\begin{figure}[!h]
\centering
\vspace{0mm}
 \includegraphics[width=4.5in]{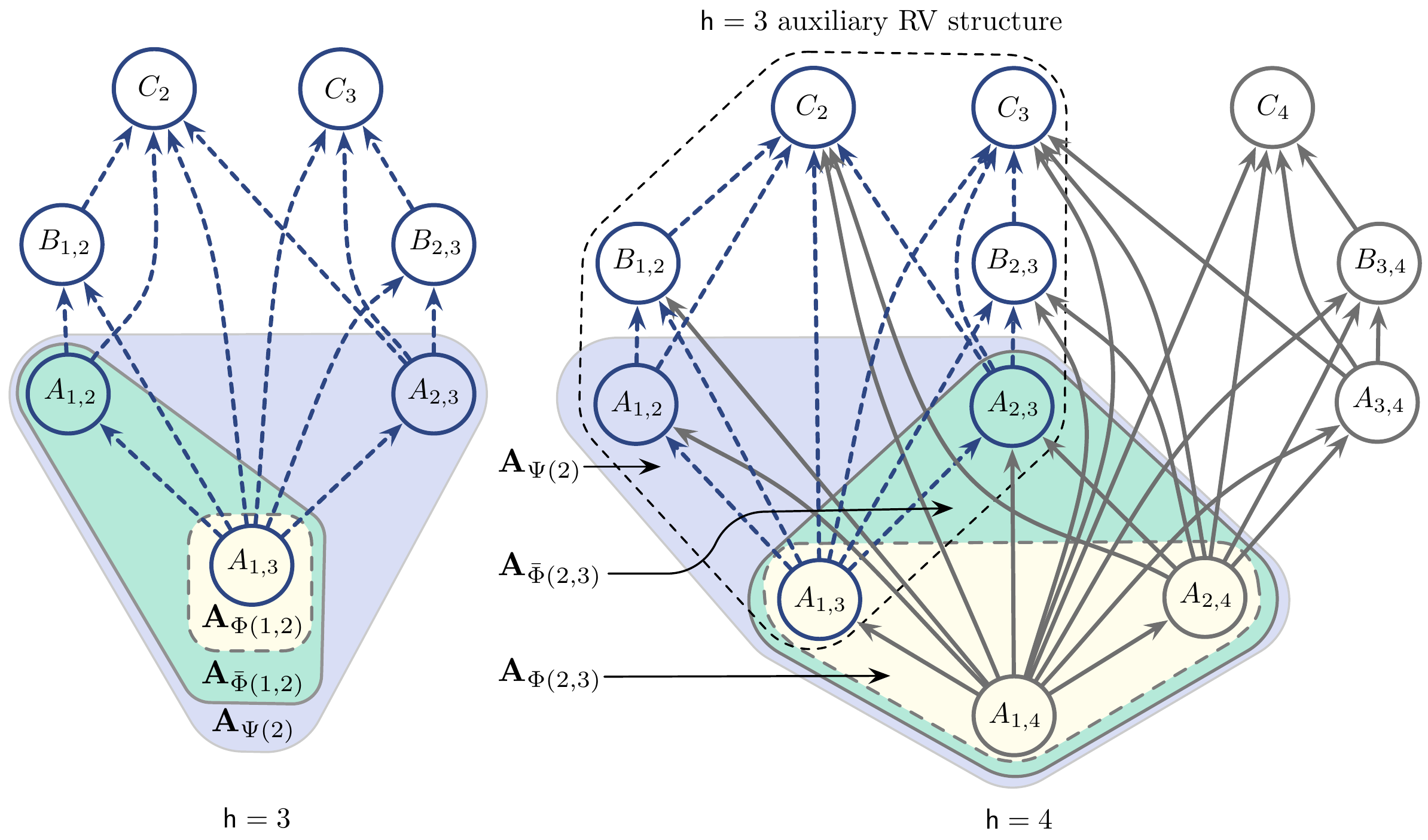}
 \vspace{-2mm}
\caption{An illustration of the structure of auxiliary random variables when $\h=3,4$.}
 \label{Fig-4}
 \vspace{-4mm}
\end{figure}
%---------------------------------------------------------------------------------------------------------------------
\begin{itemize}
\item There are three different groups of Auxiliary RVs collectively indicated by the letters $A, B$, and $C$. There are $\binom{\h}{2}$ random variables $\{A_{i,j}: 1\leq i<j\leq \h\}$, $\h-1$ random variables $\{B_{i,i+1}:1\leq i <\h\}$, and $\h-1$ random variables $\{C_i:1\leq i \leq \h\}$.
\item For $1\leq i<j\leq \h$, auxiliary RV $A_{i,j}$ represents message generated at Node $i$ intended for Node $j$ and hence can be used by Nodes $i,i+1,\ldots,j$. 
\item  For each $1\leq i<j\leq \h$, we impose the following Markov chain.
\begin{align} 
A_{i,j} \mkv \mbf A_{\Phi(i,j)} \mkv\left\{A_{i',j'}: 1\leq i'<j'\leq \h\right\}\setminus \{A_{i,j}\}, \label{eqn-ARVReln1}
\end{align}
where we let 
\begin{align}
\Phi(i,j) &\triangleq \left\{ (i',j'): i' \leq i < j \leq j' \} \setminus \{(i,j)\right\}\\
\mbf A_{S} &= \{A_s\}_{s\in S}, \quad S\subseteq \{(i,j): 1\leq i <j\leq \h\}
\end{align}
Note that $\mbf A_{\Phi(i,j)}$ represent messages that are generated by nodes prior to Node $i$ and are intended for nodes situated after Node $j$. Hence, each Node $k$, $i\leq k \leq j$, plays a role in communicating $\mbf A_{\Phi(i,j)}$ to their intended nodes, Nodes $i,\ldots,j$ will have access to the codewords corresponding to $\mbf A_{\Phi(i,j)}$. Hence, we can allow arbitrary correlation between  $A_{i,j}$ and $A_{\Phi(i,j)}$. For each $(i',j') \notin \overline{\Phi}(i,j)\triangleq \Phi(i,j)\cup\{(i,j)\}$, there is at least one $k'$ such that $i\leq k' \leq j$ and Node $k'$ does not have access to $A_{i',j'}$. Hence, only those joint pmfs that satisfy these Markov chains are amenable for constructing codebooks. Fig.~\ref{Fig-4} presents an illustration of $\mbf A_{\Phi(1,2)}$ and $A_{\Phi(2,3)}$ when $\h=3$ and $\h=4$, respectively. In the figure, the subset of auxiliary RVs that are connected to each auxiliary RV  $A_{i,j}$ collectively form $\mbf A_{\Phi(i,j)}$. Further, \eqref{eqn-ARVReln1} also implies that for any $1\leq i< j\leq \h$ and $1\leq i'< j'\leq \h$ such that $(i,j)\neq (i',j')$,
\begin{align}
A_{i,j} \mkv \mbf A_{\Phi(i,j)\cap \Phi(i',j')} \mkv A_{i',j'}. 
\end{align} 
Pictorially, what this condition translates to is that two auxiliary RVs $A_{i,j}$ and $A_{i',j'}$ are conditionally independent given the subset of all auxiliary RVs that are connected to both of them.
\item For each $i=1,\ldots,\h-1$, auxiliary RV $B_{i,i+1}$ is generated by Node $i$ using all its available messages and is intended for Node $i+1$ one hop away. At a first glance, it may seem that $A_{i,i+1}$ and $B_{i,i+1}$ play the same role, and hence only one of them need to be present. However, this is not true and Example~\ref{Example-1} below sheds light on why these auxiliary RVs play distinct roles. 

As will become clear at the end of this section, in the allied action-generation problem, random variables $\{A_{i,j}: 1\leq i < j \leq \h\}$ are determined prior to the generation of any action sequence, i.e., $\hat X_2^n,\ldots, \hat X_{\h}^n$.  However, for each $1\leq i<\h$, $B_{i,i+1}$ will be identified after the action $\hat X_i^n$ is generated. This will become evident as we describe the joint pmf of the actions and the auxiliary RVs, and the order of codebook generation. 
\item Lastly, the $\h-1$ auxiliary RVs $\{C_2,\ldots,C_{\h}\}$ are introduced to quantify the use of local randomness required at each node in the network. The messages corresponding to the codewords of these auxiliary RVs are not communicated between nodes.
\end{itemize}

Now that we have loosely defined the roles of the auxiliary RVs, we define the joint pmf of the actions and auxiliary RVs $Q_{A_{1,2},\ldots,A_{\h-1,\h},B_{1,2},\ldots,B_{\h-1,\h},C_2,\ldots,C_\h,X_1,\ldots X_\h}$ that we aim to emulate to be
%\begin{align}
%Q_{A_{1,2}\cdots A_{\h-1,\h}}Q_{X_1|\mbf A_{\Psi(1)}}Q_{B_{1,2}|X_{1} \mbf A_{\overline \Phi(1,2)}}\,{\prod\limits_{j=2}^{\h}}  \left(Q_{C_j| \mbf A_{\Psi(i)}B_{j-1,j}}Q_{X_{j}| \mbf A_{\Psi(i)}B_{j-1,j} C_j}Q_{B_{j,j+1}|X_{j} \mbf A_{\overline \Phi(i,i+1)}}\right),\label{eqn-AuxRVs1}
%\end{align}
\begin{align}
Q_{A_{1,2}\cdots A_{\h-1,\h}}Q_{X_1|\mbf A_{\Psi(1)}}\,{\prod\limits_{j=1}^{\h-1}}  \left(Q_{B_{j,j+1}|X_{j} \mbf A_{\overline \Phi(j,j+1)}} Q_{C_{j+1}| \mbf A_{\Psi(j+1)}B_{j,j+1}}Q_{X_{j+1}| \mbf A_{\Psi(j+1)}B_{j,j+1} C_{j+1}}\right),\label{eqn-AuxRVs1}
\end{align}
where $Q_{A_{1,2},\ldots, A_{\h-1,\h}}$ satisfies the conditions described by \eqref{eqn-ARVReln1}. In the above equation, 
\begin{align}
\overline\Phi(i,j)&\triangleq \Phi(i,j) \cup \{(i,j)\},\\
\Psi(i)&\triangleq \{(i',j'): i'\leq i\leq j', i' \neq j'\}.
\end{align}
Note that $A_{\overline\Phi(i,j)}$ is exactly the set of all $A$-auxiliary RVs that Nodes $i,\ldots,j$ have access to, and $A_{\Psi(i)}$ represents all the auxiliary RVs that are generated in Nodes $1,\ldots,i$ that are intended for Nodes $i,\ldots, \h$. An illustration of $\Phi(1,2)$, $\overline\Phi(1,2)$, $\Psi(2)$ for $h=3$ and $\Phi(2,3)$, $\overline\Phi(2,3)$, $\Psi(3)$ for $h=4$ can be found in Fig.~\ref{Fig-4}.
Note that the choice of the auxiliary RVs in \eqref{eqn-AuxRVs1} must be such that
\begin{align}
Q_{X_1\cdots X_\h}=\mathsf Q_{X_1\cdots X_\h}.
\end{align}
We would like to remark that ideally, it is preferable that there be only one RV per hop that encapsulates the role of the message that is conveyed on a hop. However, we do not have the necessary tools to establish that, and the joint pmf in \eqref{eqn-AuxRVs1} is the most general structure of RVs for which we are able to develop an achievable scheme. Here's an example to illustrate the difference between $A_{i,i+1}$ and $B_{i,i+1}$ in this joint pmf. 
\begin{example}\label{Example-1}
Let $\h=3$. Suppose that we build a scheme with auxiliary RVs $A_{1,2}, A_{1,3}, A_{2,3}$, i..e, we set $B_{1,2}$ and $B_{2,3}$ as constant RVs. Then the joint pmf that we can emulate is given by
\begin{align}
Q_{A_{1,2}A_{1,3},A_{2,3}} Q_{X_1|A_{1,2}A_{1,3}} Q_{X_2|A_{1,2}A_{1,3},A_{2,3}}Q_{X_3|A_{1,3},A_{2,3}}.
\end{align}
Since we have $A_{1,2} \mkv A_{1,3} \mkv A_{2,3}$ we see that the joint pmf can be rearranged as
\begin{align}
Q_{A_{1,3}} Q_{X_1,A_{1,2}|A_{1,3}} Q_{X_2|A_{1,2}A_{1,3},A_{2,3}}Q_{A_{2,3}, X_3|A_{1,3}}.
\end{align}
In other words, we have $I(X_1; X_3|A_{1,3}) = 0$. Therefore, when $B_{1,2}$ and $B_{2,3}$ are set as constant RVs,  $X_1$ and $X_3$ must be conditionally independent given $A_{1,3}$, which is a restriction on the choice of $A_{1,3}$.

Now, consider the case that $A_{1,2}$ and $A_{2,3}$ are constant RVs. Then the joint pmf that we can emulate is given by
\begin{align}
&Q_{A_{1,3}} Q_{X_1|A_{1,3}} Q_{B_{1,2} | X_1,A_{1,3}} Q_{X_2|B_{1,2},A_{1,3}}Q_{B_{2,3}|X_{2},A_{1,3}}Q_{X_3|B_{2,3},A_{1,3}}\\
&=Q_{A_{1,3},X_1,B_{1,2}} Q_{X_2|B_{1,2},A_{1,3}}Q_{B_{2,3}|X_{2},A_{1,3}}Q_{X_3|B_{2,3},A_{1,3}}.
\end{align}
It can be seen that when $A_{1,2}$ and $A_{2,3}$ are constant RVs, $X_1 \mkv A_{1,3} \mkv X_3$ need not hold. Hence, employing non-trivial $(B_{1,2}, B_{2,3})$ or $(A_{1,2},A_{2,3})$ allows for different choices for $A_{1,3}$, which in turn could potentially translate into different resource requirements. $\hfill${\tiny$\blacksquare$}
\end{example}
While the above example illustrates the difference, a discussion on the need for the $B$ auxiliary RVs is presented at the end of this work in Section~\ref{sec-essentiality of Bs}. 

Note that we eventually require that the actions be generated in a distributed fashion at various network nodes. This requirement is incorporated into the structure of the joint pmf of \eqref{eqn-AuxRVs1} via the following conditional independence property.
\begin{remark} \label{rem-2-distributedgen}
For any joint pmf of \eqref{eqn-AuxRVs1}, we have for $i=1,\ldots,\h$:
\begin{align}
X_i \mkv (A_{\Psi(i)},B_{i-1,i}, B_{i,i+1}) &\mkv (X_1,\ldots, X_{i-1}, X_{i+1},\ldots,X_\h, A_{1,2},\ldots, A_{\h-1,\h}, B_{1,2},\ldots, B_{\h-1,\h}).
\end{align}
Further, for $j=1,\ldots, \h-1$,
\begin{align}
X_{j+1} &\mkv (A_{\Psi(j+1)}, X_{j}) \mkv (X_1,\ldots, X_{i-1},A_{1,2},\ldots, A_{\h-1,\h})\\
C_{j+1} & \mkv (A_{1,2},\ldots, A_{\h-1,\h}, B_{j,j+1}) \mkv (X_1,\ldots, X_j).
\end{align}
Thus, if we provide each Node $i$ with an instance of auxiliary RVs $(A_{\Psi(i)},B_{i-1,i}, B_{i,i+1})$ that are jointly correlated according to the marginal derived from \eqref{eqn-AuxRVs1}, then we can generate actions $(X_1,\ldots, X_\h)$ in a \emph{distributed} fashion.$\hfill${\tiny$\blacksquare$}
\end{remark}

We now present the precise codebook structure and construction that we use to emulate \eqref{eqn-AuxRVs1}. 

\subsubsection{Codebook Construction (\textsf{Task} 1)}
Since we have to develop codebooks that incorporate the specific structure of auxiliary RVs, we use the following ordering of pairs of indices to construct the codebooks. We define for $i<j$, $i'<j'$ and $(i,j) \neq (i',j')$, the following relation to define the codebooks in Fig.~\ref{Fig-4}.
\begin{align}
(i,j) \succ (i',j')\quad \Leftrightarrow \quad i< i'  \textrm{ or }  (i=i' \textrm{ and } j\geq j').
\end{align}
This relation induces the following total ordering on the pairs of node indices.
\begin{align}
(1,\h) \succ (1,\h-1) \succ \cdots \succ (1,2) \succ (2,\h) \succ \cdots\succ (2,3) \succ (3,\h) \succ \cdots \succ (3,4) \succ \cdots \succ (\h-1,\h).
\end{align}

The codebooks are constructed using the above order starting from the leftmost index pair. To define the codebooks, we define the rates for each codewords as in Table~\ref{tab:0-rates}. Notice that we assign two rates for each of the codebooks for $A$- and $B$-auxiliary RVs and only one for the $C$-codebook. For each of the $A$- and $B$-auxiliary RV codebooks, one of the rates (with the superscript $+$) will correspond to communicated messages, and the other (with the superscript $-$), will not. The rates corresponding to the superscript $-$ will eventually be interpreted as common randomness.
\begin{table}[th!]
\caption{Codebook Parameters and Notation}\label{tab:0-rates}\vspace{-4mm}
\begin{center}
\begin{tabular}[c]{||c|c|c||}
\hline
Auxiliary RV & \multicolumn{2}{c||}{Rates, alphabets of indices, and codebook indices}\\
\hline
\multirow{3}{*}{$\begin{array}{c}
A_{i,j}\\ (1\leq i < j \leq \h) \end{array}$} & \multicolumn{2}{c||}{$(\mu_{i,j}^\ps, \mu_{i,j}^\ms) \in (0,\infty)\times (0,\infty)$}\\
\cline{2-3}
& $\begin{array}{r}
\mathcal{M}_{i,j}^\ps \triangleq \llbracket 1, 2^{n\mu_{i,j}^+}\rrbracket \\
\mathcal{M}_{i,j}^\ms \triangleq \llbracket 1, 2^{n\mu_{i,j}^-}\rrbracket
\end{array}$ & $\begin{array}{rl} \mathcal{M}_{i,j}^{\pms} &\hspace{-2.5mm}\triangleq \mathcal{M}_{i,j}^\ps\times \mathcal{M}_{i,j}^\ms\\
{m}_{i,j}^\pms &\hspace{-2.5mm}\triangleq (m_{i,j}^\ps, m_{i,j}^\ms)\in \mathcal{M}_{i,j}^{\pms}
\end{array}$\\
\hline\hline
\multirow{3}{*}{$\begin{array}{c}
B_{i,i+1}\\ (1\leq i < \h) \end{array}$} & \multicolumn{2}{c||}{$(\kappa_i^\ps, \kappa_i^\ms) \in (0,\infty)\times (0,\infty)$}\\
\cline{2-3}
& 
 $\begin{array}{r}
\mathcal{K}_{i}^\ps \triangleq \llbracket 1, 2^{n\kappa_i^+}\rrbracket\\
\mathcal{K}_{i}^\ms \triangleq \llbracket 1, 2^{n\kappa_i^-}\rrbracket
\end{array}$ 
& $\begin{array}{rl} 
\mathcal{K}_{i}^{\pms} &\hspace{-2.5mm}\triangleq \mathcal{K}_{i}^\ps\times \mathcal{K}_{i}^\ms \\
{k}_i^\pms &\hspace{-2.5mm}\triangleq (k_i^\ps, k_i^\ms)\in \mathcal{K}_{i}^{\pms}
\end{array}$\\
\hline\hline
\multirow{3}{*}{$\begin{array}{c}
C_i\\ (1< i  \leq \h) \end{array}$} & \multicolumn{2}{c||}{$\lambda_i \in (0,\infty)$}\\
\cline{2-3}
&  \multicolumn{2}{c||}{$\begin{array}{rl}
\mathcal{L}_{i} &\hspace{-2.5mm}\triangleq \llbracket 1, 2^{n\lambda_i}\rrbracket\\
l_i &\hspace{-2.5mm}\in \mc L _i
\end{array}$}\\
\hline\hline
\end{tabular}
\end{center}
\vspace{-2mm}
\end{table}

Let us introduce this last set of notation to make our description easier to follow.
\begin{align}
\mathcal{M}_{S}^\pms &\triangleq\substack{\displaystyle\Times\\ (i',j')\in S} \Big(\mathcal{M}_{i',j'}^\ps\times \mathcal{M}_{i',j'}^\ms\Big),  \quad S\subseteq\{(i,j):1\leq i < j \leq \h\},\\
\mbf{m}_{S}^\pms &\triangleq \{ m_{s}^\pms\}_{s\in S}, \quad S\subseteq\{(i,j):1\leq i < j \leq \h\},\\
\boldsymbol{m}^\pms&\triangleq ( m_{1,2}^\pms,\ldots, m^\pms_{\mathsf h-1,\mathsf h}),\\
\boldsymbol{m}^\ps&\triangleq ( m_{1,2}^\ps,\ldots, m^\ps_{\mathsf h-1,\mathsf h}),\\
\boldsymbol{m}^\ms&\triangleq ( m_{1,2}^\ms,\ldots, m_{\mathsf h-1,\mathsf h}^\ms).
\end{align}

We now proceed to discuss the exact steps in the codebook construction.
\begin{itemize}
\item[\textbf{\texttt{A1}}] For each $m_{1,\h}^\pms\in \mc M_{1,\h}^+\times \mathcal{M}_{1,\h}^-$, generate codeword $A_{1,\h}^n(m_{1,\h}^\pms)$ randomly using  $Q_{A_{1,\h}}$. 
\item[\textbf{\texttt{A2}}] For each $1\leq i<j\leq \h$ and $1\leq i'<j'\leq \h$ such that $(i,j) \succ (i',j')$, the codebook for $A_{i,j}$ is constructed before the codebook for $A_{i',j'}$. By design, the codebook for $A_{i',j'}$ is constructed after the codebooks for $A_{i'',j''}$, $(i'',j'')\in \Phi(i',j')$.
\item[\textbf{\texttt{A3}}] For each $1\leq i<j\leq \h$,  $\mbf m_{\overline{\Phi}(i,j)}^\pms\in \mc M^\pms_{\overline \Phi(i,j)}$, generate codeword $A_{i,j}^n(\mbf m_{\overline{\Phi}(i,j)}^\pms)$ randomly using $Q_{A_{i,j}|\mbf A_{\Phi(i,j)}}$, and previously chosen $\mbf A_{{\Phi}(i,j)}^n(\mbf m_{\Phi(i,j)}^\pms)$.
\item[\textbf{\texttt{A4}}] For each $1\leq i< \h$, $(\mbf m_{\overline{\Phi}(i,i+1)}^\pms, k_i^\pms)\in\mc M^\pms_{\overline\Phi(i,i+1)}\times \mc K_i^\pms$, generate codeword $B_{i,i+1}^n(\mbf m_{\overline\Phi(i,i+1)}^\pms,k_i^\pms)$ using $Q_{B_{i,i+1}|\mbf A_{\overline\Phi(i,i+1)}}$, and previously chosen $\mbf A_{\overline{\Phi}(i,i+1)}^n(\mbf m_{\overline\Phi(i,i+1)}^\pms)$.
\item[\textbf{\texttt{A5}}] For each $1<i \leq \h$, $(\mbf m_{\Psi(i)}^\pms,\mbf k_{i-1}^\pms,l_i)\in\mc M_{\Psi(i)}^\pms \times \mathcal{K}_{i-1}^\pms\times \mc{L}_i$, generate codeword      $C_{i}^n(\mbf m_{\Psi(i)}^\pms,k_{i-1}^\pms,l_i)$ using $Q_{C_{i}|\mbf A_{\Psi(i)}B_{i-1,i}}$, and previously chosen $\mbf A_{{\Psi}(i)}^n(\mbf m_{\Psi(i)}^\pms)$ and $B_{i-1,i}^n(\mbf m_{\Psi(i)}^\pms, k_{i-1}^\pms)$.
\end{itemize}

\begin{figure}[!t]
\centering
\vspace{0mm}
 \includegraphics[width=4.5in]{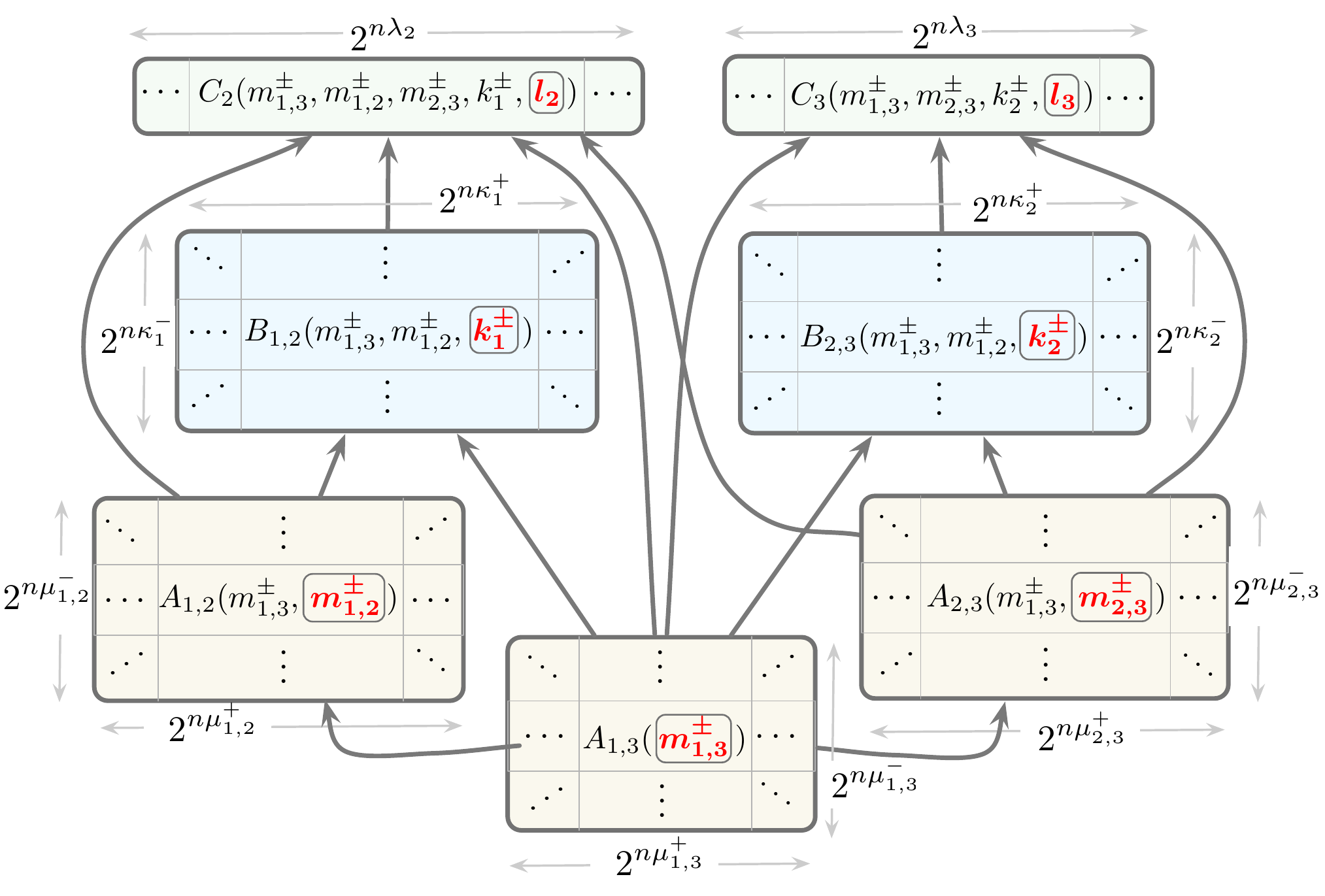}
 \vspace{-3mm}
\caption{An illustration of the structure of codebooks when $\h=3$.}
 \label{Fig-4a}
 \vspace{-5mm}
\end{figure}

An illustration of the structure of the codebooks for $\h=3$ is given in Fig.~\ref{Fig-4a}, where an incoming arrow indicates that a codebook is constructed conditionally on all the codewords of the codebooks from which there are incoming arrows.  For example, the codebook for $A_{1,2}(m_{1,3},m_{1,2})$ is constructed conditionally on the codeword $A_{1,3}(m_{1,3})$ in the $A_{1,3}$-codebook, since $\mbf A_{\Phi(1,2)}=A_{1,3}$, and the codebook for $C_2$ is constructed conditionally on the codebooks of $(\mbf{A}_{\Psi(2)}, B_{1,2})=(A_{1,3}, A_{1,2}, A_{2,3}, B_{1,2})$. In each codebook, the index/indices in black indicate the codebook index, and the index/indices in red indicate the codeword index within the codebook. For example, for $C_2$ codebooks there are $2^{n\left(\sum\limits_{(i',j')\in\Psi(2)} (\mu_{i',j'}^++\mu_{i',j'}^-)+\kappa_{1}^++\kappa_1^-\right)}$ codebooks constructed, each with $2^{n\mathsf r\lambda_2}$ codewords, and $(\mbf{m}^\pms_{\Psi(2)},k^\pms_1)=(m^\pms_{1,3},m^{\mathsmaller \pm}_{1,2}, m_{2,3}^\pms, k^\pms_1)$ provides the $C_2$ codebook index whereas $l_2$ indicates the codeword index within this codebook.

%---------------------------------------------------------------------------------------------------------------------
%---------------------------------------------------------------------------------------------------------------------

For the sake of simplicity, we introduce the following notation.
\begin{align}
\boldsymbol{A}^n(\boldsymbol{m}^\pms)&\triangleq\big(A_{1,2}^n(\mbf m^\pms_{\overline\Phi(1,2)}),\ldots,A_{\mathsf h-1,\mathsf h}^n(\mbf m^\pms_{\overline\Phi(\mathsf h-1,\mathsf h)})\big),&\\\
B_{i-1,i}^n(\boldsymbol{m}^\pms,k^\pms_{i-1})&\triangleq B_{i-1,i}^n(\mbf m^\pms_{\overline{\Phi}(i-1,i)},k^\pms_{i-1}), &\quad i=2,\ldots,\h,\\
C_{i}^n(\boldsymbol{m}^\pms,k^\pms_{i-1},l_{i})&\triangleq C_{i}^n(\mbf m^\pms_{{\Psi(i)}},k^\pms_{i-1},l_{i}), &\quad i=2,\ldots,\h.
\end{align}
The following example (as well as Fig.~\ref{Fig-4a}) presents an illustration of the above notation.
\begin{example}
For $\h=3$, the above notation translates to the following. 
\begin{align}
\boldsymbol{A}^n(\boldsymbol{m}^\pms)&\triangleq \Big(A_{1,3}^n(m_{1,3}^\ps,m_{1,3}^\ms),\,\, A_{1,2}^n(m_{1,2}^\ps,m_{1,2}^\ms,m_{1,3}^\ps,m_{1,3}^\ms),\,\, A_{2,3}^n(m_{2,3}^\ps,m_{2,3}^\ms,m_{1,3}^\ps,m_{1,3}^\ms)\Big),\\
B_{1,2}^n(\boldsymbol{m}^\pms,k^\pms_{1}) &\triangleq B_{1,2}^n(m_{1,2}^\ps,m_{1,2}^\ms,m_{1,3}^\ps,m_{1,3}^\ms, k_{1}^\ps,k_{1}^\ms),\\
B_{2,3}^n(\boldsymbol{m}^\pms,k^\pms_{2}) &\triangleq B_{2,3}^n(m_{2,3}^\ps,m_{2,3}^\ms,m_{1,3}^\ps,m_{1,3}^\ms, k_{2}^\ps,k_{2}^\ms),\\
C_2^n (\boldsymbol{m}^\pms,k^\pms_{1},l_{2})&\triangleq C_{2}^n(m_{1,2}^\ps,m_{1,2}^\ms,m_{2,3}^\ps,m_{2,3}^\ms,m_{1,3}^\ps,m_{1,3}^\ms, k_{1}^\ps,k_{1}^\ms, l_2),\\
\qquad\quad C_2^n (\boldsymbol{m}^\pms,k^\pms_{1},l_{2})&\triangleq C_{2}^n(m_{2,3}^\ps,m_{2,3}^\ms,m_{1,3}^\ps,m_{1,3}^\ms, k_{2}^\ps,k_{2}^\ms, l_3). 
\end{align}
$\qquad\qquad\qquad\qquad\qquad\qquad\qquad\qquad\qquad\hfill\textrm{\tiny$\blacksquare$}$
\end{example}

Note that we have so far neither specified the rates $\mu_{i,j}^\ps, \mu_{i,j}^\ms$, $\kappa_{i,i+1}^\ps, \kappa_{i,i+1}^\ms$, and $\lambda_i$ in the above description, nor have we described how the codewords are going to be selected for generating the actions. In the following, we will identify the required rates so that appropriate channel resolvability code design techniques can be employed to generate the actions. 

\subsubsection{Identifying Codebook Rates (\textsf{Task} 1)}
\begin{figure}[!h]
\centering
\vspace{0mm}
 \includegraphics[width=6.5in]{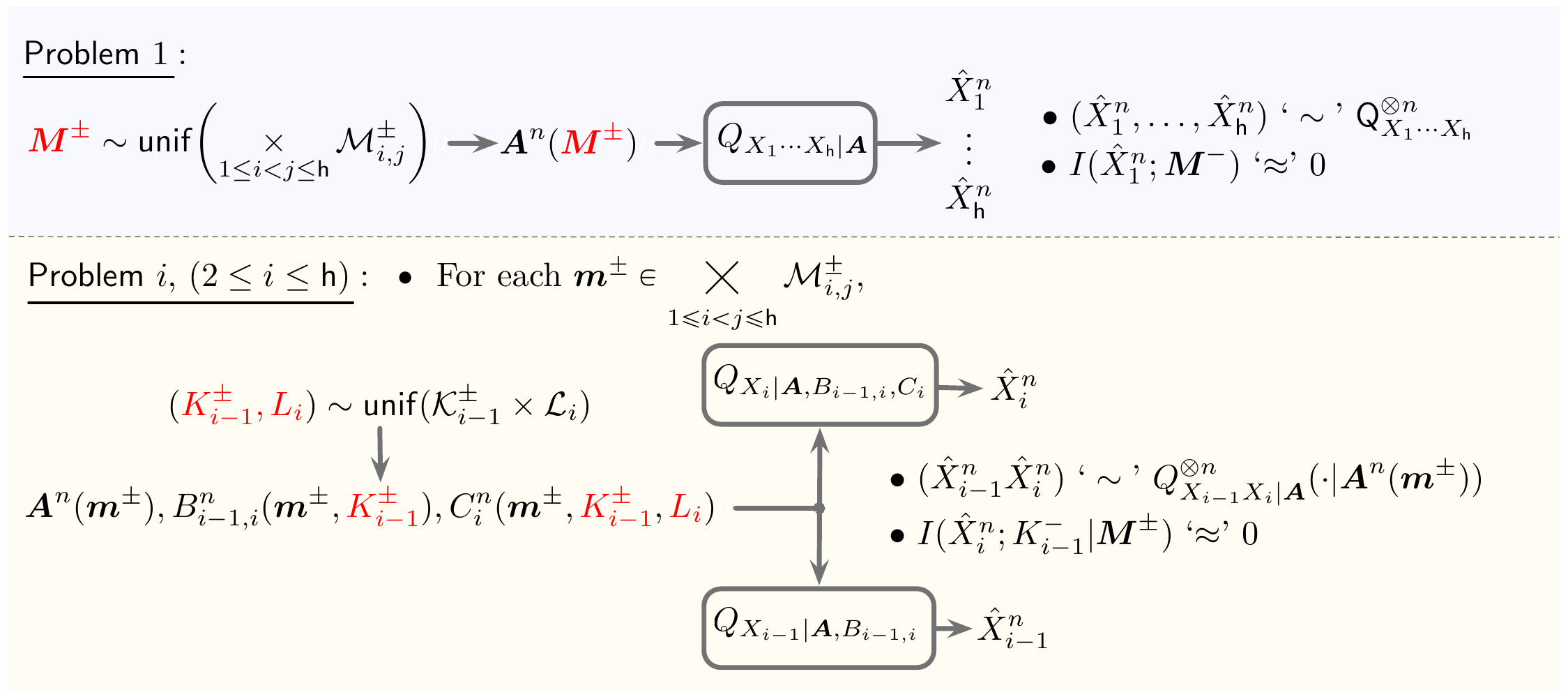}
 \vspace{-1mm}
\caption{The $\h$ subproblems.} \label{Fig-5}
 \vspace{-1mm}
\end{figure}

In order to identify the rates for the various codebooks, let us divide the allied action-generation problem into $\h$ subproblems whose solutions will be pieced together to form a solution for the allied action-generation problem.  Consider $\h$ subproblems illustrated in Fig.~\ref{Fig-5}, and formally defined below.

%---------------------------------------------------------------------------------------------------------------------
%---------------------------------------------------------------------------------------------------------------------
\begin{definition} $\mathsf{Problem }\,\,1$ pertains to characterization of the rates required for codebooks corresponding to auxiliary RVs $A_{1,2},\ldots, A_{\h-1,\h}$. Let for a given realization of the codebooks for $\{A_{i,j}: 1\leq i<j\leq \h\}$, 
\begin{align}
\widehat{Q}^{({1})}_{\hat{X}_1^n\cdots\hat{X}_\h^n}(\cdot)%&\triangleq \Exp\left[ Q_{X_1\cdots X_\h \mid \mbf{A}}^{\otimes n}(\cdot | \mbf{A}^n(\mbf M^\pms))\right]\bigg|_{\boldsymbol M^\pms\sim \mathsf{unif}{\left(\M_{1,2}^\pms\times\cdots \times \M_{\h-1,\h}^\pms\right)}}\label{eqn-Prob1distbn} \\
&\triangleq \frac{\sum\limits_{\mbf m^\pms}Q_{X_1\cdots X_\h \mid \mbf{A}}^{\otimes n}(\cdot | \mbf{A}^n(\mbf m^\pms))} {2^{n((\mu_{1,2}^++\mu_{1,2}^-)+\cdots+(\mu_{\h-1,\h}^++\mu_{\h-1,\h}^-))}}\label{eqn-Prob1distbn}\\&= \frac{\sum\limits_{\mbf m}{{\scriptstyle Q^{\otimes n}_{X_1|\mbf{A}}(\cdot|\mbf A^n (\mbf m^\pms))}}\prod\limits_{j=2}^\h \frac{ Q_{X_jX_{j-1}|\mbf A}^{\otimes n} (\cdot \mid \mbf{A}^n(\mbf m^\pms))}{Q_{X_{j-1}|\mbf{A}}^{\otimes n} (\cdot \mid \mbf{A}^n(\mbf m^\pms))}}{2^{n((\mu_{1,2}^++\mu_{1,2}^-)+\cdots+(\mu_{\h-1,\h}^++\mu_{\h-1,\h}^-))}} \label{eqn-Prob1prods}
\end{align}
denote the pmf of the output $(\hat{X}_1^n,\ldots,\hat{X}_\h^n)$ from the channel $Q_{X_1\cdots X_\h | A_{1,2}\cdots A_{\h-1,\h}}$ when the codewords are selected uniformly randomly. Note that the expectation in \eqref{eqn-Prob1distbn} is only over the uniform selection of the codewords for a given realization of the codebook. $\mathsf{Problem}$ 1 aims to derive conditions on $(\mu_{1,2}^\ps,\mu_{1,2}^\ms,\ldots, \mu_{\h-1,\h}^\ps, \mu_{\h-1,\h}^\ms)$ such that:
\begin{align}
\lim_{n\rightarrow \infty}\left(\Exp\left[\dkl(\widehat{Q}^{({1})}_{\hat{X}_1^n\cdots\hat{X}_\h^n}\parallel\mathsf{Q}_{{X}_1\cdots\hat{X}_\h}^{\otimes n})\right] \right)&=0 \label{eqn-subprobcond1.1},\\
\lim_{n\rightarrow \infty}\left(\sum_{\mbf m^\ms}\frac{\Exp\left[ \dkl(\widehat{Q}^{({1})}_{\hat{X}_1^n| \mbf M^\ms}(\cdot|\mbf m^\ms) \parallel \widehat{Q}^{({1})}_{\hat{X}_1^n})\right] }{2^{n(\mu_{1,2}^\ms+\cdots+\mu_{\h-1,\h}^\ms)}}\right)&=0,\label{eqn-subprobcond1.2}
\end{align}
where $\widehat{Q}^{({1})}_{\hat{X}_1^n| \mbf M^\ms}(\cdot|\tilde{\mbf m}^\ms) $ is the conditional pmf of $\hat X_1^n$ given $\mbf M^\ms = \tilde{\mbf m}^\ms$, and equals
\begin{align}
\widehat{Q}^{({1})}_{\hat{X}_1^n| \mbf M^-}(\cdot|\tilde{\mbf m}^\ms) &\triangleq \sum\limits_{\mbf m^\pms:\mbf m^-=\tilde{\mbf m}^\ms}\frac{Q_{X_1\cdots X_\h \mid \mbf{A}}^{\otimes n}(\cdot | \mbf{A}^n(\mbf m^\pms))} {2^{n(\mu_{1,2}^\ps+\cdots+\mu_{\h-1,\h}^\ps)}}.\label{eqn-Q1defn}
\end{align} \hfill{\textrm{\tiny{$\blacksquare$}}}
\end{definition}
Note that $\mathsf{Problem}\,1$ relates to only $A$-codebooks, and poses two constraints: 
\begin{itemize}
\item the constraint in \eqref{eqn-subprobcond1.1} requires that the outputs be statistically indistinguishable from the outputs of DMS $\mathsf Q_{X_1\cdots X_\h}$, which is a natural constraint to impose for we would like to eventually translate a solution for the allied action-generation problem to one for the strong coordination problem; and 
\item the constraint in \eqref{eqn-subprobcond1.2}, though may seem unintuitive at the moment, is essential in translating the solution of the allied action-generation problem to one for the strong coordination problem. This constraint ensures that $\mbf M^\ms\triangleq (M^{\ms}_{1,2},\ldots, M^{\ms}_{\h-1,\h})$ and $\hat X_1^n$ are (nearly) statistically independent. Later in this section, the scheme for strong coordination will equate $\mbf M^\ms$ with the randomness common to all nodes, and since common randomness and the specified action are independent of one another, we must impose this constraint explicitly while designing the codebooks. 
\end{itemize}
The following result provides a sufficient condition on the rates of the $A$-codebooks to meet the two constraints of $\mathsf{Problem}$ 1 given in \eqref{eqn-subprobcond1.1} and \eqref{eqn-subprobcond1.2}.
\begin{theorem}\label{thm-1}
Suppose that the auxiliary RV codebooks are constructed using Steps \textbf{\texttt{A1}}-\textbf{\texttt{A5}}. For each $S\subseteq \mc F \triangleq \{(i,j):1\leq i < j\leq \h\}$, let $\mathcal{J}_S \triangleq \left\{ (i,j): \overline\Phi(i,j)\cap S \neq \emptyset\right\}$. Then, \eqref{eqn-subprobcond1.1} is met provided for each $S\subseteq \mc F$ such that $(1,\h)\notin S$, 
\begin{align}
\sum_{s\in \mc F \setminus S} (\mu_s^\ps + \mu_s^\ms) & >  I\big(X_1,\ldots,X_\h; A_{\mathcal{J}_S^c}\big),\label{eqn-Req1}
\end{align}
and \eqref{eqn-subprobcond1.2} is met provided for each $S\subseteq \mc F$ such that $(1,\h)\notin S$, 
\begin{align}
\sum_{s\in \mc F \setminus S} \mu_s^\ps   &>  I\big(X_1; A_{\mathcal{J}_S^c}\big).\label{eqn-Req1.2}
\end{align}
\end{theorem}
\begin{IEEEproof}
See Appendix~\ref{App-1}. 
\end{IEEEproof}
While at first glance, it seems that $2\binom{\h}{2}$ rate conditions need to be met for $\mathsf{Problem}$ 1, as illustrated below, many equations can be redundant. 
\begin{remark}
Consider the three hop-setting, i.e., $\h=4$.  In this setting, the rate constraints imposed by \eqref{eqn-Req1} for $S=\{(1,3)\}$, $\{(1,2), (1,3)\}$, $\{(1,3),(2,3)\}$ and $\{(1,2),(1,3),(2,3)\}$ are given by 
\begin{align}
\mu_{1,4}^\ps + \mu_{1,4}^\ms+\mu_{2,4}^\ps + \mu_{2,4}^\ms+\mu_{3,4}^\ps + \mu_{3,4}^\ms+\mu_{1,2}^\ps + \mu_{1,2}^\ms+\mu_{2,3}^\ps + \mu_{2,3}^\ms&> I(\{X_j\}_{j=1}^4; A_{1,4}, A_{2,4},A_{3,4}),\\
\mu_{1,4}^\ps + \mu_{1,4}^\ms+\mu_{2,4}^\ps + \mu_{2,4}^\ms+\mu_{3,4}^\ps + \mu_{3,4}^\ms+\mu_{2,3}^\ps + \mu_{2,3}^\ms&> I(\{X_j\}_{j=1}^4; A_{1,4}, A_{2,4},A_{3,4}),\\
\mu_{1,4}^\ps + \mu_{1,4}^\ms+\mu_{2,4}^\ps + \mu_{2,4}^\ms+\mu_{3,4}^\ps + \mu_{3,4}^\ms+\mu_{1,2}^\ps + \mu_{1,2}^\ms&> I(\{X_j\}_{j=1}^4; A_{1,4}, A_{2,4},A_{3,4}),\\
\mu_{1,4}^\ps + \mu_{1,4}^\ms+\mu_{2,4}^\ps + \mu_{2,4}^\ms+\mu_{3,4}^\ps + \mu_{3,4}^\ms&> I(\{X_j\}_{j=1}^4; A_{1,4}, A_{2,4},A_{3,4}).
\end{align}
Note that the last is the only non-trivial constraint, since the first three constraints are implied by it. $\hfill${\tiny$\blacksquare$}
\end{remark}

A simple application of Pinsker's and Jensen inequalities to \eqref{eqn-subprobcond1.1} yields the following result.
\begin{remark}\label{rem-Pins1}
When the rate constraints given by \eqref{eqn-Req1} are met, we also have
\begin{alignat}{3}
\lim_{n\rightarrow\infty}\Exp\bigll \widehat{Q}^{({1})}_{\hat{X}_1^n\cdots\hat{X}_\h^n}&- \mathsf Q^{\otimes n}_{X_1\cdots X_\h} \bigll_1 =\lim_{n\rightarrow \infty}\Exp\biggll \sum\limits_{\mbf m^\pms }\frac{Q_{X_1\cdots X_\h \mid \mbf{A}}^{\otimes n}(\cdot | \mbf{A}^n(\mbf m^\pms))} {2^{\,\sum\limits_{i,j: 1\leq i < j \leq \h} n(\mu_{i,j}^\ps+\mu_{i,j}^\ms)}} -  \mathsf Q^{\otimes n}_{X_1\cdots X_\h} \biggll_1 \label{eqn-Rem1-1}
\\&=\lim_{n\rightarrow\infty}
\Exp\Bigll \frac{\sum\limits_{\mbf m^\pms}{{\scriptstyle Q^{\otimes n}_{X_1|\mbf{A}}(\cdot|\mbf A^n (\mbf m^\pms))}}\prod\limits_{j=2}^\h \frac{ Q_{X_jX_{j-1}|\mbf A}^{\otimes n} (\cdot \mid \mbf{A}^n(\mbf m^\pms))}{Q_{X_{j-1}|\mbf{A}}^{\otimes n} (\cdot \mid \mbf{A}^n(\mbf m^\pms))}}{2^{(n(\mu_{1,2}^\ps+\mu_{1,2}^\ms)+\cdots+n(\mu_{\h-1,\h}^\ps+\mu_{\h-1,\h}^\ms))}}  - \mathsf{Q}^{\otimes n}_{X_1\cdots X_\h}\Bigll_1 =0, \label{eqn-rem1}
\end{alignat}
where the expectation operator is over only the codebooks for $A_{i,j}$'s.$\hfill${\tiny$\blacksquare$}
\end{remark}
Thus, by choosing the rates of the codebooks for $A_{i,j}$'s satisfying the constraints of Theorem~\ref{thm-1}, we are guaranteed the existence of codebooks with which we can generate the $\h$ actions. At this point, it might seem that we are done with the generation of the $\h$ actions, and can proceed with identifying the indices used for codeword selection with the appropriate resources, and inverting the operation at Node $1$. However, this is not the case since we have to devise a scheme that generates the actions in a  \emph{distributed} fashion (cf. \textsf{Task} 1 of Section~\ref{Sec-Prelims}, and Fig.~\ref{Fig-3}). We do not yet have distributed generation of actions, i.e.,  the channel in Problem 1 of Fig.~\ref{Fig-5} does not decompose into $\h$ parallel channels, since
\begin{align}
Q_{X_1,\ldots, X_\h | \mbf A } = Q_{X_1 | \mbf A} \prod_{i=2}^n Q_{X_i | \mbf A, X_{i-1}}.
\end{align}
In order to decompose the channel $Q_{X_1,\ldots, X_\h | \mbf A }$ into parallel channels, we need to use the $B$ and $C$ codebooks, and we do that via the following  $\h-1$ subproblems. 

\begin{definition}
For $i=2,\ldots\h$, $\mathsf{Problem } \,\,i$  pertains to the characterization of the rates corresponding to the codebooks for auxiliary RVs $B_{1,2},\ldots, B_{\h-1,\h}$ and $C_2,\ldots, C_\h$.  Let for a given realization of all the codebooks,  $i\in\{2,\ldots,\h\}$, and $\mbf m^{\pms}\in \M_{1,2}^\pms\times \cdots \times \M_{\h-1,\h}^\pms$,
\begin{align}
{\widehat{Q}^{({i}, \mbf m^\pms)}_{\hat{X}_{i-1}^n\hat{X}_i^n|\boldsymbol{A}^n}}(\cdot)%&\triangleq  \Exp\left[ Q^{\otimes n}_{X_{i-1} X_i |\mbf AB_{i-1,i}C_{i}} (\cdot | {\scriptstyle \mbf A^n(\mbf m^\pms), B_{i-1,i}^n(\mbf m^\pms, K_{i-1}^\pms), C_{i}^n(\mbf m^\pms, K_{i-1}^\pms,L_i)})\right] \Bigg|_{\substack{(K_{i-1}^\pms,L_i)\sim \mathsf{unif} (\mc K_{i-1}^\pms\times \mc{L}_i)}}\notag\\
&= \frac{\sum\limits _{\mbf k_{i-1}^\pms,l_i} {Q}_{{X}_{i-1}{X}_i|\boldsymbol{A}B_{i-1,i} C_i}^{\otimes n}(\cdot| \mbf A^n(\boldsymbol{m}^\pms), B_{i-1,i}^n (\mbf m^\pms,k_{i-1}^\pms),  C_{i}^n(\mbf m^\pms, k_{i-1}^\pms,l_i))}{2^{n(\kappa_{i-1}^\ps+\kappa_{i-1}^\ms+\lambda_i)}} \label{eqn-Qhati+1defn}
\end{align}
denote the pmf of outputs $(\hat{X}_{i-1}^n,\hat{X}_i^n)$  conditioned on $\mbf A^n=\mbf{A}^n(\mbf m^\pms)$ (i.e., when averaged over $B_{i-1,i}$ and $C_i$ codewords corresponding to $\mbf M^\pms=\mbf m^\pms$ alone). Here, ${Q}_{{X}_{i-1}{X}_{i}|\mbf A }^{\otimes n}(\cdot | \boldsymbol A^n (\boldsymbol m))$ is computed using the design p.m.f chosen in \eqref{eqn-AuxRVs1}, and the realized codeword $\mbf A^n(\mbf m^\pms)$. Similarly, for $i\in\{2,\ldots,\h\}$, $\mbf m^{\pms}\in \M_{1,2}^\pms\times \cdots \times \M_{\h-1,\h}^\pms$ and $ k_{i-1}^\ms \in \mc K_{i-1}^\ms$, let
\begin{align}
{\widehat{Q}^{({i}, \mbf m^\pms)}_{\hat{X}_{i-1}^n|\boldsymbol{A}^n,K^{\ms}_{i-1}}}(\cdot| k_{i-1}^\ms)&\triangleq  \frac{\sum\limits _{\substack{(l_i, \tilde{ k}_{i-1}^\pms):\tilde{k}_{i-1}^\ms={ k}_{i-1}^\ms}} \hspace{-3mm}{Q}_{{X}_{i-1}|\boldsymbol{A}B_{i-1,i} C_i}^{\otimes n}(\cdot| \mbf A^n(\boldsymbol{m}^\pms), B_{i-1,i}^n (\mbf m^\pms,\tilde{ k}_{i-1}^\pms),  C_{i}^n(\mbf m^\pms, \tilde{k}_{i-1}^\pms,l_i))}{2^{n(\kappa_{i-1}^\ps+\lambda_i)}}
\end{align}
denote the  conditional pmf of $\hat X_{i-1}^n$ conditioned on $\mbf A^n=\mbf{A}^n(\mbf m^\pms)$ and $B_{i-1,i}^n = B_{i-1,i}^n (\mbf m^\pms,{ k}_{i-1}^\pms)$ (i.e., when averaged over $B_{i-1,i}$ and $C_i$ codewords corresponding to $\mbf M^\pms=\mbf m^\pms$ and $K_{i-1}^\ms =  k_{i-1}^\ms$ alone). 
Note that from \eqref{eqn-AuxRVs1}, for $i=2,\ldots,\h$, $X_{i-1}$ is conditionally independent of $C_i$ given $\mbf {A}, B_{i-1}$. Hence, we also have the following:
\begin{align}
{\widehat{Q}^{({i}, \mbf m^\pms)}_{\hat{X}_{i-1}^n|\boldsymbol{A}^n,K^{\ms}_{i-1}}}(\cdot| k_{i-1}^\ms)&=  \frac{\sum\limits _{\substack{ \tilde{ k}_{i-1}^\pms:\tilde{k}_{i-1}^\ms={ k}_{i-1}^\ms}} \hspace{-3mm}{Q}_{{X}_{i-1}|\boldsymbol{A}B_{i-1,i}}^{\otimes n}(\cdot| \mbf A^n(\boldsymbol{m}^\pms), B_{i-1,i}^n (\mbf m^\pms,\tilde{ k}_{i-1}^\pms))}{2^{n\kappa_{i-1}^\ps}}.
\end{align}

For $i=2,\ldots,\h$, $\mathsf{Problem}$ $i$ aims to derive conditions on $\kappa_{i-1}^\ps,\kappa_{i-1}^\ms, \lambda_i$ such that:
\begin{align}
&\lim_{n\rightarrow \infty} \frac{\sum\limits_{\mbf{m}^\pms}\Exp\left[\dkl\left(\widehat{Q}^{({i}, \mbf m^\pms)}_{\hat{X}_{i-1}^n\hat{X}_i^n|\boldsymbol{A}^n} \bigll  {Q}_{{X}_{i-1}{X}_{i}|\mbf A }^{\otimes n}\big(\cdot | \boldsymbol A^n (\boldsymbol m^\pms)\big)\right)\right]}{2^{n(\mu_{1,2}^\ps+\mu_{1,2}^\ms+\cdots+\mu_{\h-1,\h}^\ps+\mu_{\h-1,\h}^\ms)}}=0. \label{eqn-subprobcond2}\\
&\lim_{n\rightarrow \infty} \frac{\sum\limits_{\mbf{m}^\pms, k_{i-1}^\ms}\Exp\left[\dkl\left(\widehat{Q}^{({i}, \mbf m^\pms)}_{\hat{X}_{i-1}^n|\boldsymbol{A}^n, K_{i-1}^\ms} (\cdot| k_{i-1}^\ms) \bigll  {Q}_{{X}_{i-1}|\mbf A }^{\otimes n}\big(\cdot | \boldsymbol A^n (\boldsymbol m^\pms)\big)\right)\right]}{2^{n((\mu_{1,2}^\ps+\mu_{1,2}^\ms)+\cdots+(\mu_{\h-1,\h}^\ps+\mu_{\h-1,\h}^\ms)+ \kappa_{i-1}^\ms)}}=0. \label{eqn-subprobcond2-1}
\end{align}$\hfill${\tiny$\blacksquare$}
\end{definition}
Just like $\mathsf{Problem}$ 1, $\mathsf{Problem}$ $i$, $i=2,\ldots,\h$, also poses two constraints, each of which is similar to the corresponding constraint in Problem 1. 
\begin{itemize}
\item The first constraint enables us to approximate the conditional pmfs that appear within the product in \eqref{eqn-Prob1prods} by conditional pmfs in \eqref{eqn-Qhati+1defn} derived from outputs of suitable channel resolvability codes. 
\item The second constraint ensures for each $\mbf m^\pms$ the action $\hat X_{i-1}^n$ generated by averaging over the $B_{i-1,i}$ and $C_{i}$ codebooks is nearly independent of $K_{i-1}^\ms$. Again, the need for this constraint will be manifest when we transform the scheme for the allied action-generation problem into a scheme for strong coordination, at which time, we will view $K_{i-1}^\ms$ as a part of randomness common to all the nodes.
\end{itemize}
The following result characterizes sufficient conditions on the codebook rates for \eqref{eqn-subprobcond2} and \eqref{eqn-subprobcond2-1} to hold.
\begin{theorem}\label{thm-2}
Fix  $i\in\{2,\ldots, \h\}$. Suppose that the auxiliary RV codebooks are constructed using \textbf{\texttt{A1}}-\textbf{\texttt{A5}}. Then, \eqref{eqn-subprobcond2} is met provided the rates are chosen such that:
\begin{align}
\kappa_{i-1}^\ps +\kappa_{i-1}^\ms +\lambda_i& > I(X_{i-1}, X_i; B_{i-1,i}, C_{i}|\boldsymbol{A})\label{eqn-Prob2-h+1cond1}\\
\kappa_{i-1}^\ps+\kappa_{i-1}^\ms  & > I(X_{i-1}, X_i; B_{i-1,i}  |\boldsymbol{A}),\label{eqn-Prob2-h+1cond4}
\end{align}
and \eqref{eqn-subprobcond2-1} is met provided:
\begin{align}
\kappa_{i-1}^\ps  & > I(X_{i-1}; B_{i-1,i}  |\boldsymbol{A}).\label{eqn-Prob2-h+1cond5}
\end{align}
\end{theorem}
\begin{IEEEproof}
See Appendix~\ref{App-2}. 
\end{IEEEproof}
Analogous to Remark~\ref{rem-Pins1}, an application of Pinsker's and  Jensen inequalities to \eqref{eqn-subprobcond2} yields the following.
\begin{remark}\label{rem-Pins2}
When the rate constraints \eqref{eqn-Prob2-h+1cond1}-\eqref{eqn-Prob2-h+1cond4} are met for each $i=2,\ldots,\h$, we are guaranteed that
\begin{align}
\lim_{n\rightarrow \infty}\frac{\sum\limits_{\mbf m^\pms} \Exp\bigll {\widehat{Q}^{({i},\mbf m^\pms)}_{\hat{X}_{i-1}^n\hat{X}_i^n|\boldsymbol{A}^n}}- Q_{X_{i-1}X_i | \mbf{A}}^{\otimes n}(\cdot| \mbf A^n(\mbf m^\pms))\bigll_1}{2^{n((\mu_{1,2}^\ps+\mu_{1,2}^\ms)+\cdots+(\mu_{\h-1,\h}^\ps+\mu_{\h-1,\h}^\ms))}} =0, \label{eqn-rem2}
\end{align}
where $\widehat{Q}^{({i+1}, \mbf m)}_{\hat{X}_{i-1}^n\hat{X}_i^n|\boldsymbol{A}^n(\boldsymbol{m})}$ is defined in \eqref{eqn-Qhati+1defn}, and the outer  $\Exp$ operator operates only over the codebooks for the auxiliary RVs $A_{1,2},\ldots,A_{\h-1,\h}$.$\hfill${\tiny$\blacksquare$}\end{remark}
We are now ready to combine together the solutions for the $\h$ problems to obtain a code for generating the $\h$ sources from the auxiliary RVs via their codebooks.

\subsubsection{A Solution for the Allied Action-generation Problem (\textsf{Task} 1)}
Before we present the exact coding strategy, we first make use of the following result that enables replacing ideal conditional pmfs in the product term of \eqref{eqn-rem1} with those obtained from channel resolvability codes using the solutions to $\mathsf{Problem}$ $i$, $i=2,\ldots, \h$.  
\begin{lemma}\label{lem-1}
For the random codebook construction given in \textbf{\texttt{A1}}-\textbf{\texttt{A5}}, if the rate constraints given in Theorems~\ref{thm-1} and~\ref{thm-2} are met, then, in addition to \eqref{eqn-subprobcond1.2}, \eqref{eqn-rem1}, \eqref{eqn-subprobcond2-1}, and \eqref{eqn-rem2},  the following also holds. 
\begin{align}
 \Exp \left(\sum_{x_1^n,\ldots, x_{\h}^n}\Bigg|\mathsf Q^{\otimes n}_{X_1\cdots X_\h}(x_1^n,\ldots, x_{\h}^n) - \sum\limits_{\mbf m^\pms} \frac{Q^{\otimes n}_{X_1|\mbf A}(x_1^n| \mbf A^n (\mbf m^\pms)) \prod\limits_{j=2}^\h \frac{ \widehat{Q}^{(j,\mbf m^\pms)}_{\hat X_{j-1}^n\hat X_{j}^n | \mbf A^n}(x_{j-1}^n,x_j^n)}{ \widehat{Q}^{(j,\mbf m^\pms)}_{\hat X_{j-1}^n | \mbf A^n}(x_{j-1}^n)}}{2^{n((\mu_{1,2}^\ps+\mu_{1,2}^\ms)+\cdots+(\mu_{\h-1,\h}^\ps+\mu_{\h-1,\h}^\ms))}}  \Bigg| \right)\mathop{\longrightarrow}^{n\rightarrow \infty}0.\label{eqn-lem1-3}
\end{align} 
\end{lemma}
\begin{IEEEproof}
See Appendix~\ref{App-3}. 
\end{IEEEproof}
Let us take a closer look at the subtracted term with the summation in \eqref{eqn-lem1-3}. Suppose that the codebooks for the auxiliary RVs are constructed according to  \textbf{\texttt{A1}}-\textbf{\texttt{A5}}. Let $\boldsymbol M^\pm\sim \mathsf{unif}{\left( \mathop{\displaystyle\Times}\limits_{1\leq i<j\leq \mathsf h} \mathcal M_{i,j}^\pm\right)}$ be used to select codewords for $A_{1,2},\ldots, A_{\h-1,\h}$ from the codebooks. Then, the following observations hold.
\begin{itemize}
\item $Q^{\otimes n}_{\hat X_1|\mbf A}(x_1^n| \mbf A^n (\mbf M^\pms))$ is the channel output when $\mbf A^n (\mbf M^{\pms})$ is fed into the channel $Q_{X_1 | \mbf A}$. Note that by \eqref{eqn-AuxRVs1}, $X_1 \mkv (A_{1,2},\ldots, A_{1,\h}) \mkv \boldsymbol A$. Hence, even though we use the channel $Q_{X_1 | \mbf A}$, it is effectively $Q_{X_1 | A_{1,2},\ldots, A_{1,\h}}$, and hence it is only the codewords $A_{1,2}^n (\mbf M^{\pms}), \ldots, A_{1,\h}^m (\mbf M^{\pms})$ that determine this distribution. 
\item $\mathsf{Problem}$ $j$, $j=2,\ldots, \h$, induces a joint correlation between $\mbf M^{\pms}$, $K^{\pms}_{j-1}$, $L_j$ and actions at Node $j-1$ and $j$. Let $\widehat 
Q_{\mbf M^{\pms}, K^{\pms}_{j-1},L_j, \hat{X}_{j-1}^n, \hat{X}_{j}^n}^{[j]}$ denote this joint pmf. Note that by the setup of $\mathsf{Problem}$ $j$  (also see Fig.~\ref{Fig-5}), we have 
\begin{align}
\widehat Q_{\mbf M^{\pms}, K^{\pms}_{j-1},L_{j}}^{[j]} \sim \mathsf{unif}\left( \Big(  \mathop{\Times}\limits_{1\leq i<i'\leq \mathsf h} \mathcal M_{i,i'}^\pms\Big) \times \mc K^{\pms}_{j-1} \times \mc L_j\right).
\end{align}
Now, $Q^{\otimes n}_{ X_1|\mbf A}(x_1^n| \mbf A^n (\mbf m^\pms)) \frac{ \widehat{Q}^{(2,\mbf m^\pms)}_{\hat X_{1}^n\hat X_{2}^n | \mbf A^n}(x_{1}^n,x_2^n)}{ \widehat{Q}^{(2,\mbf m^\pms)}_{\hat X_{1}^n | \mbf A^n}(x_1^n)}$ is exactly the joint pmf of actions at Nodes 1 and 2 obtained by: (a) setting the action for Node 1 as the channel output when $\mbf A^n (\mbf m^{\pms})$ is fed into the channel $Q_{X_1 | \mbf A}$; (b) by generating an instance of $K_1^\ms \sim \mathsf{unif}(\mc K_1^\ms)$ independently and using Node 1's action, the realized $\mbf M^{\pms}$, and the generated $K_1^\ms$ to select an instance of $K_1^\ps\sim\widehat Q_{K^{\ps}_1 |  \hat{X}_1^n, \mbf M^{\pms}, K_1^{\ms}}^{[2]}$; and finally (c) by generating an instance of $L_2\sim \mathsf{unif}(\mc L_2)$ independently, and then generating Node 2's action using the indices, and the conditional pmf $\widehat Q_{\hat{X}_2^n| \mbf M^{\pms}, K^{\pms}_1,L_2}^{[2]}$ induced by $\mathsf{Problem}$ 2. Thus,
\begin{align}
 \frac{ \widehat{Q}^{(2,\mbf m^\pms)}_{\hat X_{1}^n\hat X_{2}^n | \mbf A^n}(x_{1}^n,x_2^n)}{ \widehat{Q}^{(2,\mbf m^\pms)}_{\hat X_{1}^n | \mbf A^n}(x_{1}^n)}&=  \sum_{k_1^\ps,k_1^\ms, l_2} \frac{  \widehat Q_{K^{\ps}_1 |  \hat{X}_1^n \mbf M^{\pms} K_1^{\ms}}^{[2]}(k_1^\ps|x_1^n,\mbf m^\pms, k_1^\ms) \widehat Q_{\hat{X}_2^n| \mbf M^{\pms}K^{\pms}_1L_2}^{[2]}(x_2^n |  \mbf m^{\pms}, k^{\pms}_1,l_2)}{|\mc K_1^\ms| |\mc L_2|}.  \end{align}
Since $\mathsf{Problem}$ $j$, $j>2$ is similar in setup to $\mathsf{Problem}$ 2, we can also see that for $j=3,\ldots,\h$,
\begin{align}
 \frac{ \widehat{Q}^{(j,\mbf m^\pms)}_{\hat X_{j-1}^n\hat X_{j}^n | \mbf A^n}(x_{j-1}^n,x_j^n)}{ \widehat{Q}^{(j,\mbf m^\pms)}_{\hat X_{j-1}^n | \mbf A^n}(x_{j-1}^n)}&=   \sum_{k_{j-1}^\pms, l_{j}} \frac{\Bigg( \begin{array}{l} \widehat Q_{K^{\ps}_{j-1} |  \hat{X}_{j-1}^n \mbf M^{\pms} K_{j-1}^{\ms}}^{[{j}]}(k_{j-1}^\ps|x_{j-1}^n,\mbf m^\pms, k_{j-1}^\ms) \\ \qquad\qquad\times \widehat Q_{\hat{X}_{j}^n| \mbf M^{\pms}K^{\pms}_{j-1}L_{j}}^{[{j}]}(x_{j}^n |  \mbf m^{\pms}, k^{\pms}_{j-1},l_{j})\end{array}\Bigg)}{|\mc K_{j-1}^\ms| |\mc L_{j}|}.  \end{align}
\item Piecing together the $\h-1$ conditional pmfs, we see that  $Q^{\otimes n}_{X_1|\mbf A}(x_1^n| \mbf A^n (\mbf m^\pms)) \prod\limits_{j=2}^\h \frac{ \widehat{Q}^{(j,\mbf m^\pms)}_{\hat X_{j-1}^n\hat X_{j}^n | \mbf A^n}(x_{j-1}^n,x_j^n)}{ \widehat{Q}^{(j,\mbf m^\pms)}_{\hat X_{j-1}^n | \mbf A^n}(x_j^n)}$ is the joint pmf of the $\h$ actions that are generated when the first node's action is generated as described above, and all subsequent nodes use a procedure similar to above description for the second node but with the conditional pmfs derived from the corresponding problems.
\end{itemize} 
Lastly, Lemma~\ref{lem-1} presents the conditions for the joint pmf of the actions generated by the above scheme to be close to the target pmf, leading us to the following scheme for the allied action-generation problem.
\begin{itemize}
\item[\textbf{\texttt{B1}}] Pick an auxiliary RV pmf meeting \eqref{eqn-AuxRVs1} with the additional choice that $C_{i}=X_i$, $i>1$.  Let $\{(\mu_{i,j}^\ps,\mu_{i,j}^\ms): 1\leq i< j \leq \h\}$, $\{(\kappa_{i}^\ps, \kappa_{i}^\ms): 1\leq i <\h\}$, and $\{\lambda_i: 1< i \leq \h\}$ satisfy \eqref{eqn-Req1}, \eqref{eqn-Req1.2}, and \eqref{eqn-Prob2-h+1cond1}-\eqref{eqn-Prob2-h+1cond5}.
\item[\textbf{\texttt{B2}}] Generate codebooks for $\{A_{i,j}: 1\leq i< j \leq \h\}$, $\{B_{i,i+1}: 1\leq i <\h\}$ and $\{C_i: 1< i \leq \h\}$ using Steps \textbf{\texttt{A1}}-\textbf{\texttt{A5}}.
\item[\textbf{\texttt{B3}}] Let for $i=2,\ldots, \h$, $\widehat Q_{\mbf M^{\pms}, K^{\pms}_{i-1},L_{i}, \hat{X}_{i-1}^n, \hat{X}_i^n}^{[i]}$ be the joint pmf induced by codebooks and actions induced by $\mathsf{Problem}$ $i$.
\item[\textbf{\texttt{B4}}] For $1\leq i< j \leq \h$ generate independent instances of RVs $M_{i,j}^\pms \sim \mathsf{unif}( \llbracket 1, 2^{n\mu_{i,j}^\ps} \rrbracket \times \llbracket 1, 2^{n\mu_{i,j}^\ms} \rrbracket  )$. Let $\mbf m^\pms$ be the realization of $\mbf M^\pms$.
\item[\textbf{\texttt{B5}}] Let $\hat X_1^n$ be the output of the channel $Q_{X_1 | \mbf A}$ when the input to the channel is $\mbf A^n (\mbf m^\pms)$. Let  $\hat x_1^n$ be the realization of $\hat X_1^n\sim  Q_{X_1 | \mbf A}^{\otimes n}( \cdot | \mbf A^n(\mbf m^\pms)) = Q_{X_1 | A_{1,2} \cdots A_{1,\h}}^{\otimes n} ( \cdot | A_{1,2}(\mbf m^\pms)\cdots A_{1,\h}(\mbf m^\pms))$. 
\item[\textbf{\texttt{B6}}] For $\h>i\geq 1$, we repeat the following three steps in the given order. 
\item[\textbf{\texttt{B7}}] Select $K_{i}^\ms \sim \mathsf{unif}( \llbracket 1, 2^{n\kappa_{i}^\ms} \rrbracket)$ independently and let $k_i^\ms$ be its realization. Generate an instance $k_i^\ps$ of $K_i^\ps\in \llbracket 1, 2^{n\kappa_{i}^\ps} \rrbracket$ such that $K_i^\ps \sim \widehat{Q}^{[i+1]}_{K_{i}^\ps| \hat X_i^n, \mbf M^{\pms},K_i^\ms}(\cdot | \hat{x}^n_i, \mbf m^\pms, k_i^\ms)$. 
\item[\textbf{\texttt{B8}}] Let $L_{i+1}\sim  \mathsf{unif}( \llbracket 1, 2^{n\mathsf r_{i+1}} \rrbracket) $ and $l_{i+1}$ be the realization of $L_{i+1}$.
\item[\textbf{\texttt{B9}}] Declare $\hat X_{i+1}^n \triangleq C_{i+1}^n(\mbf m^\pms, k_{i}^\ps, k_{i}^\ms, l_{i+1})$.
\end{itemize}
The following observations concerning the above scheme are now in order.
\begin{itemize}
\item Codewords for $\{A_{i,j}:1\leq i < j \leq \h\}$ are selected prior to the generation of any action, and in specific, the codewords for $\{A_{i,i+1}: 1\leq i <\h\}$ are selected prior to the generation of the $\h$ sources. However, the index and thereby the codeword for $B_{i,i+1}$ are selected \emph{after} $X_i^n$ is generated. This allows auxiliary RVs $B_{i,i+1}$ and $A_{i,i+1}$ to play distinct roles. The discussion at the end of Section~\ref{subsec-Markov} sheds more light on the need for $B_{i,i+1}$.
\item Selecting $K_i^\ms$ independent of the realization of  $\hat X_i^n$ at Step \textbf{\texttt{B7}} is possible only because of the fact that we have ensured that \eqref{eqn-subprobcond2-1} holds by choosing rates that meet \eqref{eqn-Prob2-h+1cond5}.
\item RVs $\mbf {M}^\pms$, $\{K_{i}^\pms:1\leq i <\h\}$ and $\{L_i: 2\leq i \leq \h\}$ are not independent. While $\mbf{M}^\pms$ and each $K_i^\pms$ are statistically independent, $\mbf M^\pms$, $K_i^\pms$ and $K_{i+1}^\pms$ together can possibly be dependent. This joint pmf is defined implicitly by the selection at Step \textbf{\texttt{B7}}.
\end{itemize}
We are now ready to state the achievable scheme for strong coordination using the scheme in \textbf{\texttt{B1}}-\textbf{\texttt{B9}} for the allied action-generation problem.

\subsubsection{A Scheme for Strong Coordination and the Resources Required (\textsf{Task}  2 and \textsf{Task} 3)}\label{Subsec-strngcoordcode}
Now that we have essentially completed the design of a scheme that generates the $\h$ actions, we are done with \textsf{Task} 1 described in Fig.~\ref{Fig-3}. Before we present the details of  \textsf{Task} 2 of Fig.~\ref{Fig-3} that relates to identifying the resources used for corresponding codebook rates, we detail \textsf{Task} 3 that relates to inverting the operation at Node 1 to generate the messages from the specified action. The strong coordination scheme derived from the above action-generation scheme is as follows. 
\begin{itemize}
\item[\textbf{\texttt{C1}}] Pick a pmf meeting \eqref{eqn-AuxRVs1} with the additional constraint that $C_{i}=X_i$ for $i>1$.  Let $\{(\mu_{i,j}^\ps,\mu_{i,j}^\ms): 1\leq i< j \leq \h\}$, $\{(\kappa_{i}^\ps, \kappa_{i}^\ms): 1\leq i <\h\}$, and $\{\lambda_i: 1< i \leq \h\}$ satisfy \eqref{eqn-Req1}, \eqref{eqn-Req1.2}, and \eqref{eqn-Prob2-h+1cond1}-\eqref{eqn-Prob2-h+1cond5}.

\item[\textbf{\texttt{C2}}] Generate codebooks for $\{A_{i,j}: 1\leq i< j \leq \h\}$, $\{B_{i,i+1}: 1\leq i <\h\}$ and $\{C_i: 1< i \leq \h\}$ using Steps \textbf{\texttt{A1}}-\textbf{\texttt{A5}}.
\item[\textbf{\texttt{C3}}] Let $\widehat Q_{\hat X_1^n, \mbf M^{\pms}}^{[1]}$ be the joint p.m.f induced by the codebooks  in  $\mathsf{Problem}$ 1 (see \eqref{eqn-Q1defn}). For $i=2,\ldots, \h$, let $\widehat Q_{\mbf M^{\pms}, K^{\pms}_{i-1}1,L_{i}, \hat{X}_{i-1}^n, \hat{X}_i^n}^{[j]}$ be the joint pmf induced by codebooks and actions in $\mathsf{Problem}$ $i$.
\item[\textbf{\texttt{C4}}] Generate an instance $(\{m_{i,j}^\ms\}_{1\leq i <j \leq \h}, \{k_{i}^\ms\}_{1\leq i <\h})$ of RVs\ $(\{M_{i,j}^\ms\}_{1\leq i <j \leq \h}, \{K_{i}^\ms\}_{1\leq i <\h})$ such that 
\begin{align}
 (\{M_{i,j}^\ms\}_{1\leq i <j \leq \h}, \{K_{i}^\ms\}_{1\leq i <\h})\sim \mathsf{unif}\left(\mathop{\Times}_{1\leq i <j\leq \h} \llbracket 1, 2^{n\mu_{i,j}^\ms}\rrbracket \times \mathop{\Times}_{1\leq \ell <\h} \llbracket 1, 2^{n\kappa_\ell^\ms}\rrbracket\right). \label{eqn-commonrandompart}
\end{align} 
These RVs are assumed to be extracted from the common randomness available to all nodes. \color{black}
 \item[\textbf{\texttt{C5}}] Generate an instance $\{m_{i,j}^\ps: 2\leq i< j \leq \h\}$ of $\{M_{i,j}^\ps\}_{2\leq i< j \leq \h} \sim \mathsf{unif}\left(\mathop{\displaystyle\Times}\limits_{2\leq i< j \leq \h} \llbracket 1, 2^{n\mu_{i,j}^\ps} \rrbracket \right)$. 
 
 \item[\textbf{\texttt{C6}}]  Given $X_1^n = x_1^n$, generate an instance of indices $(M_{1,2}^\ps,\ldots, M_{1,\h}^\ps)$ such that 
 \begin{align}
  (M_{1,2}^\ps,\ldots, M_{1,\h}^\ps) \sim \widehat{Q}^{[1]}_{M_{1,2}^\ps\cdots M_{1,\h}^\ps|\hat X_1^n,M_{1,2}^\ms,\ldots,M_{1,\h}^\ms} (\cdot | x_1^n, m_{1,2}^\ms,\ldots,m_{1,\h}^\ms).
  \end{align}
   Let $\mbf m^\pms$ be the instance of the realized and generated indices in Steps \textbf{\texttt{C4}}-\textbf{\texttt{C6}}.
\item[\textbf{\texttt{C7}}]  For $i\geq 1$ we repeat the following three steps in the given order. 
\item[\textbf{\texttt{C8}}] Generate a realization $k_i^\ps$ of $K_i^\ps\in \llbracket 1, 2^{n\kappa_{i}^\ps} \rrbracket$ such that $K_i^\ps \sim \widehat{Q}^{[i+1]}_{K_{i}^\ps| \hat X_i^n, \mbf M^{\pms},K_i^\ms}(\cdot | \hat{x}^n_i, \mbf m^\pms, k_i^\ms)$. 
\item[\textbf{\texttt{C9}}] Node $i$ forwards $\{m_{i',j'}^\ps: i'\leq i<j'\}$,  and $k_{i}^\ps$ to Node $i+1$.
\item[\textbf{\texttt{C10}}] Let $l_{i+1}$ be a realization of $L_{i+1}\sim  \mathsf{unif}( \llbracket 1, 2^{n\lambda_{i+1}} \rrbracket)$ selected independent of all other RVs.
\item[\textbf{\texttt{C11}}] Declare $\hat X_{i+1}^n \triangleq C_{i+1}^n(\mbf m^\pms, k_{i}^\ps, k_i^\ms, l_{i+1})$.
\end{itemize}
What remains now is the task of associating the rates of the various indices to that of the resources (i.e., communication rates, and (local and common) randomness rates), which is the \textsf{Task} 2 described in Fig.~\ref{Fig-3}. Note that this association is dependent on the mode of intermediate note operation illustrated in Fig.~\ref{Fig-2}. We first present the association for the functional and unrestricted cases, and then for the action-dependent mode. They are as follows.

(a) \textbf{Functional Mode:}
In this mode, local randomness at each node (other than Node 1) is only used to generate the corresponding actions. This imposes two constraints: 
\begin{itemize}
\item[1.] since the  generation of intermediate node actions uses local randomness, and since $B_{i,i+1}$, $i>1$ generally depends on the action (see \eqref{eqn-AuxRVs1} and Step \textbf{\texttt{C8}}),  $B_{i,i+1}$ has to be set as a constant RV.  Since $B_{1,2}$ can be subsumed in $A_{1,2}$, without loss of generality, we may assume that $B_{1,2}$ is also a constant. Thus, $\kappa_{i} = 0$, $i=1,\ldots, \h-1$; and 
\item[2.] the randomness involved in determining the indices $\{M_{i,j}^\ps: 1<i<j\leq \h\}$ in Step  \textbf{\texttt{C5}} must come from either the incoming message or the common randomness. Consider $M_{i,j}^\ps$ for $1<i<j$. If a part of randomness comes from communication from Node $i-1$ and a part from common randomness, it must be that Node $i-1$ is also aware of $M_{i,j}^\ps$. Thus, Node $i-1$ is also aware of the exact codeword chosen for $A_{i,j}$. Proceeding inductively, we can argue that Node $1$ must be aware of the codeword selected for $A_{i,j}$. Thus, for each $j=3,\ldots,\h$, we can embed auxiliary RVs $A_{2,j},\ldots,A_{j-1,j}$ into auxiliary RV $A_{1,j}$ without affecting the communication, local randomness or common randomness requirements. Thus, for $2\leq i <j$, we can set $A_{i,j}$ as constant RVs, and $\mu_{i,j}^\ps = \mu_{i,j}^\ms=0$. 
\end{itemize}
Thus, the only auxiliary RVs that remain in the system are $A_{1,2},\ldots, A_{1,\h}$ and $C_i=X_i$ for $i=2,\ldots,\h$, and therefore we need to only assign resources for indices $M_{1,2}^\ps, M_{1,2}^\ms, \ldots, M_{1,\h}^\ps, M_{1,\h}^\ms$ and $L_2,\dots, L_\h$. Notice that common randomness is used to generate $M_{1,2}^\ms,\ldots, M_{1,\h}^\ms$  in Step \textbf{\texttt{C4}}, and hence,
\begin{align}
\mathsf R_c \triangleq \mu_{1,2}^\ms + \cdots + \mu_{1,\h}^\ms. \label{eqn-FuncRc}
\end{align}
Indices $M_{1,2}^\ps,\ldots, M_{1,\h}^\ps$ are determined at Node $1$ and for $\ell=2,\ldots, \h$, $M_{1,\ell}^\ps$ is communicated to Node $\ell$ by hop-to-hop communication (see Step \textbf{\texttt{C9}}). Thus,
\begin{align}
\mathsf{R}_\ell \triangleq  \mu_{1,\ell+1}^\ps + \cdots + \mu_{1,h}^\ps, \quad 1\leq \ell <\h. \label{eqn-FuncRl}
\end{align}

Now to identify the rate of local randomness required at each node, notice that Nodes $2,\ldots, \h$ generate their actions using Step  \textbf{\texttt{C11}}, which in turn requires $L_2,\ldots, L_\h$ to be generated in Step \textbf{\texttt{C10}}. Since these indices are not communicated, without loss of generality, we can assume that they are generated using local randomness in Step  \textbf{\texttt{C10}}, and hence,
\begin{align}
\rho_i \triangleq \lambda_i, \quad 1<i\leq\h.
\end{align}

Lastly, to identify the rate of local randomness required at Node $1$, observe that the local randomness at Node $1$, unlike other nodes, is not used to generate the action. Instead, it is used to select the codewords for $(A_{1,2},\ldots, A_{1,h})$ by selecting $(M_{1,2}^\ps,\ldots, M_{1,\h}^\ps)$ at Step  \textbf{\texttt{C5}}. The amount of local randomness required to make this selection is exactly quantified by Theorem~\ref{thm-reversechannelres} of Appendix~\ref{app-revchan1}, and is given by:

\begin{align}
\rho_1 \triangleq \sum_{j=2}^\h \mu_{1,j}^\ps - I(X_1; A_{1,2},\ldots,A_{1,\h}).\label{eqn-Funcrho1}
\end{align}
A summary of the common and local randomness required to implement the various steps in the functional mode is presented in Table~\ref{tab:1-func}.

\begin{table}[th!]
\caption{Allocation of resources to indices in the functional mode}\label{tab:1-func}
\vspace{-2mm}
\begin{center}
\begin{tabular}[c]{||c|c|c|c|c||}
\hline
Step &\makecell{Operation}  &Codebook indices &   \makecell{Required randomness} &  \makecell{Resource used \\ for  the operation} \\ \hline \hline
\textbf{\texttt{C4}}  & \makecell{Select indices for $A_{1,2},\ldots, A_{1,\h}$} &$\{M_{1,j}^\ms: 1<j\leq\h\}$ &   $ \mu_{1,2}^\ms+\cdots+\mu_{1,\h}^\ms $ &$\mathsf R_c$\\ \hline
\textbf{\texttt{C6}} &\makecell{Generate indices for \\ $A_{1,2},\ldots,A_{1,\h}$ at Node 1} & $M_{1,2}^\ps, \ldots, M_{1,\h}^\ps$ &   $ \sum\limits_{j=2}^\h \mu_{1,j}^\ps - I(X_1; \{A_{1,\ell}\}_{\ell=2}^\h) $ &$\rho_1$\\ \hline
\textbf{\texttt{C10}} & \makecell{For $2\leq \ell \leq\h$, select index\\ for  $C_{\ell}$ at Node $\ell$} &\makecell{$L_{\ell}$}  &    \makecell{$\lambda_{\ell}$ } &\makecell{$\rho_{\ell}$}\\ \hline
\end{tabular}
\end{center}
\vspace{-1mm}
\end{table}

(b) \textbf{Unrestricted Mode:}  Since there are no restrictions in this mode, we need to assign resources for indices $\{(M_{i,j}^\ps, M_{i,j}^\ms): 1\leq i<j\leq \h\}$, $\{(K_{i}^\ps, K_i^\ms):1\leq i<\h\}$ and $L_2,\dots, L_\h$. As before, in Step \textbf{\texttt{C4}}, the common randomness is used to select $\{ M_{i,j}^\ms: 1\leq i<j\leq \h\}$, $\{K_i^\ms:1\leq i<\h\}$, and  therefore,
\begin{align}
\mathsf R_c &\triangleq   (\kappa_{1}^\ms+\cdots+ \kappa_{\h-1}^\ms)+ \sum_{(i,j):1\leq i <j\leq \h} \mu_{i,j}^\ms. \label{eqn-UnresRc}
\end{align}
The communication requirements are determined in Step \textbf{\texttt{C9}}. For each $i=1,\ldots, \h-1$ and $i <\ell \leq \h$, index $M_{i,\ell}^\ps$ is determined by Node $i$ and is forwarded by hop-by-hop communication over to Node $\ell$. Further, for each $i=1,\ldots, \h-1$, Node $i$ also determines index $K_{i}^\ps$ and forwards it to Node $i+1$. Hence, the communication rate requirements between adjacent nodes are given by
\begin{align}
\mathsf{R}_i \triangleq \kappa_{i}^\ps + \sum_{i',j':\,i' \leq i < j'} \mu_{i',j'}^\ps, \quad 1\leq i <\h.  \label{eqn-UnresRi}
\end{align}

Now, to determine the amount of local randomness at each node, we see that at Node 1 we need to use local randomness to generate indices $M_{1,2}^\ps,\ldots, M_{1,\h}^\ps$ and $K_{1}^\ps$ corresponding to auxiliary RVs $A_{1,2},\ldots, A_{1,\h}$ and $B_{1,2}$, respectively, using the given realization of $X_1^n$, and the common realization. The amount of local randomness required to generate these indices is quantified by Theorem~\ref{thm-reversechannelres} of Appendix~\ref{app-revchan1}. 

For $2\leq i < \h$, the local randomness required at Node $i$ has three purposes: (1) to select messages $\{M_{i, j}^\ps: j>i\}$, which requires a rate of $\mu_{i,j}^\ps$ (see \textbf{\texttt{C5}});  (2) to generate $K_{i}^\ps$ at Step \textbf{\texttt{C8}}, which requires a rate of $\kappa_{i}^\ps - I (X_i; B_{i,i+1}|\mbf A)$ quantified by Theorem~\ref{thm-reversechannelres1} of Appendix~\ref{app-revchan2}; and lastly, (3) to generate $L_i$ for use in generating the Node $i$'s action, which requires a rate $\lambda_i$.  At Node $\h$, local randomness is only needed to generate $L_\h$ to output action $\hat X_\h^n$, which requires a rate $\lambda_\h$ specified in Step \textbf{\texttt{C2}}. Combining these arguments, we see that the rates of local randomness at nodes are given by:
\begin{align}
\rho_\ell &\triangleq \left\{ \begin{array}{ll} \sum_{j=2}^\h \mu_{1,j}^\ps+\kappa_{1}^\ps - I (X_1; A_{1,2}, \ldots, A_{1,\h}, B_{1,2}), & \ell =1 \\ 
\kappa_{\ell}^\ps - I(X_\ell; B_{\ell,\ell+1} | \mbf{A})+ \lambda_\ell+ \sum_{j=\ell+1}^\h \mu_{\ell,j}^\ps, & 1< \ell <\h \\ 
\lambda_\h, & \mathsf \ell=\h\end{array} \right..  \label{eqn-Unresrhoi}
\end{align}
A summary of the common and local randomness required to implement steps in the unrestricted mode is presented in Table~\ref{tab:3-unrest}.

\begin{table}[th!]
\caption{Allocation of resources to indices in the unrestricted mode}\label{tab:3-unrest}\vspace{-2mm}
\begin{center}
\begin{tabular}[c]{||c|c|c|c|c||}
\hline
Step &\makecell{Operation}  &Codebook indices &   \makecell{Required randomness} &  \makecell{Resource used \\ for  the operation} \\ \hline \hline
\textbf{\texttt{C4}}  & \makecell{Select indices for \\ $\{A_{i,j}: 1\leq i<j\leq\h\}$\\ and $\{B_{i,i+1}:1\leq i <\h\}$} &\makecell{$\{M_{i,j}^\ms: 1\leq i<j\leq\h\}$\\ $\{K_i^\ms: 1\leq i <\h\} $} &   $ \sum\limits_{i,j:1\leq i<j\leq\h} \mu_{i,j}^\ms + \sum\limits_{\ell=1}^{\h-1} \kappa_{\ell}^\ms$ &$\mathsf R_c$\\ \hline
\textbf{\texttt{C5}}  &\makecell{For $1<\ell<\h$, select indices\\ for $\{A_{\ell,j}: \ell<j\leq\h\}$ at Node $i$} & $\{M_{\ell,j}^\ps: \ell<j\leq\h\}$ &  $ \sum\limits_{j=\ell+1}^\h \mu_{\ell,j}^\ps $ &$\rho_\ell$\\ \hline
\textbf{\texttt{C6}} &\makecell{Generate indices for \\ $A_{1,2},\ldots,A_{1,\h}$} & $M_{1,2}^\ps, \ldots, M_{1,\h}^\ps$ &   $ \sum\limits_{j=2}^\h \mu_{1,j}^\ps - I(X_1; \{A_{1,\ell}\}_{\ell=2}^\h) $ &$\rho_1$\\ \hline
\textbf{\texttt{C8}} &\makecell{For $1\leq\ell<\h$, generate indices\\ for  $B_{\ell,\ell+1}$ at Node $\ell$} & $K_{\ell}^\ps$ &   $\kappa_{\ell}^\ps - I (X_\ell; B_{\ell,\ell+1}|\mbf A)$ &$\rho_\ell$\\ \hline
\textbf{\texttt{C10}} & \makecell{For $2\leq \ell\leq \h$, select index\\ for $C_{\ell}$ at Node $\ell$} &\makecell{$L_{\ell}$}  &    \makecell{$\lambda_{\ell}$ } &\makecell{$\rho_{\ell}$}\\ \hline
\end{tabular}
\end{center}\vspace{-2mm}
\end{table}

(c) \textbf{Action-dependent Mode:} 
In this mode, as in the case of the functional mode, an intermediate node's local randomness must be used only to generate the corresponding actions, but not to generate the next-hop messages (see Fig.~\ref{Fig-2}). Therefore, the randomness required to implement Steps  \textbf{\texttt{C5}} and  \textbf{\texttt{C8}} must be extracted from either the incoming message and the common randomness. Since the incoming message itself cannot use the local randomness of the node that generated it, the randomness usable from the incoming message must indeed originate from Node 1. Thus, in this mode: 
\begin{itemize}
\item we can, just as in the functional mode, embed auxiliary RVs $A_{2,3},\ldots, A_{\h-1,\h}$ into auxiliary RVs $A_{1,2},\ldots, A_{1,\h}$. Thus, without loss of generality, we may assume that for  $1<i<j\leq \h$,  $A_{i,j}$ is a constant RVs, and $\mu_{i,j}^\ps = \mu_{i,j}^\ms = 0$.
\item for $1\leq\ell<\h$, the randomness required for generating the index $K_{\ell}^\ps$ at Node $\ell$ (during Step \textbf{\texttt{C8}})  is quantified by Theorem~\ref{thm-reversechannelres1} of Appendix~\ref{app-revchan2} to  be $\kappa_{\ell}^\ps - I (X_\ell; B_{\ell,\ell+1}|\mbf A)$. This randomness can be, without loss of generality, assumed to be obtained from Node 1's local randomness via hop-by-hop communication.
\end{itemize} 
However, the amount of local and common randomness needed for the following three steps are identical to those needed in the unrestricted setting:
\begin{itemize}
\item in Step \textbf{\texttt{C4}} for selecting $\{M_{1,\ell}^\ps: 1<\ell \leq \h\}$ and $\{K_{i}^\ps: 1\leq i < \h\}$ at  intermediate nodes;
\item in Step \textbf{\texttt{C6}} for generating $\{M_{1,2}^\ps, \ldots, M_{1,\h}^\ps\}$ at Node 1 using the realized $X_1^n$; and 
\item in Step \textbf{\texttt{C10}}: selecting $\{L_{2},\ldots, L_\h\}$ at Nodes $2,\ldots,\h$.
\end{itemize}
Thus, the amount of common randomness required in this mode is given by 
\begin{align}
\mathsf R_c \triangleq  (\kappa_{1}^\ms+\cdots+ \kappa_{\h-1}^\ms)+(\mu_{1,2}^\ms+ \cdots + \mu_{1,\h}^\ms).\label{eqn-ADRc}
\end{align}
The indices to be communicated between Node $i$ and Node $i+1$ include $\{M_{1,\ell}: i+1\leq \h\}$ and the part of Node 1's randomness used to implement Step  \textbf{\texttt{C8}} at Node $\ell$ for $i+1\leq \h$. Hence,
 \begin{align}
 \mathsf R_i = \sum_{\ell \geq i+1} (\mu_{1,\ell}^\ps + \kappa_{\ell}^\ps -  I (X_\ell; B_{\ell,\ell+1}|\mbf A)).
 \end{align}
Lastly, since the local randomness of Nodes $2,\ldots, \h$ are used only to generate their respective actions, 
\begin{align}
\rho_i \triangleq \lambda_i, \quad i=2,\ldots, \h.
\end{align} 
However, the local randomness in Node 1 must be used to determine the indices $M_{1,2}^\ps,\ldots, M_{1,\h}^\ps$ in Step \textbf{\texttt{C6}}, and $K_1^\ps,\ldots, K_{\h-1}^\ps$ in Step \textbf{\texttt{C8}}. Hence,
\begin{align}
\rho_1 \triangleq \sum_{j=2}^\h \mu_{1,j}^\ps - I (X_1; A_{1,2}, \ldots, A_{1,\h}) +  \sum_{\ell =1}^{\h-1} (\kappa_{\ell}^\ps -  I (X_\ell; B_{\ell,\ell+1}|\mbf A)). \label{eqn-ADrho1}
\end{align}
A summary of the common and local randomness required to implement various steps in the action-dependent mode is presented in Table~\ref{tab:2-actdep}.

\begin{table}[h!]
\caption{Allocation of  resources to indices in the action-dependent mode}\label{tab:2-actdep}\vspace{-2mm}
\begin{center}
\begin{tabular}[c]{||c|c|c|c|c||}
\hline
Step &\makecell{Operation}  &Codebook indices &   \makecell{Required randomness} &  \makecell{Resource used \\ for  the operation} \\ \hline \hline
\textbf{\texttt{C4}}  & \makecell{Select indices for $\{A_{1,j}: 1<j\leq\h\}$\\ and $\{B_{i,i+1}:1\leq i <\h\}$} &\makecell{$\{M_{i,j}^\ms: 1\leq i<j\leq\h\}$\\ $\{K_i^\ms: 1\leq i <\h\} $} &   $ \sum\limits_{j:1<j\leq\h} \mu_{1,j}^\ms + \sum\limits_{\ell=1}^{\h-1} \kappa_{\ell}^\ms$ &$\mathsf R_c$\\ \hline
\textbf{\texttt{C6}} &\makecell{Generate indices for $A_{1,2},\ldots,A_{1,\h}$} & $M_{1,2}^\ps, \ldots, M_{1,\h}^\ps$ &   $ \sum\limits_{j=2}^\h \mu_{1,j}^\ps - I(X_1; \{A_{1,\ell}\}_{\ell=2}^\h) $ &$\rho_1$\\ \hline
\textbf{\texttt{C8}} &\makecell{For $1\leq\ell<\h$, generate indices\\ for  $B_{\ell,\ell+1}$ at Node $\ell$} & $K_{\ell}^\ps$ &   $\kappa_{\ell}^\ps - I (X_\ell; B_{\ell,\ell+1}|\mbf A)$ &\makecell{$\rho_1$}\\ \hline
\textbf{\texttt{C10}} & \makecell{For $2\leq \ell\leq\h$, select index\\ for $C_{\ell}$ at Node $\ell$} &\makecell{$L_{\ell}$}  &    \makecell{$\lambda_{\ell}$ } &\makecell{$\rho_{\ell}$}\\ \hline
\end{tabular}
\end{center}\vspace{-2mm}
\end{table}

\subsubsection{Inner Bound}
Now that we have completed the three steps, we can implicitly state the general inner bound to the capacity region that is achievable by the above scheme as follows. 
\begin{itemize}
\item The portion of the capacity region achievable by the functional mode version of the above scheme is given by the conditions in \eqref{eqn-FuncRc}-\eqref{eqn-Funcrho1} along with the rate-transfer arguments allowed by Lemma~\ref{eqn-ADratetransfer}.
\item The portion of the capacity region achievable by the action-dependent mode version of the above scheme is given by the conditions in \eqref{eqn-ADRc}-\eqref{eqn-ADrho1} along with the rate-transfer arguments allowed by Lemma~\ref{eqn-ADratetransfer}.
\item In the unrestricted mode, an inner bound to the capacity region is given by the conditions in \eqref{eqn-UnresRc}-\eqref{eqn-Unresrhoi} along with the rate-transfer arguments allowed by Lemma~\ref{eqn-URratetransfer}.
\end{itemize}

\subsection{Functional-mode Capacity Region}
We begin this section with the capacity result characterizing the tradeoffs among common randomness, local randomness and communication rates to establish strong coordination using exclusively the functional mode of intermediate node operation.  
 \begin{theorem}\label{Res:MainFuncAchThm}
A rate point $(\mathsf R_c, \mathsf R_1,\ldots, \mathsf R_{\h-1}, \rho_1,\ldots, \rho_\h)$  is achievable with the functional mode \emph{if and only if} there exist auxiliary RVs $Z_2,\ldots, Z_h$ jointly correlated with the actions according to pmf $Q_{X_1,\ldots, X_\h, Z_{2}, Z_{3},\ldots, Z_{\h}}$ such that:
\begin{itemize}
\item[1.] $Q_{X_1,\ldots, X_\h} =\mathsf Q_{X_1,\ldots, X_\h}$;
\item[2.] $Q_{X_1,\ldots, X_\h, Z_{2}, Z_{3},\ldots, Z_{\h}} = Q_{Z_{2}, Z_{3},\ldots, Z_{\h}} Q_{X_1 | Z_{2} ,Z_{3},\ldots, Z_{\h}}\prod\limits_{\ell=2}^\h Q_{X_\ell | Z_{\ell},\ldots, Z_{\h}}$; and
\item[3.] for all $i=1,\ldots, \h-1$,
\begin{align}
%\rho_1 &> \mathsf{R}_1 - I(X_1; A_{1,2},\ldots, A_{1,\h})\label{eqn-funcCR1}\\
\mathsf{R}_i& \geq I (X_1; Z_{i+1}, \ldots, Z_{\h}),\label{eqn-FuncRR1stcond}\\
\mathsf R_c + \mathsf{R}_i+ \sum_{s\in S} \rho_s &\geq I (\{X_l\}_{l=1}^\h; \{Z_{\ell}\}_{\ell=i+1}^\h) + H(X_S |\{Z_{\ell}\}_{\ell=i+1}^\h), \quad S\subseteq \{i+1,\ldots, \h\},\label{eqn-FuncRR2ndcond}\\
\mathsf R_c +  \rho_1 + \sum_{s\in T} \rho_s &\geq  I (X_2,\ldots, X_\h; Z_{2}, \ldots, Z_{\h} , X_T | X_{1}), \quad T\subseteq \{2,\ldots, \h\}. \label{eqn-FuncRR3rdcond}
\end{align}
\end{itemize}
 \end{theorem}
 \begin{IEEEproof}
We begin with the achievability part.  Given $Q_{X_1\cdots X_\h Z_{2} Z_{3}\cdots Z_{\h}}$, consider the achievable scheme of Section~\ref{subsec-GenInnerBnd} with the following choices:
\begin{align}
A_{i,j} &\triangleq\textrm{constant}, \quad 1<i<j\leq\h,\\
B_{i,i+1}&\triangleq\textrm{constant}, \quad 1\leq i<\h,\\
A_{1,\ell }&\triangleq Z_\ell, \,\,\qquad\quad 1< \ell<\h,\\
C_{\ell }&\triangleq X_\ell, \,\qquad\quad 1< \ell<\h.
\end{align}
For this choice, the decomposition of \eqref{eqn-AuxRVs1}  aligns with the decomposition in condition 2 of Theorem~\ref{Res:MainFuncAchThm}. Using the assignments \eqref{eqn-FuncRc}-\eqref{eqn-Funcrho1} for the functional setting, we see that the following rate region is achievable.
\begin{align}
\mathsf R_c &\geq  \mu_{1,2}^\ms + \ldots + \mu_{1,\h}^\ms, \\
\mathsf R_i& \geq  \mu_{1,i+1}^\ps + \ldots + \mu_{1,\h}^\ps\,, \quad 1\leq i <\h, \\
\rho_i & \geq  \left \{\begin{array}{ll} \mu_{1,2}^\ps + \ldots + \mu_{1,\h}^\ps - I (X_1; Z_{2},\ldots, Z_{\h})  & i=1\\ H(X_i | Z_{i},\ldots, Z_{\h}) & i>1 \end{array}\right.,
\end{align}
where the code parameters $\mu_{1,2}^\ps, \ldots, \mu_{1,\h}^\ps, \mu_{1,2}^\ms, \ldots, \mu_{1,\h}^\ms$ can take non-negative values meeting the following conditions derived in Theorems~\ref{thm-1} and \ref{thm-2}:
\begin{align}
\displaystyle\sum\limits_{k=i}^\h\, (\mu_{1,i}^\ps + \mu_{1,i}^\ms) &\geq  I (X_1,\ldots, X_\h; Z_{i},\ldots, Z_{\h}),\quad 1<i \leq \h,\\
\displaystyle\sum\limits_{k=i}^\h \mu_{1,i}^\ps &\geq  I (X_1; Z_{i},\ldots, Z_{\h}),\qquad\,\,\,\quad\quad 1<i \leq \h.
\end{align}
 Now, applying the rate-transfer arguments Lemma~\ref{eqn-ADratetransfer} to the above region, we see that the achievable region includes the following region.
\begin{align}
\mathsf R_c& \geq  \mu_{1,2}^\ms + \ldots + \mu_{1,\h}^\ms+\delta_1+\ldots + \delta_h, \\
\mathsf R_i& \geq  \mu_{1,i+1}^\ps + \ldots + \mu_{1,\h}^\ps+\ve_{i+1}+\ldots \ve_{\h}, \qquad\qquad\qquad\qquad\qquad 1\leq i <\h, \\
\rho_1 &\geq  \mu_{1,2}^\ps + \ldots + \mu_{1,\h}^\ps - I (X_1; Z_2,\ldots, Z_{\h})-\delta_1 +\ve_2+\ldots +\ve_{\h},  \\
\rho_i & \geq   H(X_i | Z_{i},\ldots, Z_{\h})-\delta_i -\ve_i, \qquad\qquad\quad\qquad\qquad\qquad\qquad1< i \leq \h,\\
\displaystyle\sum\limits_{k=i}^\h \,(\mu_{1,i}^\ps + \mu_{1,i}^\ms) &\geq  I (X_1,\ldots, X_\h; Z_{i},\ldots, Z_{\h}),\,\,\,\,\qquad\qquad\quad\qquad\qquad\qquad\qquad 1<i \leq \h,\\
\displaystyle\sum\limits_{k=i}^\h \mu_{1,i}^\ps &\geq  I (X_1; Z_{i},\ldots, Z_{\h}),\qquad\qquad\,\quad\qquad\qquad\qquad\qquad\qquad\qquad 1<i \leq \h,
\end{align}
where in addition to the non-negativity constraints of the code parameters, we also impose
\begin{align}
\delta_j&\geq 0, \quad1\leq j \leq \h,\\
\ve_j&\geq 0, \quad1<j\leq \h.
\end{align}
In the above, $\delta_i$ denotes the portion of common randomness that is used only by Node $i$ as its local randomness (see the first rate-transfer condition of Lemma~\ref{eqn-ADratetransfer}), and $\ve_i$ denotes the portion of local randomness of Node 1 that is communicated to Node $i$ to be used as its local randomness (see the second rate-transfer condition of Lemma~\ref{eqn-ADratetransfer}). Finally, an application of Fourier-Motzkin elimination to dispose of the code and rate-transfer parameters yields the required result.

Now, for the converse part, let $\mathbf{R}\triangleq (\mathsf{R}_c, \mathsf{R}_1,\ldots,\mathsf{R}_{\h-1},\rho_1, \rho_2,\ldots, \rho_{\h})$ be achievable. Fix $\ve>0$ and an $\ve$-code of length $n$ operating at $\mathbf{R}$ that outputs $\hat{X}_i^n$ at Node $i$, $i>1$.  Then, 
\begin{align}
\vnm{\mathsf{Q}_{X_1}^{\otimes n}Q_{\hat{X}_2^n\cdots \hat{X}_\h^n| X_1^n}-\mathsf Q_{X_1\cdots {X}_\h}^{\otimes n}}_1 \leq \ve. \label{eqn-vardistconstmet}
\end{align}
For notational ease, denote $\hat{X}_1^n \triangleq X_1^n$. Since~\eqref{eqn-vardistconstmet} holds, we infer from~\cite[Sec.~V.$\,$A]{PC-GenCorRVs} that for any $S\subseteq\{1,\ldots,\h\}$ and $i\in \{1,\ldots,\h\}$,
\begin{align}
H(\{\hat{X}_{j}^n\}_{j\in S}) &\geq { \sum\limits_{k=1}^n} \,H(\{\hat{X}_{j,k}\}_{j\in S}) -n\delta'_{n,\ve} ,\label{eqn-vardistconstmetconseq}\\
H(\{\hat{X}_{j}^n\}_{j\in S}| \hat X_i^n) &\geq { \sum\limits_{k=1}^n} \,H(\{\hat{X}_{j,k}\}_{j\in S}|\hat X_{i,k}) -n\delta'_{n,\ve}, \label{eqn-vardistconstmetconseq1}
\end{align}
for some $\delta'_{n,\ve}\rightarrow 0$ as $\ve\rightarrow 0$. 
%Then, we can find RVs $\bar{X}_1^n\cdots \bar{X}_\h^n$ such that 
%\begin{align}
% P_{\bar{X}_j^n}-P_{X_j^n}&=0, \quad j=1,\ldots,\h\\
% \Pr[\bar{X}_j^n\neq \hat{X}_j^n]&<\ve, \quad j=2,\ldots,\h\\
%\Pr[\bar{X}_1^n\neq {X_1^n}]&<\ve
%\end{align}
%WLOG, we may also assume that these RVs are correlated as
%\begin{align}
%P_{M_c M_1\cdots M_{\h-1} X_1^n\hat{X}_2^n\cdots \hat{X}_2} \bigg(P_{\bar{X}_1^n | X_1^n} \prod_{j=2}^n P_{\bar{X}_j^n | \hat X_j^n} \bigg) \label{eqn-AuxRVchoice}
%\end{align}
%By construction, we have 
%\begin{align}
%H(\bar{X}_1^n\cdots \bar{X}_\h^n)= \sum_{i=1}^n H(\bar{X}_{1_i}\cdots \bar{X}_{\h_i}) - n \tilde{\delta}_{n,\ve}
%\end{align}
%for some $\delta_{n,\ve}\rightarrow 0$ as $n\rightarrow \infty$. 
Then, for any $i\in \llbracket 1,\h-1\rrbracket$,
\begin{align}
n\mathsf{R}_i &\geq H(\mathsf{I}_i) \geq H(\mathsf{I} _i|\mathsf{M}_c)\\
& \geq I(\hat X_1^n; \mathsf{I}_i|\mathsf{M}_c) \\ 
&\stackrel{(a)}{=} I(\hat{X}_1^n; \mathsf{I}_i,\ldots, \mathsf{I}_{\h-1}|\mathsf{M}_c)\\
&\stackrel{(b)}{=}I(\hat{X}_1^n; \mathsf{M}_c, \mathsf{I}_i,\ldots, \mathsf{I}_{\h-1})\\
&\stackrel{(c)}{=} {\sum\limits_{k=1}^n}\, I({\hat{X}_{1,k}};\mathsf{M}_c, \{\mathsf{I}_j\}_{j=i}^{\h-1}, {\hat{X}_1}^{k-1})\\
&\stackrel{(d)}{=}{\sum\limits_{k=1}^n}\, I({\hat{X}_{1,k}};\{Z_{j,k}\}_{j=i+1}^{\h})\\
&\stackrel{(e)}{=} nI\big({\hat{X}_{1,U}}; \{Z_{j,U}\}_{j=i+1}^{\h}\mid U\big)\\
&\stackrel{(f)}{=} nI\big({\hat{X}_{1,U}};\{Z_{j}^*\}_{j=i+1}^{\h}\big),
\end{align}
where 
\begin{itemize}
\item[$(a)$] follows due to the functional mode of message generation; 
\item[$(b)$] because $\hat{X}_1^n=X_1^n$, and $\mathsf{M}_c$ are independent; 
\item[$(c)$] because $\hat{X}_1^n=X_1^n$ is i.i.d.;
\item[$(d)$] by defining $Z_{j,k} \triangleq (\mathsf{M}_c,\mathsf{I}_{j-1},\hat X_1^{k-1})$ for $2\leq j\leq \h$ and $1\leq k \leq n$; 
\item[$(e)$] by defining time-sharing RV $U~\sim\mathsf{unif}(\llbracket 1,n\rrbracket)$; and 
\item[$(f)$] by setting $Z_{j}^*\triangleq (U,{Z_{j,U}})$  for $2\leq j\leq \h$, and since $\hat{X}_{1,U}$ and $U$ are independent.  
\end{itemize} 
Lastly, note that by choice, the following Markov chains hold
 \begin{align}
& \hat X_{1,U}\mkv \{Z_k^*\}_{k=2}^{\h}\mkv (\hat X_{2,U},\dots,\hat X_{\h,U}),\\
&\hat{X}_{i,U} \mkv \{Z_k^*\}_{k=i}^{\h} \mkv (\{Z_j^*\}_{j=2}^{i-1},\{\hat{X}_{j,U}: j\neq i\}), \quad 1<i\leq \h.
 \end{align}

Next, pick $i\in\{1,\ldots,\h-1\}$ and  subset $S\subseteq \llbracket i+1,\h\rrbracket$. Now, consider the following argument.
\begin{align}
n\big(\mathsf{R}_c +\mathsf{R}_i + {\sum\limits_{s\in S}} \,\rho_s\big) &\geq H(\mathsf{M}_c, \mathsf{I}_i, \{\mathsf{M}_{L_s}\}_{s\in S})\\
&\stackrel{(a)}{=} H(\mathsf{M}_c, \mathsf{I}_i,\ldots, \mathsf{I}_{\h-1},  \{\mathsf{M}_{L_s}\}_{s\in S}\{\hat{X}^n_s\}_{s\in S})\\
&\stackrel{}{\geq} I(\hat X_1^n,\ldots, \hat X_{\h}^n; \mathsf{M}_c, \mathsf{I}_i,\ldots, \mathsf{I}_{\h-1}, \{\mathsf{M}_{L_s}\}_{s\in S},\{\hat{X}^n_s\}_{s\in S})  \\
&\stackrel{(b)}{\geq} I(\hat X_1^n,\ldots, \hat X_{\h}^n; \mathsf{M}_c, \mathsf{I}_i,\ldots, \mathsf{I}_{\h-1}, \{\hat{X}^n_s\}_{s\in S})  \\
&\stackrel{(c)}{\geq} {\sum\limits_{k=1}^n}\,I\big(\{\hat X_{\ell,k}\}_{\ell=1}^\h; \mathsf{M}_c, \mathsf{I}_i,\ldots, \mathsf{I}_{\h-1},\{\hat X_\ell^{k-1}\}_{\ell=1}^\h, \{\hat{X}^n_s\}_{s\in S}\big) -n\delta'_{n,\ve} \\
&\stackrel{}{\geq} {\sum\limits_{k=1}^n}\,I\big(\{\hat X_{\ell,k}\}_{\ell=1}^\h; \{Z_{j,k}\}_{j=i+1}^{\h},\{\hat{X}_{s,i}\}_{s\in S}\big) -n\delta'_{n,\ve} \\
&\stackrel{(d)}{=}  nI\big(\{\hat X_{\ell,U}\}_{\ell=1}^\h; \{Z_{j,U}\}_{j=i+1}^{\h},\{\hat{X}_{s,U}\}_{s\in S}\mid U\big)-n \delta'_{n,\ve}\\
&\stackrel{(e)}{\geq}  n I\big(\{\hat X_{\ell,U}\}_{\ell=1}^\h; \{Z_{j}^*\}_{j=i+1}^{\h},\{\hat{X}_{s,U}\}_{s\in S}\big)  -n \delta''_{n,\ve}\\
&= nH(\{\hat{X}_{s,U}\}_{s\in S}) \hspace{-0.3mm}+\hspace{-0.3mm} nI(\{\hat X_{\ell,U}\}_{\ell=1}^\h;  \{Z_{j}^*\}_{j=i+1}^{\h}|\{\hat{X}_{s,U}\}_{s\in S})  \hspace{-0.3mm}-\hspace{-0.3mm}n \delta''_{n,\ve},
\end{align} 
where \begin{itemize}
\item[$(a)$] follows due to functional mode of message generation, and because $H(\hat{X}_j^n|\mathsf{M}_c ,\mathsf{I}_{j-1}, \mathsf{M}_{L_j})=0$ for $j>1$; 
\item[$(b)$] follows by dropping $\{\mathsf{M}_{L_s}\}_{s\in S}$ from the mutual information term;
\item[$(c)$] uses  the chain rule, and then \eqref{eqn-vardistconstmetconseq} to get rid of the conditioning; 
\item[$(d)$] follows by introducing the uniform time-sharing RV $U$; and lastly, 
\item[$(e)$] follows by defining $\delta''_{n,\ve} \triangleq \delta'_{n,\ve}+ I(\{\hat X_{\ell,U}\}_{\ell=1}^\h; U)$, and by noting that \eqref{eqn-vardistconstmet} ensures that $(\hat{X}_1^n, \ldots, \hat X_{\h}^n)$ are nearly i.i.d.,  which then implies that 
\begin{align}
I(\hat X_{1,U},\ldots, \hat X_{1,U}; U)= {H(\hat X_{1,U},\ldots, \hat X_{1,U})- H(\hat X_{1,U},\ldots, \hat X_{1,U}|U)}\stackrel{\ve\rightarrow 0}{\longrightarrow} 0. \label{eqn-Uindep}
\end{align}
\end{itemize}
Next,  consider the following argument for $S\subseteq \{2,\ldots,\h\}$.
\begin{align}
n\big(\mathsf{R}_c +\rho_1+ {\sum\limits_{s\in S}} \,\rho_s\big) &\geq H(\mathsf{M}_c,  {\mathsf{M}}_{L_1}\{\mathsf{M}_{L_s}\}_{s\in S})\\
&\geq I(\hat{X}_1^n, \ldots, \hat{X}_\h^n; \mathsf{M}_c,  {\mathsf{M}}_{L_1}\{\mathsf{M}_{L_s}\}_{s\in S})\\
&\stackrel{(a)}{\geq} I(\{\hat X_{\ell}^n\}_{\ell=2}^\h; \mathsf{M}_c,  {\mathsf{M}}_{L_1},\{\mathsf{M}_{L_s}\}_{s\in S}| \hat X_1^n)\\
&\stackrel{(b)}{=} I(\{\hat X_{\ell}^n\}_{\ell=2}^\h; \mathsf{M}_c,  {\mathsf{M}}_{L_1},\{\mathsf{M}_{L_s}\}_{s\in S},\{ \mathsf I_\ell\}_{\ell=1}^{\h-1}| \hat X_1^n)\\
&\stackrel{(c)}{=} I(\{\hat X_{\ell}^n\}_{\ell=2}^\h; \mathsf{M}_c,  {\mathsf{M}}_{L_1},\{\mathsf{M}_{L_s}\}_{s\in S}, \{ \mathsf I_\ell\}_{\ell=1}^{\h-1}, \{\hat{X}_s^n\}_{s\in S}| \hat X_1^n)\\
&\stackrel{(d)}{\geq} \sum_{k=1}^n I(\{\hat X_{\ell,k}\}_{\ell=2}^\h; \mathsf{M}_c,  {\mathsf{M}}_{L_1}, \{ \mathsf I_\ell\}_{\ell=1}^{\h-1}, \{\hat{X}_s^n\}_{s\in S},\{\hat X_\ell^{k-1}\}_{\ell=1}^\h, \hat X_1^n | \hat X_{1,k})-n\delta'_{n,\ve}\\
&\stackrel{}{\geq} \sum_{k=1}^n I(\{\hat X_{\ell,k}\}_{\ell=1}^\h; \mathsf{M}_c, \{ \mathsf I_\ell\}_{\ell=1}^{\h-1}, \{\hat{X}_{s,k}\}_{s\in S}, \hat X_1^{k-1} | \hat X_{1,k})-n\delta'_{n,\ve}\\
&\stackrel{(e)}{\geq} \sum_{k=1}^nI(\{\hat X_{\ell,k}\}_{\ell=1}^\h; \{Z_{j,k}\}_{j=2}^{\h},\{\hat{X}^n_{s,k}\}_{s\in S}| \hat X_{1,k}) -n\delta'_{n,\ve} \\
&\stackrel{(f)}{\geq}n I(\{\hat X_{\ell,U}\}_{\ell=1}^\h; \{Z_{j,U}\}_{j=2}^{\h},\{\hat{X}^n_{s,U}\}_{s\in S}| \hat X_{1,U}, U)-n\delta'_{n,\ve}  \\
&\stackrel{(g)}{\geq}n I(\{\hat X_{\ell,U}\}_{\ell=1}^\h; \{Z_{j}^*\}_{j=2}^{\h},\{\hat{X}^n_{s,U}\}_{s\in S}| \hat X_{1,U})-n\tilde\delta_{n,\ve} ,
\end{align} 
where \begin{itemize}
\item[$(a)$] follows because common and local randomness are independent of the action specified at Node 1; 
\item[$(b)$] holds since $\mathsf I_1$ is a function of $(\mathsf M_c, {\mathsf M}_{L_1}, \hat X_1^n)$, and due to the functional mode of operation at other nodes;
\item[$(c)$] follows because $H (\hat X_j^n | \mathsf M_c, {\mathsf M}_{L_j}, \mathsf I_{j-1})=0$ for $j=2,\ldots,\h$; 
\item[$(d)$] uses the chain rule, then drops $\{\mathsf M_{L_s}\}_{s\in S}$, and lastly employs \eqref{eqn-vardistconstmetconseq1} to get rid of the conditioning; 
\item[$(e)$] follows by introducing the auxiliary RVs defined earlier;
\item[$(f)$] follows by introducing the uniform time-sharing RV $U$; and lastly, 
\item[$(g)$] follows by defining $\tilde \delta_{n,\ve} \triangleq \delta'_{n,\ve}+ I(\{\hat X_{\ell,U}\}_{\ell=1}^\h; U| \hat X_{1,U})$, and from \eqref{eqn-Uindep}, we see that 
\begin{align}
I(\{\hat X_{\ell,U}\}_{\ell=1}^\h; U| \hat X_{1,U})\leq I(\{\hat X_{\ell,U}\}_{\ell=1}^\h; U) \stackrel{\ve\rightarrow 0}{\longrightarrow} 0. \label{eqn:timesharingvarremove}
\end{align}
\end{itemize}
Lastly, note that since we have the correct structure for the auxiliary RVs, we can restrict their cardinalities using  Carath\'{e}odory's theorem. The proof is then complete by limiting $\ve\rightarrow 0$, and by invoking the continuity of the information functional, the compactness of the space of joint pmfs of the actions and the auxiliary RVs, and the facts that $\delta''_{n,\ve}\rightarrow 0$ and $\tilde \delta_{n,\ve}\rightarrow 0$. 
 \end{IEEEproof}

 \begin{remark}\label{rem-3}
Unlike usual source-coding outer-bound proofs, the joint pmf of $ (\hat X_{1,U},\ldots,\hat X_{\h,U})$ \emph{depends} on $\ve$. However,  $\vnm{Q_{\hat X_{1,U},\ldots, \hat X_{\h,U}} -  \mathsf Q_{X_1\cdots X_\h}}_1 \rightarrow 0$ as $\ve\rightarrow 0$.$\hfill${\tiny$\blacksquare$}
\end{remark}
 
The above result automatically yields the tradeoffs between  between common randomness rate and communication rates  when local randomness is absent at Nodes $2,\ldots,\h$, since in this setting, the only mode intermediate nodes can operate in is the functional mode. Both the achievability and the converse arguments of Theorem~\ref{Res:MainFuncAchThm} can be modified appropriately to show the following.
\begin{remark}
When $\rho_2=\cdots=\rho_\h=0$, the tradeoffs between common randomness rate and the rates of communication are given by:
\begin{align}
\mathsf R_\ell &\geq I(X_1; X_{\ell+1},\ldots, X_\h),  \quad \ell =1,\ldots, \h-1,\\
\mathsf R_c + \mathsf R_\ell &\geq  H(X_{\ell+1},\ldots, X_\h),\,\,\quad\quad  \ell =1,\ldots, \h-1,\\
\mathsf R_c + \rho_1 &\geq  H(X_{2},\ldots, X_\h|X_1). 
\end{align}$\hfill${\tiny$\blacksquare$}
\end{remark}

 We conclude this section with the following result that argues that when the common randomness shared by all the nodes is sufficiently large, it is sufficient to focus on strong coordination schemes where intermediate nodes operate in the functional mode. Hence, the tradeoffs defined by Theorem~\ref{Res:MainFuncAchThm} approximate the strong coordination capacity region at large rates of common randomness. 

 \begin{theorem}\label{Thm-GeneralLargeRc}
Suppose that the rate of common randomness available to all nodes is sufficiently large, i.e., $\mathsf R_c> H(X_2,\ldots, X_{\h}| X_1)$. Then, the requirements for local randomness rates and communication rates are decoupled, and are given by:
\begin{align}
\rho_i &\geq 0, \qquad\qquad\qquad\qquad\quad i=1,\ldots,\h,\\
\mathsf R_i &\geq I(X_1; X_{i+1},\ldots, X_{\h}), \quad i=1,\ldots,\h-1.
\end{align}
Further, in this setting, it suffices to focus on functional schemes alone.
\end{theorem}
\begin{IEEEproof}
 For the achievability, consider the functional scheme of  Theorem~\ref {Res:MainFuncAchThm} with the following choices:
\begin{align}
A_{1,i} &= X_{i}, \quad i=2,\ldots, \h,\\
\mu_{1,i}^\ps&= \left\{ \begin{array}{ll} I(X_1; X_{i}| X_{i+1},\ldots, X_{\h}),&  i=2,\ldots, \h-1\\ I (X_1; X_\h), & i=\h \end{array} \right.,\\
\mu_{1,i}^\ms&= H(X_{i}| X_{i+1},\ldots, X_{\h}, X_1), \,\,\,\quad i=2,\ldots, \h.
\end{align} 
For this achievable scheme, we see from \eqref{eqn-FuncRc} and \eqref{eqn-FuncRl} that the rate of common randomness required is $\sum_{i=2}^\h \mu_{1,i}^\ms = H(X_2,\ldots, X_{\h}| X_1)$, and the communication rate between Node $i$ and Node $i+1$ is $I(X_1; X_{i+1},\ldots, X_{\h})$. Further, for this choice, we do not require that Nodes $1,\ldots, \h$ possess any local randomness, thus establishing the achievability of the region claimed. 

The optimality of this scheme is evident from the cut-set argument that for $1\leq i<\h$, the rate between Node $i$ and Node $i+1$ can be no smaller than the smallest rate in a one-hop network where the first node is specified the action $X_1^n$ and the second node requires the action $(X_{i+1}^n,\ldots, X_{\h}^n)$. For the one-hop network, the smallest rate of communication is given by~\cite[Theorem 10]{PC-HP-TC-CordCap}, and equals $I(X_1;X_{i+1},\ldots, X_\h).$
\end{IEEEproof}

\subsection{Markov Actions}\label{subsec-Markov}
In this section, we study the specific setting where the set of $\h$ actions for a Markov chain that is aligned with the network topology, i.e., we have 
\begin{align}
X_1\mkv X_2\mkv \cdots \mkv X_{\h-1}\mkv X_\h. \label{eqn-MarkovActns}
\end{align}
For this specific class of actions, while we do not have a complete characterization of the optimal tradeoffs among the common and local randomness required at the node nodes and the communication rates on the links, we derive two partial characterizations that correspond to the two extreme cases of common randomness rates. More specifically, we quantify the required rates of local randomness and the communication rates on each link of the network when:
\begin{itemize}
\item[(a)] common randomness is \emph{sufficiently large}, i.e., $\mathsf R_c > H(X_{2},\ldots,X_{\h}| X_1)$; and 
\item[(b)]  common randomness is absent; 
 \end{itemize}
 The first setting when the common randomness is sufficiently large is a direct consequence of Theorem~\ref{Thm-GeneralLargeRc}, and the corresponding result for Markov actions is as follows.

 \begin{remark}\label{Rem-GeneralLargeRcMarkov}
The requirements on local randomness rates and communication rates when the rate of common randomness available to all nodes is $\mathsf R_c > H(X_2,\ldots, X_{\h}| X_1)$ are given by:
\begin{align}
\rho_i &\geq 0,\qquad\qquad\qquad\quad i=1,\ldots,\h,\\
\mathsf R_i &\geq I(X_1; X_{i+1}), \quad i=1,\ldots,\h-1.
\end{align}
Further, it suffices to use functional schemes to achieve the above lower bounds.$\hfill${\tiny$\blacksquare$}
\end{remark}
The main result in this section is the following result characterizing the trade-offs when there is no common randomness.
\begin{theorem}\label{Thm-Mkv3}
Strong coordination is achievable at local randomness rates $(\rho_1,\ldots, \rho_\h)$ and communication rates $(\mathsf R_1,\ldots,\mathsf R_{\h-1})$  and zero common randomness rate (i.e., in the absence of common randomness) provided there exist auxiliary RVs $Z_1,\ldots, Z_{\h-1}$ such that 
\begin{align}
X_1 \mkv Z_1 \mkv X_2 \mkv Z_2 \mkv \cdots \mkv X_{\h-1} \mkv Z_{\h-1} \mkv X_\h, \label{eqn:long chain}
\end{align}
and for each $1\leq i \leq j \leq \h$,
\begin{align}
\mathsf R_i + \sum_{k=i+1}^j \rho_{k} &\geq \left\{\begin{array}{ll}  I(X_i,X_{i+1}; Z_i), & i=j<\h \\
H(X_{i+1},\ldots, X_j| X_i)+  I(X_i; Z_i)+I(X_{j+1}; Z_j | X_j),&i<j<\h\\ H(X_{i+1},\ldots,X_{\h}| X_i) + I(X_i; Z_i), & i<j=\h\\ I(X_\h; Z_\h), & i=j=\h\end{array}\right.,\label{eqn:RR1-Mkv3}\\
\sum_{k=1}^j \rho_{k} &\geq 
\left\{\begin{array}{ll} I(X_2;Z_1|X_1), & j=1\\ H(X_{2},\ldots, X_j| X_1)+I(X_{j+1}; Z_j | X_j),&1<j<\h\\ H(X_{2},\ldots,X_{\h}| X_1),  & j=\h\end{array}\right..\label{eqn:RR2-Mkv3}
\end{align}
\end{theorem}
\begin{IEEEproof}
For the achievable part of the proof, pick $Q_{X_1,\ldots,X_\h,Z_1,\ldots,Z_{\h-1}}$ such that  the above Markov chain holds, and $Q_{X_1,\ldots,X_\h}=\mathsf Q_{X_1,\ldots,X_\h}$. To build a code using this joint pmf, we adapt the code design of Section~\ref{subsec-GenInnerBnd} with the following assignments:
\begin{align}
A_{i,j} &\triangleq \textrm{constant},  \quad 1\leq i <j\leq \h,\\
B_{i,i+1} &\triangleq Z_i,  \,\,\quad\quad\qquad 1\leq i < \h,\\
C_i &\triangleq X_i,   \,\,\quad\quad\qquad 1< i \leq\h.
\end{align}
such that the joint pmf of actions and the auxiliary RVs are:
\begin{align}
Q_{X_1}Q_{B_{1,2}|X_{1} }\,{\prod\limits_{j=2}^{\h}}  \left(Q_{C_j|B_{j-1,j}}Q_{X_{j}|B_{j-1,j} C_j}Q_{B_{j,j+1}|X_{j}}\right),
\end{align} 
with $Q_{X_1,\ldots, X_\h} = \mathsf Q_{X_1,\ldots, X_\h}$. Note that this assignment meets the decomposition specified in \eqref{eqn-AuxRVs1}. Now, from the analysis in Section~\ref{subsec-GenInnerBnd}, and  specifically from Theorem~\ref{thm-2}  we see that we can build a strong coordination code with the following  codebook parameters.
\begin{align}
(\mu_{i,j}^\ps, \mu_{i,j}^\ms) &\triangleq (0,0), & \quad 1\leq i <j\leq \h,\\
(\kappa_{i}^\ps, \kappa_{i}^\ms) &\triangleq \left(I(X_i X_{i+1}; B_{i,i+1}), 0\right) = \left(I (X_i X_{i+1}; Z_i), 0\right) ,&\quad 1\leq i <\h,\\
\lambda_i &\triangleq I(X_{i-1},X_i; C_i | B_{i-1,i}) = H(X_i | Z_{i-1}), &\quad 1< i \leq \h.
\end{align} 
Now, using the assignments (for the unrestricted mode of intermediate-node operation) in Section~\ref{Subsec-strngcoordcode}, we infer that a code can be built with the following common, local and communication rates.
\begin{align}
\mathsf R_c &\triangleq 0,\\
\rho_j &\triangleq  \left\{\begin{array}{ll} I(X_2; Z_1 | X_1), &j=1\\  I(X_{j+1}; Z_j | X_j)+H(X_{j}|Z_{j-1}), & 1<j<\h \\ H(X_\h | Z_{\h-1}), & j=\h \end{array}\right.,\\
\mathsf R_j &\triangleq I(X_j, X_{j+1} ; Z_j ), \quad\quad 1\leq j <\h.
\end{align}
Since the rate-transfer from  $\rho_j$ to $\rho_{j-1}$ is allowable by communicating local randomness from Node $j-1$ to Node $j$ (Lemma~\ref{eqn-URratetransfer}), we see that the following randomness and communication rates also suffice to achieve strong coordination.
\begin{align}
\mathsf R_c &\triangleq 0,\\
\rho_j &\triangleq  \left\{\begin{array}{ll} I(X_2; Z_1 | X_1)+\delta_2, &j=1\\  I(X_{j+1}; Z_j | X_j)+ H(X_{j}|Z_{j-1})-\delta_j+\delta_{j+1}, & 1<j<\h \\ H(X_\h | Z_{\h-1})-\delta_j, & j=\h \end{array}\right.,\\
\mathsf R_j &\triangleq I(X_j, X_{j+1} ; Z_j )+\delta_j, \quad\quad 1\leq j <\h,
\end{align}
where the rate-transfer variables $\delta_i$'s are subject to non-negativity constraints. A routine Fourier-Motzkin elimination to dispose of the rate-transfer variables yields the requisite rate region.

Now, to prove the converse, suppose that there exists a scheme requiring a local randomness rate of $\rho_i +\ve$ at Node $i$ and a communication rate of $\mathsf R_i+\ve$ from Node $i$ to Node $i+1$ such that the joint pmf of the actions given the first node's action satisfies
\begin{align}
\vnm{Q_{\hat{X}_1^n\cdots \hat{X}_\h^n}-\mathsf Q_{X_1\cdots {X}_\h}^{\otimes n}}_1 \leq \ve,
\end{align}
where we let $\hat X_1^n = X_1^n$ to be the action specified at Node $1$. Then, for any $1\leq j<\h$, 
\begin{align}
n\sum_{k=1}^j\rho_k &\geq H(\mathsf M_{L_1}, \ldots, \mathsf M_{L_j})\\
&\geq H(\mathsf M_{L_1}, \ldots, \mathsf M_{L_j} | \hat X_1^n) \\
&\geq I(\{\hat X_\ell^n\}_{\ell=2}^{j+1}; \{\mathsf{M}_{L_\ell}\}_{\ell=1}^j| \hat X_1^n) \\
&\stackrel{(a)}{\geq} I(\{\hat X_\ell^n\}_{\ell=2}^{j+1};\{\mathsf{I}_\ell\}_{\ell=1}^j,\{\mathsf{M}_{L_\ell}\}_{\ell=1}^j | \hat X_1^n) \\
&= \sum_{k=1}^n I(\{\hat X_{\ell,k}^n\}_{\ell=2}^{j+1}; \{\mathsf{I}_\ell\}_{\ell=1}^j,\{\mathsf{M}_{L_\ell}\}_{\ell=1}^j \mid \hat X_1^n,  \{\hat X_\ell^{k-1}\}_{\ell=2}^j) \\
&\stackrel{(b)}{\geq} \sum_{k=1}^n I(\{\hat X_{\ell,k}^n\}_{\ell=2}^{j+1}; \{\mathsf{I}_\ell\}_{\ell=1}^j, \{\hat X_{\ell,k}\}_{\ell=2}^j\mid \hat X_1^n, \{\hat X_\ell^{k-1}\}_{\ell=2}^j)  \\
&\stackrel{(c)}{\geq} \sum_{k=1}^n I(\{\hat X_{\ell,k}^n\}_{\ell=2}^{j+1}; \{\mathsf{I}_\ell\}_{\ell=1}^j, \{\hat X_{\ell,k}\}_{\ell=2}^j, \hat X_1^n, \{\hat X_\ell^{k-1}\}_{\ell=2}^{j+1} | \hat X_{1,k}) -n\delta'_{n,\ve} \\
&\geq \sum_{k=1}^n I(\{\hat X_{\ell,k}^n\}_{\ell=2}^{j+1};\mathsf{I}_j, \{\hat X_{\ell,k}\}_{\ell=2}^j | \hat  X_{1,k}) -n\delta'_{n,\ve} \\
&\stackrel{(d)}{=} \sum_{k=1}^n I(\{\hat X_{\ell,k}^n\}_{\ell=2}^{j+1};Y_j, \{\hat X_{\ell,k}\}_{\ell=2}^j | \hat X_{1,k}) -n\delta'_{n,\ve} \\
&\stackrel{(e)}{=} n I(\{\hat X_{\ell,U}^n\}_{\ell=2}^{j+1};Y_j, \{\hat X_{\ell,U}\}_{\ell=2}^j | \hat  X_{1,U}, U) -n\delta'_{n,\ve} \\
&\stackrel{(f)}{\geq} n I(\{\hat X_{\ell,U}^n\}_{\ell=2}^{j+1};\bar Y_{j}, \{\hat X_{\ell,U}\}_{\ell=2}^j | \hat X_{1,U}) -n\tilde\delta_{n,\ve}\\
& = nH(\{\hat X_{\ell,U}^n\}_{\ell=2}^{j}| \hat X_{1,U})+ nI(\hat{X}_{j+1,U}; \bar Y_j |\{\hat X_{\ell,U}^n\}_{\ell=1}^{j})-n\tilde\delta_{n,\ve}\\
& \stackrel{(g)}{=} nH(\{\hat X_{\ell,U}^n\}_{\ell=2}^{j} | \hat X_{1,U})+ nI(\hat{X}_{j+1,U}; \bar Y_j, \{\hat X_{\ell,U}^n\}_{\ell=1}^{j-1} | \hat{X}_{j,U})-n\bar\delta_{j,n,\ve}\\
&\geq nH(\{\hat X_{\ell,U}^n\}_{\ell=2}^{j} | \hat X_{1,U})+ nI(\hat{X}_{j+1,U}; \bar Y_j | \hat{X}_{j,U})-n\bar\delta_{j,n,\ve},
\end{align}
where
\begin{itemize}
\item[$(a)$] follows because in the absence of common randomness,  $\mathsf I_1$ is a function of $X_1^n$ and $\mathsf M_{L_1}$, and for $i=2,\ldots, j$, $\mathsf I_j$ is a function of $\mathsf I_{j-1}$ and $\mathsf M_{L_j}$; 
\item[$(b)$] follows from two steps: 1) introducing action variables $\{\hat X_{\ell,k}\}_{\ell=2}^j$, since they are functions of $\{\mathsf I_\ell\}_{l=1}^{j-1}$ and $\{\mathsf M_{L_\ell}\}_{\ell=2}^j$; and then 2) by dropping  $\{\mathsf M_{L_\ell}\}_{\ell=2}^j$; 
\item[$(c)$] follows from \eqref{eqn-vardistconstmetconseq1}, since the actions are nearly i.i.d.; 
\item[$(d)$] by defining $Y_j \triangleq \mathsf I_j$; 
\item[$(e)$] by introducing a time-sharing variable $U$ that is uniform over $\{1,\ldots, n\}$; 
\item[$(f)$] from by setting $\bar Y_{j} \triangleq (Y_{j}, U)$ and defining
\begin{align}
\tilde\delta_{n,\ve}\triangleq \delta'_{n,\ve} + I(\hat{X}_{2,U},\ldots, \hat X_{\h,U}; U| X_{1,U}),  \label{eqn:remUconditioning}
\end{align}
which due to \eqref{eqn:timesharingvarremove},  is guaranteed  to vanish as we let $\ve \rightarrow 0$; and finally, 
\item[$(g)$] follows by defining for $j=1,\ldots,\h-1$,
\begin{align}
\bar\delta_{j,n,\ve} \triangleq \tilde{\delta}_{n,\ve}+ I(\hat X_{j+1,U};  \hat{X}_{1,U},\ldots,\hat X_{j-1,U} | \hat X_{j,U}), \label{eqn-bardeltadefn}
\end{align}
which, due to the Markovity of the actions \eqref{eqn-MarkovActns} and Remark~\ref{rem-3}, is also guaranteed to vanish as $\ve \rightarrow 0$.
\end{itemize}
While this establishes \eqref{eqn:RR2-Mkv3} for $j<\h$, the argument for when $j=\h$ follows from the above by setting $\mathsf I_\h$ and $Y_{\h}$ as constant RVs. 

Now, to prove \eqref{eqn:RR1-Mkv3}, we proceed as follows. Let $1\leq i\leq j<\h$. Then,
\begin{align}
\hspace{-2mm}n\left(\mathsf R_i + \sum_{k=i+1}^j\rho_k\right) &\geq H(\mathsf{I}_i, \{\mathsf M_{L_\ell}\}_{\ell=i+1}^j ) \\
&\geq H(\{\mathsf{I}_\ell\}_{\ell=i}^{j}, \{\mathsf M_{L_\ell}\}_{\ell=i+1}^j ) \\
&\geq I(\{\hat{X}_\ell^n\}_{\ell=i}^{j+1}; \{\mathsf{I}_\ell\}_{\ell=i}^{j}, \{\mathsf M_{L_\ell}\}_{\ell=i+1}^j ) \\
&\geq I(\hat X_i^n; \{\mathsf{I}_\ell\}_{\ell=i}^{j}, \{\mathsf M_{L_\ell}\}_{\ell=i+1}^j )+I(\{\hat{X}_\ell^n\}_{\ell=i+1}^{j+1}; \{\mathsf{I}_\ell\}_{\ell=i}^{j}, \{\mathsf M_{L_\ell}\}_{\ell=i+1}^j \mid \hat X_i^n) \\
&\geq I(\hat X_i^n; \mathsf I_i)+I(\{\hat{X}_\ell^n\}_{\ell=i+1}^{j+1}; \{\mathsf{I}_\ell\}_{\ell=i}^{j}, \{\mathsf M_{L_\ell}\}_{\ell=i+1}^j \mid \hat X_i^n) \\
&= I(\hat X_i^n; \mathsf I_i)+I(\{\hat{X}_\ell^n\}_{\ell=i+1}^{j+1}; \{\mathsf{I}_\ell\}_{\ell=i}^{j}, \{\mathsf M_{L_\ell}\}_{\ell=i+1}^j,\{\hat{X}_\ell^n\}_{\ell=i+1}^{j} \mid\hat X_i^n) \\
&\geq I(\hat X_i^n; \mathsf I_i)+I(\{\hat{X}_\ell^n\}_{\ell=i+1}^{j+1}; \mathsf{I}_j,\{\hat{X}_\ell^n\}_{\ell=i+1}^{j} \mid\hat X_i^n) \\
&= \sum_{k=1}^n \Big(I(\hat X_{i,k}; \mathsf I_i | \hat X_i^{k-1})+I(\{\hat{X}_{\ell,k}\}_{\ell=i+1}^{j+1}; \mathsf{I}_j,\{\hat{X}_\ell^n\}_{\ell=i+1}^{j} \mid\hat X_i^n, \{\hat{X}_\ell^{k-1}\}_{\ell=i+1}^{j+1}) \Big)\\
&\hspace{-3mm}\stackrel{\eqref{eqn-vardistconstmetconseq},\eqref{eqn-vardistconstmetconseq1}}{\geq} \sum_{k=1}^n \Big(I(\hat X_{i,k}; \mathsf I_i)+I(\{\hat{X}_{\ell,k}\}_{\ell=i+1}^{j+1}; \mathsf{I}_j,\{\hat{X}_{\ell,k}\}_{\ell=i+1}^{j} \mid\hat X_{i,k}) \Big)-2n \delta'_{n,\ve}\\
&\stackrel{(a)}{\geq} nI(\hat X_{i,U}; \bar Y_i) +nI(\{\hat{X}_{\ell,U}\}_{\ell=i+1}^{j+1}; \bar Y_j, \{\hat{X}_{\ell, U}\}_{\ell=i+1}^{j} \mid\hat X_{i,U})-2n \tilde\delta_{n,\ve}\\
&= nI(\hat X_{i,U}; \bar Y_i) +n H(\{\hat{X}_{\ell, U}\}_{\ell=i+1}^{j} | \hat X_{i,U}) + nI(\hat{X}_{j+1,U}; \bar Y_j |  \{\hat{X}_{\ell, U}\}_{\ell=i}^{j})-2n \tilde\delta_{n,\ve}, \notag\\
&\stackrel{(b)}{\geq} n\big(I(\hat X_{i,U}; \bar Y_i) \hspace{-0.5mm}+ \hspace{-0.5mm} H(\{\hat{X}_{\ell, U}\}_{\ell=i+1}^{j} | \hat X_{i,U}) + I(\hat{X}_{j+1,U}; \bar Y_j, \{\hat{X}_{\ell, U}\}_{\ell=i}^{j-1} |  \hat{X}_{j, U})-2 \bar\delta_{j,n,\ve}\big)\\
&\stackrel{}{\geq} n\big(I(\hat X_{i,U}; \bar Y_i) + H(\{\hat{X}_{\ell, U}\}_{\ell=i+1}^{j} | \hat X_{i,U}) + I(\hat{X}_{j+1,U}; \bar Y_j |  \hat{X}_{j, U})-2 \bar\delta_{j,n,\ve}\big),
\end{align}
where: 
\begin{itemize}
\item[$(a)$] follows from a time-sharing argument with auxiliary RVs $\{\bar Y_j:1\leq j< \h\}$ defined previously; and 
\item[$(b)$] follows by the use of $\bar\delta_{j,n,\ve}$ defined in \eqref{eqn-bardeltadefn}.
\end{itemize}
 Also, as before, the proof of \eqref{eqn:RR1-Mkv3} for  $j=\h$ follows by setting $\mathsf I_\h$ and $ Y_\h$ as constants. 
 
 We are left to establish one last fact, which is the Markov condition to be met by the actions and the auxiliary RVs. Note that as per the definitions of the auxiliary RVs, we do not have the chain $X_1\mkv \bar{Y}_1 \mkv X_2 \mkv\cdots\mkv \bar Y_{\h-1} \mkv X_\h$. 
This follows from the fact that our choice of the auxiliary RV $\bar Y_j = \mathsf I_j$ ensures that  we have conditional independence of actions at adjacent nodes given the message communicated over the hop connecting the two nodes (i.e., $X_{j,U} \mkv \bar Y_j \mkv X_{j+1,U}$ for all $j=1,\ldots,\h-1$); however, we do not have conditional independence of messages conveyed on adjacent hops conditioned on the action of the node in-between (i.e.,  we do not have $\bar Y_j \mkv X_{j+1,U} \mkv \bar Y_{j+1}$).  Note however that the information functionals in \eqref{eqn:RR1-Mkv3} and \eqref{eqn:RR2-Mkv3} only contain one auxiliary RV.  Hence, it is possible to define a new set of auxiliary RVs that would both satisfy the long chain in the claim and  preserve the information functionals. To do so, define RVs  $\tilde X_k$, $k=1,\ldots, \h$, and $Z_j$, $j=1,\ldots, \h-1,$ such that their joint pmf is given by
\begin{align}
Q_{\tilde X_1,\ldots,\tilde X_\h} & \triangleq Q_{\hat X_{1,U}, \ldots, \hat X_{\h,U}},\\
Q_{Z_1,\ldots, Z_{\h-1}|\tilde X_1,\ldots,  \tilde X_\h }(z_1,\ldots, z_{\h-1}|x_1,\ldots, x_h) &\triangleq \prod_{j=1}^{\h-1} Q_{\bar Y_{j} | \hat X_{j,U}\hat X_{j+1,U}}(z_j|x_j, x_{j+1}) .
\end{align}
Note that  we have $\tilde X_1 \mkv Z_1 \mkv\tilde  X_2 \mkv \cdots \mkv Z_{\h-1} \mkv \tilde X_\h$, and further for $1\leq i < j \leq \h$,
\begin{align}
H(\tilde X_{i+1},\ldots,\tilde X_j | \tilde X_i) &= H(\hat X_{i+1, U},\ldots,\hat X_{j,U} | \hat X_{i,U}),   \\
I(\tilde X_{j+1}; Z_j | \tilde X_j) &= I(\hat X_{j+1,U}; \bar Y_j | \hat X_{j,U}),\\
I(\tilde X_{j}; Z_j) &= I(\hat X_{j,U}; \bar Y_j).
\end{align}
The proof is completed by limiting the size of the auxiliary RVs $\{Z_j\}_{j=1}^{\h-1}$, and then by limiting $\ve \rightarrow 0$, which ensures that $Q_{\tilde X_1,\ldots,\tilde X_\h}  \rightarrow \mathsf Q_{X_1,\ldots, X_\h}$, and each of infinitesimals in $\{\delta_{j,n,\ve}:1\leq j<\h\}$ vanishes. 
\end{IEEEproof}

We conclude this section with a short discussion on the indispensability of auxiliary RVs $\{B_{i,i+1}:i=1,\ldots\h-1\}$ using the above result. 

\subsection{The essentiality of $\{B_{i,i+1}:1\leq i <\h\}$}\label{sec-essentiality of Bs}
The auxiliary RVs $\{A_{i,j}: 1\leq i < j \leq \h\}$ have a natural purpose --  Node $i$ uses $A_{i,j}$ to coordinate its actions with that of Node $j$. However, the need for $\{B_{i,i+1}: 1\leq i <\h\}$ is technical, and arises from the fact that not all joint pmfs for $\{A_{i,j}: 1\leq i < j \leq \h\}$ can be realized as a coding scheme. For example, their joint pmf must satisfy the chains in \eqref{eqn-ARVReln1}. For simpilicity, let's focus on the following setting. Let random variables $V_1$ and $V_2$ be jointly correlated according to $\mathsf Q_{V_1,V_2}$ that has full support and suppose that $I(V_1;V_2)>0$. Let us focus on strong coordination on a line network with $\h=3$ nodes, where actions $X_1=V_1$, $X_2=(V_1,V_2)$, and $X_3=V_2$. Let us focus on the setting where there is no common information, and each node has sufficient local randomness, say $\rho_i > H(V_1,V_2)$ for each $i=1,2,3$. Since, we have $X_1 \mkv X_2 \mkv X_3$, we can see that the rates for communication required for strong coordination specified by Theorem~\ref{Thm-Mkv3} are as follows.
\begin{align}
\mathsf R_i &\geq H(V_i), \quad i\in\{1,2\}.
\end{align}
An achievable code for the corner point of the above region can be constructed by setting $A_{1,2}=A_{1,3}=A_{2,3}=\textrm{constant}$, and by choosing $B_{1,2}=X_1=V_1$ and $B_{2,3}=V_2=X_3$. We will now show that it is impossible to attain the corner point $(\mathsf R_1, \mathsf R_2) = H(V_1, V_2)$ by use of only $A_{1,2}, A_{1,3}, A_{2,3}$. 

Using the rate expressions for the unrestricted mode of operation at Node 2 given in Section~\ref{Subsec-strngcoordcode}, we see that we can build a code with common randomness rate $\mathsf R_c =0$, local randomness rates $\rho_1=\rho_2=\rho_3=H(V_1,V_2)$, and communication rates $\mathsf R_1 = H(V_1)$ and $\mathsf R_2 = H(V_2)$ with only auxiliary RVs $A_{1,2}, A_{1,3}, A_{2,3}$ if there exists a joint pmf $Q_{X_1X_2X_3A_{1,2}, A_{1,3}, A_{2,3}} $ such that
\begin{align}
A_{1,2} &\mkv A_{1,3}\mkv A_{2,3},\label{eqn-codeconstmkv}\\
V_1=X_1 &\mkv (A_{1,2}, A_{1,3}) \mkv (X_2,X_3)=(V_1,V_2),\label{eqn-MkvCndR1}\\
V_2 =X_3 &\mkv (A_{1,3}, A_{2,3}) \mkv (X_1,X_2)=(V_1,V_2).\label{eqn-MkvCndR2}
\end{align}
provided the codebook rates satisfy the following conditions imposed by the unrestricted mode of operation and by Theorems~\ref{thm-1} and~\ref{thm-2}.
\begin{align}
\mathsf R_c &\stackrel{\eqref{eqn-UnresRc}}{=} \mu_{1,2}^\ms+\mu_{2,3}^\ms + \mu_{1,3}^\ms = 0,\\
\mathsf R_1 &\stackrel{\eqref{eqn-UnresRi}}{=} \mu_{1,2}^\ps+ \mu_{1,3}^\ps= H(V_1),\\
\mathsf R_2 &\stackrel{\eqref{eqn-UnresRi}}{=} \mu_{2,3}^\ps+ \mu_{1,3}^\ps= H(V_2),\\
\mu_{1,3}^\ps & \geq I(X_1,X_2,X_3; A_{1,3}) = I(V_1,V_2; A_{1,3}),\\
\mu_{1,3}^\ps + \mu_{1,2}^\ps & \geq I(X_1,X_2,X_3; A_{1,3}) = I(V_1,V_2; A_{1,2},A_{1,3}),\label{eqn-R1ArgBnd}\\
\mu_{1,3}^\ps + \mu_{2,3}^\ps & \geq I(X_1,X_2,X_3; A_{1,3}) = I(V_1,V_2; A_{1,3},A_{2,3}),\label{eqn-R2ArgBnd}\\
\mu_{1,3}^\ps + \mu_{1,2}^\ps+ \mu_{2,3}^\ps & \geq I(X_1,X_2,X_3; A_{1,3}) = I(V_1,V_2; A_{1,2}, A_{1,3},A_{2,3}).
\end{align}
 Then, it must be true that
 \begin{align}
I(V_2;A_{1,3} | V_1)=0,\label{eqn-A13condindep}
\end{align}
since
\begin{align}
I(V_2; A_{1,2} A_{1,3} | V_1) &= I(V_1,V_2; A_{1,2} A_{1,3} V_1) - H(V_1)\stackrel{\eqref{eqn-MkvCndR1}}{=}  I(V_1,V_2; A_{1,2} A_{1,3}) - H(V_1)\\
&\stackrel{\eqref{eqn-R1ArgBnd}}{\leq} \mathsf R_1 - H(V_1) = 0.
\end{align}
Similarly, $I(V_1;A_{1,3} | V_2)=0$, since 
\begin{align}
I(V_1; A_{1,3} A_{2,3} | V_2) &= I(V_1,V_2; A_{1,3} A_{2,3} V_2) - H(V_2)\stackrel{\eqref{eqn-MkvCndR2}}{=}  I(V_1,V_2; A_{1,3} A_{2,3}) - H(V_2)\\
&\stackrel{\eqref{eqn-R2ArgBnd}}{\leq} \mathsf R_2 - H(V_2) = 0.
\end{align}
Thus, we have $A_{1,3} \mkv V_1 \mkv V_2$ and $V_1 \mkv V_2 \mkv A_{1,3}$. Since $\mathsf Q_{V_1,V_2}$ has full support, it follows that for any  $(a_{1,3}, v_1, v_2) \in \A_{1,3} \times \V_1 \times \V_2$, we have
\begin{align}
Q_{A_{1,3} | V_1} (a_{1,3}|v_1)= \frac{Q_{A_{1,3} V_1 V_2}(a_{1,3},v_1,v_2)}{\mathsf Q_{V_1 V_2}(v_1, v_2)} =  Q_{A_{1,3} | V_2} (a_{1,3}|v_2).\label{eqn-A13prop}
\end{align}
Hence, for any $(a_{1,3}, v_1) \in \A_{1,3} \times \V_1$,
\begin{align*}
Q_{A_{1,3} | V_1} (a_{1,3}|v_1)&= \sum_{v_2} Q_{A_{1,3} | V_1} (a_{1,3}|v_1)\mathsf Q_{V_2}(v_2)\stackrel{\eqref{eqn-A13prop}}{=}  \sum_{v_2} Q_{A_{1,3} | V_2} (a_{1,3}|v_2)\mathsf Q_{V_2}(v_2) = Q_{A_{1,3}} (a_{1,3}).
\end{align*}
Hence, $A_{1,3}$ is independent of $V_1$. Combining this fact with \eqref{eqn-A13condindep}, we see that 
\begin{align}
I(V_1,V_2;A_{1,3})=0. \label{eqn-A13indep}
\end{align}
Since \eqref{eqn-MkvCndR1} and \eqref{eqn-MkvCndR2} imply that $V_1$ is a function of $(A_{1,2}, A_{1,3})$, and  $V_2$ is a function of $(A_{1,3}, A_{2,3})$, it then follows that 
\begin{align}
0< I(V_1;V_2) &\stackrel{\eqref{eqn-A13indep}}{=} I(V_1;V_2 ,A_{1,3}) \stackrel{\eqref{eqn-A13indep}}{=} I(V_1;V_2|A_{1,3}) \stackrel{}{\leq} I(A_{1,2}; A_{2,3} |A_{1,3}), \label{eqn-V1V2Contra}
\end{align}
which is a contradiction since \eqref{eqn-V1V2Contra} violates \eqref{eqn-codeconstmkv}. Hence, we cannot achieve this corner point with the use of $A_{1,2}, A_{1,3}, A_{2,3}$ alone. This above argument also establishes that the corner point is not achievable using the functional mode of operation at intermediate nodes, thereby highlighting the following observation.
\begin{remark}
The portion of the strong coordination capacity region achievable by schemes with functional intermediate-node operation is, in general, a \emph{strict} subset of the strong coordination capacity region.
\end{remark}

\section{Conclusion}\label{sec-conclu}
In this work, we have analyzed the communication and randomness resources required to establish strong coordination over a multi-hop line network. To derive an achievable scheme, we first build an intricate multi-layer structure of channel resolvability codes to generate the actions at all the nodes, which is then appropriately inverted to obtain a strong coordination code. While the resultant strong coordination code is not universally optimal, i.e., it is not known to achieve the optimal tradeoffs among common randomness rate, local randomness rates, and hop-by-hop communication rates, it is shown to achieve the best tradeoffs in several settings, including when all intermediate nodes operate under a functional regime, and when common randomness is plentiful. 

The need for an intricate multi-layer scheme stems from a basic limitation in our understanding of the design of strong coordination codes for general multi-terminal problems: unlike in typicality-based schemes for a multi-user setups, where we can use joint typicality as the criterion to use a received signal (at some intermediate network node) to select a codeword for transmission, we do not have a similar criterion here to translate messages from one hop to the next. Consequently, a general way to design strong coordination codes where communicated messages are non-trivially correlated is open. However, through this work, we have made partial progress in this regard, since the achievable scheme allows messages in different hops to be correlated. However, there is limited control over the joint correlation, since it is determined implicitly by the codebooks at the time of construction.

\appendix
\subsection{Proof of Theorem~\ref{thm-1}} \label{App-1}
Before we proceed, we first use the following notation to simplify the analysis.
\begin{align*}
\begin{array}{rl}
\boldsymbol{Y}&\triangleq(X_1,\ldots,X_\h)\\
\hat{\boldsymbol{Y}}&\triangleq(\hat X_1,\ldots,\hat X_\h)\\
\mathsf N&\triangleq2^{n( (\mu_{1,2}^\ps+\mu_{1,2}^\ms)+\cdots+ (\mu_{\h-1,\h}^\ps+\mu_{\h-1,\h}^\ms))}
\end{array}.
\end{align*}
Let for any $S\subseteq \mc F\triangleq \{(i,j): 1\leq i < j\leq \h\}$, let $\mathcal{J}_S \triangleq \left\{ (i,j): \overline\Phi(i,j)\cap S \neq \emptyset\right\}$. Now, to find the conditions on the rates, we proceed in a fashion similar to  \cite{MB-JK-Line2} and \cite{MB-JK-Line1}.
%\begin{figure*}[th!]
\begin{align}
\hspace{-2mm}\Exp\big[\dkl &({Q}^{({1})}_{\hat{\boldsymbol Y}^n}\hspace{-1mm}\parallel\hspace{-1mm} \mathsf{Q}_{\boldsymbol Y}^{\otimes n})\big] \hspace{-0.5mm}=\hspace{-0.5mm}\Exp\left[ \sum_{\boldsymbol{y}^n}\left(\frac{\sum_{\boldsymbol{m}^\pms} Q^{\otimes n}_{\boldsymbol{Y}| {\boldsymbol A}}(\boldsymbol{y}^n|{\boldsymbol A}^n(\mbf m^\pms) )}{\mathsf N}\right)\hspace{-0.5mm}\log\hspace{-0.5mm} \left(\frac{\sum_{\tilde{\boldsymbol{m}}^\pms} Q^{\otimes n}_{\boldsymbol{Y}| {\boldsymbol A} }(\boldsymbol{y}^n|{\boldsymbol A}^n(\tilde{\mbf m}^\pms))}{\mathsf N\mathsf Q^{\otimes n}_{\boldsymbol Y}(\boldsymbol y^n)}\right)\right]\label{eqn-Anlys1}\\
{}
&=\sum_{\boldsymbol{y}^n,\boldsymbol{m}^\pms}\Exp\Bigg[{\frac{Q^{\otimes n}_{\boldsymbol{Y}| {\boldsymbol A}}(\boldsymbol{y}^n|{\boldsymbol A}^n(\mbf m^\pms))}{\mathsf N}}\Exp\bigg[\log \bigg(\sum_{\tilde{\boldsymbol{m}}^\pms} \frac{Q^{\otimes n}_{\boldsymbol{Y}| {\boldsymbol A}  }(\boldsymbol{y}^n|{\tilde{\boldsymbol{m}}^\pms} )}{\mathsf N\mathsf Q^{\otimes n}_{\boldsymbol Y}(\boldsymbol y^n)}\bigg)\bigg| \boldsymbol{A}^n(\mbf m^\pms)\bigg]\Bigg]\label{eqn-Anlys2}\\
{}
&\leq\sum_{\boldsymbol{y}^n,\boldsymbol{m}^\pms}\Exp\bigg[{\frac{Q^{\otimes n}_{\boldsymbol{Y}| {\boldsymbol A}}(\boldsymbol{y}^n|{\boldsymbol A}^n(\mbf m^\pms))}{\mathsf N}}\log\bigg(\Exp \bigg[\sum_{\tilde{\boldsymbol{m}}^\pms} \frac{Q^{\otimes n}_{\boldsymbol{Y}| {\boldsymbol A}  }(\boldsymbol{y}^n|{\tilde{\boldsymbol{m}}^\pms} )}{\mathsf N\mathsf Q^{\otimes n}_{\boldsymbol Y}(\boldsymbol y^n)}\bigg| \boldsymbol{A}^n(\mbf m^\pms)\bigg]\bigg)\bigg]\label{eqn-Anlys3}\\
{}
&\leq\sum_{\boldsymbol{y}^n,\mbf m^\pms}\Exp\bigg[{\frac{Q^{\otimes n}_{\boldsymbol{Y}| {\boldsymbol A}}(\boldsymbol{y}^n|{\boldsymbol A}^n(\mbf m^\pms))}{\mathsf N}}\log\bigg(1+\sum_{S: (1,\h)\notin S}  \frac{Q^{\otimes n}_{\boldsymbol{Y}| {\boldsymbol A}_{\mathcal{J}^\mathsf{ c}_S}  }(\boldsymbol{y}^n|{\boldsymbol A}_{\mathcal{J}^\mathsf{ c}_S}^n(\mbf m^\pms) )}{\big(2^{n \sum_{s\notin S} (R_{s}^\ps+R_{s}^\ms)}\big)\mathsf Q^{\otimes n}_{\boldsymbol Y}(\boldsymbol y^n)}\bigg)\bigg]\label{eqn-Anlys4}\\ 
{}
&\leq \sum_{\boldsymbol{y}^n, \boldsymbol{a}^n}  Q^{\otimes n}_{\boldsymbol{YA}}(\boldsymbol{y}^n, \boldsymbol{a}^n)\log\Bigg[1+\sum_{S: (1,\h)\notin S}  \frac{Q^{\otimes n}_{\boldsymbol{Y}| {\boldsymbol A}_{\mathcal{J}^\mathsf{ c}_S}  }(\boldsymbol{y}^n|{\boldsymbol a}_{\mathcal{J}^\mathsf{ c}_S}^n)}{2^{n\left( \sum_{s\notin S} (\mu_{s}^\ps+\mu_{s}^\ms)\right)}\mathsf Q^{\otimes n}_{\boldsymbol Y}(\boldsymbol y^n)} \Bigg].\label{eqn-Anlys5}
\end{align}
%\hrule
%\vspace{-5.75mm}
%\end{figure*}

The notation and arguments for the manipulations  in  \eqref{eqn-Anlys1}-\eqref{eqn-Anlys5} are as follows:
\begin{itemize}
\item  \eqref{eqn-Anlys2} follows by the use of the law of iterated expectations, where the inner conditional expectation denotes the expectation over over all random codeword constructions $\{\mbf{A}(\tilde{\mbf m}^\pms): \tilde{\mbf m}^\pms \neq \mbf m^\pms\}$ conditioned on the codeword vector $\boldsymbol{A}(\boldsymbol{m}^\pms)$;
\item \eqref{eqn-Anlys3} follows from Jensen's inequality; and
\item \eqref{eqn-Anlys4} follows by splitting the inner sum in \eqref{eqn-Anlys3} according to the indices where $\mbf{m}^\pms$ and $\tilde{\mbf{m}}^\pms$ differ. 
Let $\Gamma(\mbf{m}^\pms,\tilde{\mbf{m}}^\pms)\triangleq \{ s\in\mc F: m^\pms_s\neq \tilde m^\pms_s\}  $. For any pair of indices $(\mbf{m}^\pms,\tilde{\mbf{m}}^\pms)$, the following hold:
\begin{itemize}
\item if $(i,j)\notin \mathcal{J}_{\Gamma(\mbf{m}^\pms,\tilde{\mbf{m}}^\pms)}$, then $A_{i,j} (\mbf m^\pms)=  A_{i,j} (\tilde{\mbf m}^\pms)$.
This is because if $(i,j)\notin \mathcal{J}_{\Gamma(\mbf{m}^\pms,\tilde{\mbf{m}}^\pms)}$, then by definition, $\mbf{m}^\pms_{\overline\Phi(i,j)} = \tilde{\mbf{m}}^\pms_{\overline\Phi(i,j)}$, and both $A_{i,j} (\mbf m^\pms)$ and $A_{i,j} (\tilde{\mbf m}^\pms)$ correspond to the codeword for $A_{i,j}$ corresponding to $\mbf{m}^\pms_{\overline\Phi(i,j)} = \tilde{\mbf{m}}^\pms_{\overline\Phi(i,j)}$. 
\item if $(i,j)\in\mathcal{J}_{\Gamma(\mbf{m}^\pms,\tilde{\mbf{m}}^\pms)}$, then the random variables $A_{i,j} (\mbf m^\pms)$, $A_{i,j} (\tilde{\mbf m}^\pms)$ are conditionally independent given $\{A_{i,j}(\mbf m^\pms): (i,j)\notin \mathcal{J}_{\Gamma(\mbf{m}^\pms,\tilde{\mbf{m}}^\pms)}\}$, which by the earlier remark, is exactly the same as $\{A_{i,j}(\tilde{\mbf m}^\pms): (i,j)\notin \mathcal{J}_{\Gamma(\mbf{m}^\pms,\tilde{\mbf{m}}^\pms)}\}$.
\end{itemize}
Combining both, we see that
\begin{align}
\boldsymbol{A}^n(\mbf m^\pms)\mkv \{A_{i,j}(\tilde{\mbf m}^\pms): (i,j)\notin \mathcal{J}_{\Gamma(\mbf{m}^\pms,\tilde{\mbf{m}}^\pms)}\}  \mkv \boldsymbol{A}^n(\tilde{\mbf m}^\pms). \label{eqn-MKVchainforcodebook}
\end{align}
Given $\mbf m^\pms$, let $ \mc H_{\mbf m^\pms, S}\triangleq \left\{ \tilde{\mbf m}^\pms: \Gamma(\tilde{\mbf m}^\pms, {\mbf m}^\pms) = S\right\}$. Then, we see that 
\begin{align}
\hspace{-4mm}\Exp\bigg[\sum_{\tilde{\boldsymbol{m}}^\pms} \frac{Q^{\otimes n}_{\boldsymbol{Y}| {\boldsymbol A}  }(\boldsymbol{y}^n|{\boldsymbol A}^n(\tilde{\mbf m}^\pms) )}{\mathsf N\mathsf Q^{\otimes n}_{\boldsymbol Y}(\boldsymbol y^n)}\bigg| \boldsymbol{A}^n(\mbf m^\pms) \bigg] &= \sum_{S} \sum_{\tilde{\mbf{m}}^\pms\in\mc H_{\mbf m^\pms, S}} \frac{\Exp\Big[Q^{\otimes n}_{\boldsymbol{Y}| {\boldsymbol A}  }(\boldsymbol{y}^n|{\boldsymbol A}^n(\tilde{\mbf{m}}^\pms) )\big| \boldsymbol{A}^n(\mbf m^\pms) \Big]}{\mathsf N\mathsf Q^{\otimes n}_{\boldsymbol Y}(\boldsymbol y^n)}. \label{eqn-SplitusingS}
{}
\end{align}
Note that if $(1,\h)\in S$, then $\mathcal J_S = \mc F$ and hence $\boldsymbol{A}^n(\mbf m^\pms)$ and $\boldsymbol{A}^n(\tilde{\mbf m}^\pms)$ are independent, and hence
\begin{align}
\sum_{S:(1,\h)\in S}\, \sum_{\tilde{\mbf{m}}^\pms\in\mc H_{\mbf m^\pms, S}} \frac{\Exp\Big[Q^{\otimes n}_{\boldsymbol{Y}| {\boldsymbol A}  }(\boldsymbol{y}^n|{\boldsymbol A}^n(\tilde{\mbf{m}}^\pms) )\big| \boldsymbol{A}^n(\mbf m^\pms) \Big]}{\mathsf N\mathsf Q^{\otimes n}_{\boldsymbol Y}(\boldsymbol y^n)} = \sum_{S:(1,\h)\in S}\, \sum_{\tilde{\mbf{m}}^\pms\in\mc H_{\mbf m^\pms, S}} \frac{1}{\mathsf N} \leq 1.
\end{align}
Further, when $(1,\h)\notin S$, then $\mathcal J_S \subsetneq \mc F$. Using the chain in \eqref{eqn-MKVchainforcodebook}, we see that when $(1,\h)\notin S$,
\begin{align}
\sum_{\tilde{\mbf{m}}^\pms\in\mc H_{\mbf m^\pms, S}} \frac{\Exp\Big[Q^{\otimes n}_{\boldsymbol{Y}| {\boldsymbol A}  }(\boldsymbol{y}^n|{\boldsymbol A}^n(\tilde{\mbf{m}}^\pms) )\big| \boldsymbol{A}^n(\mbf m^\pms) \Big]}{\mathsf N\mathsf Q^{\otimes n}_{\boldsymbol Y}(\boldsymbol y^n)} = \sum_{\tilde{\mbf{m}}^\pms\in\mc H_{\mbf m^\pms, S}}  \frac{ Q^{\otimes n}_{\boldsymbol{Y}| {\boldsymbol A}_{\mathcal{J}^\mathsf{ c}_{S}}  }(\boldsymbol{y}^n|{\boldsymbol A}_{\mathcal{J}^\mathsf{ c}_{S}}^n(\mbf m^\pms) )}{\mathsf N\mathsf Q^{\otimes n}_{\boldsymbol Y}(\boldsymbol y^n)}.
\end{align}
Combining the above arguments, we see that 
\begin{align}
\Exp\bigg[\sum_{\tilde{\boldsymbol{m}}^\pms} \frac{Q^{\otimes n}_{\boldsymbol{Y}| {\boldsymbol A}  }(\boldsymbol{y}^n|{\boldsymbol A}^n(\tilde{\mbf m}^\pms) )}{\mathsf N\mathsf Q^{\otimes n}_{\boldsymbol Y}(\boldsymbol y^n)}\bigg| \boldsymbol{A}^n(\mbf m^\pms) \bigg] & \leq 1+ 
\sum_{S: (1,\h)\notin S}\left( \sum_{\tilde{\mbf{m}}^\pms\in\mc H_{\mbf m^\pms, S}}  \frac{ Q^{\otimes n}_{\boldsymbol{Y}| {\boldsymbol A}_{\mathcal{J}^\mathsf{ c}_{S}}  }(\boldsymbol{y}^n|{\boldsymbol A}_{\mathcal{J}^\mathsf{ c}_{S}}^n(\mbf m^\pms) )}{\mathsf N\mathsf Q^{\otimes n}_{\boldsymbol Y}(\boldsymbol y^n)}\right)\\
{}
& \stackrel{(a)}{\leq}1+\sum_{S: (1,\h)\notin S}  \frac{ 2^{n \left(\sum_{s\in S} (\mu_{s}^++\mu_{s}^-)\right)}Q^{\otimes n}_{\boldsymbol{Y}| {\boldsymbol A}_{\mc J_S^c}  }(\boldsymbol{y}^n|{\boldsymbol A}_{\mathcal{J}^\mathsf{ c}_{S}}^n(\mbf m^\pms) )}{\mathsf N\mathsf Q^{\otimes n}_{\boldsymbol Y}(\boldsymbol y^n)}\\
{}
&=1+ \sum_{S: (1,\h)\notin S}  \frac{Q^{\otimes n}_{\boldsymbol{Y}| {\boldsymbol A}_{\mathcal{J}_{S}^c}  }(\boldsymbol{y}^n|{\boldsymbol A}_{\mathcal{J}^\mathsf{ c}_{S}}^n(\mbf m^\pms) )}{2^{n \sum\limits_{s\notin S} (\mu_{s}^+ +\mu_s^-)}\mathsf Q^{\otimes n}_{\boldsymbol Y}(\boldsymbol y^n)},
\end{align}
where in $(a)$ we use a counting argument that yields
\begin{align}
|\mc  H_{\mbf m^\pms, S}|   \leq 2^{n \left( \sum_{s\in S} (\mu_{s}^++\mu_s^-)\right)}.
\end{align}
\end{itemize}
 Finally,  the required rate conditions can be derived from \eqref{eqn-Anlys5} by splitting the outer sum depending on whether $(\boldsymbol{y}^n, \boldsymbol{a}^n)\in T^n_\ve[Q_{\mbf{YA}}]$ or not. The sum for atypical realizations in \eqref {eqn-Anlys5} is no more than
\begin{align}
\Pr\big[(\boldsymbol{Y}^n, \boldsymbol{A}^n)\notin T^n_\ve[Q_{\mbf{YA}}]\big] \cdot\log \left(1+2^{\h^2} \eta_{\mbf Y}^{-n}\right),
\end{align}
where $\eta_\mbf{Y}= \min\limits_{\mbf{y}\,\in\,\mathsf{support}(X_1,\ldots,X_\h)} \mathsf Q_{\mbf Y}(\mbf y).$ This term goes to zero as $n\rightarrow \infty$. The contribution from typical realizations can be made to vanish asymptotically, if for each $S\subseteq\mc F$, $\{(\mu_{i,j}^\ps, \mu_{i,j}^\ms): (i,j)\in\mc F\}$ satisfy:
\begin{align}
\sum_{s\notin S} \,(\mu_{s}^\ps+\mu_{s}^\ms)  > I\big(\mbf{Y}; A_{\mathcal{J}_S^\mathsf{ c}}\big)= I\big(X_1,\ldots,X_\h; \{A_{i,j}: (i,j)\notin {\mathcal{J}_S}\}\big).
\end{align}
That completes the proof of sufficient conditions for meeting   \eqref{eqn-subprobcond1.1}. Now, to ensure that \eqref{eqn-subprobcond1.2} is met, we note that by the random construction of the codebooks,
\begin{align}
\sum_{\mbf m^{\ms}}\frac{\Exp\left[ \dkl(\widehat{Q}^{({1})}_{\hat{X}_1^n| \mbf M^\ms}(\cdot| \mbf m^{\ms}) \parallel \widehat{Q}^{({1})}_{\hat{X}_1^n})\right]}{2^{n(\sum_{(i,j)\in \mc F}\mu_{i,j}^\ms)}} &= \Exp\left[ \dkl(\widehat{Q}^{({1})}_{\hat{X}_1^n| \mbf M^{\ms}}(\cdot | \mbf{ \underline 1}) \parallel \widehat{Q}^{({1})}_{\hat{X}_1^n})\right]\\
&=  \Exp\left[ \dkl(\widehat{Q}^{({1})}_{\hat{X}_1^n| M_\h}(\cdot | \mbf{ \underline 1}) \parallel Q^{\otimes n}_{\hat{X}_1})-\dkl(\widehat{Q}^{({1})}_{\hat{X}_1^n} \parallel Q^{\otimes n}_{\hat{X}_1})\right],
\end{align}
where $\mbf{ \underline 1}$ denotes the all-one vector of length $|\mc F| = \binom{\h}{2}$. Note that the analysis in \eqref{eqn-Anlys1}-\eqref{eqn-Anlys5}  ensures that  the second term in the equation above vanishes as we let $n$ diverge (provided \eqref{eqn-subprobcond1.2} is met). So, we proceed almost exactly as we did in the first part of this proof. 
%\begin{figure*}[th!]
\begin{align}
 \Exp\big[\dkl &(\widehat{Q}^{({1})}_{\hat{X}_1^n|\mbf{M}^\ms}(\cdot| \mbf{ \underline 1})\parallel \mathsf{Q}_{\boldsymbol X_1}^{\otimes n})\big]\\& =\Exp\left[ \sum_{x_1^n}\Bigg(\frac{\sum\limits_{\mbf{m}^\pms: \mbf m^\ms =\mbf{ \underline 1}} Q^{\otimes n}_{X_1| {\boldsymbol{{A}}}}(x_1^n|{\boldsymbol{{A}}}^n(\mbf{m}^\pms) )}{\mathsf N'}\Bigg)\log \left(\frac{\sum\limits_{\tilde{\mbf{m}}^\pms: \tilde{\mbf m}^\ms = \mbf{ \underline 1}} Q^{\otimes n}_{X_1| {\boldsymbol{{A}}} }(x_1^n|{\boldsymbol{{A}}}^n(\tilde{\mbf{m}}^\pms))}{\mathsf N'\mathsf Q^{\otimes n}_{X_1}(x_1^n)}\right)\right]\label{eqn-Anlys1.1}\\
&\leq \sum_{{x_1}^n, \boldsymbol{a}^n}  Q^{\otimes n}_{\boldsymbol{X_1A}}(x_1^n, \boldsymbol{a}^n)\log\Bigg[\sum_{S}  \frac{Q^{\otimes n}_{X_1}| {\boldsymbol A}_{\mathcal{J}^\mathsf{ c}_S}  (x_1^n |{\boldsymbol a}_{\mathcal{J}^\mathsf{ c}_S}^n)}{2^{n \left(\sum_{s\notin S} \mu_{s}^\ps\right)}\mathsf Q^{\otimes n}_{X_1}(x_1^n)} \Bigg].\label{eqn-Anlys4.1}
\end{align}
%\hrule
%\vspace{-5.75mm}
%\end{figure*}

The notation and arguments for the manipulations  in  \eqref{eqn-Anlys1.1}-\eqref{eqn-Anlys4.1} are as follows:
\begin{itemize}
\item We denote $\mathsf N' = 2^{n(\mu_{1,2}^\ps +\cdots + \mu_{\h-1,\h}^\ps)}$.
\item  \eqref{eqn-Anlys4.1} follows from steps identical to those between \eqref{eqn-Anlys1} and \eqref{eqn-Anlys5}. The sole difference is that $(X_1^n,\ldots, X_\h^n)$  is replaced by $X_1^n$, and the sums correspond to only all possible values taken by $\mbf M^{\ps}$, since we are restricted to $\mbf M^\ms = \mbf{ \underline 1}$.
\end{itemize}
As before, the sum of terms in \eqref{eqn-Anlys4.1} corresponding to atypical sequences yields a quantity no more than 
\begin{align}
\Pr\big[(X_1^n, \mbf{{A}}^n)\notin T^n_\ve[Q_{X_1\mbf{{A}}}]\big]\cdot \log \left(1+2^{\h^2} \eta_{X_1}^{-n}\right),
\end{align}
where $\eta_{X_1}= \min\limits_{\mbf{y}\,\in\,\mathsf{supp}(X_1)} \mathsf Q_{X_1}(x_1).$ Note that the above quantity vanishes as $n\rightarrow \infty$. On the other hand, the contribution from typical sequences can be made arbitrarily small if $\mu_{1,2}^\ps,\ldots, \mu_{1,\h}^\ps$ satisfy:
\begin{align}
\sum_{s\notin S} \mu_{s}^\ps  > I\big(X_1; A_{\mathcal{J}_S^\mathsf{ c}}\big)= I\big(X_1,\ldots,X_\h; \{A_{i,j}: (i,j)\notin {\mathcal{J}_S}\}\big), \quad S\subseteq \mc F.
\end{align}

\subsection{Proof of Theorem~\ref{thm-2}}\label{App-2}
We proceed in a way similar to the proof of Theorem~\ref{thm-1}. We use the following notation in this proof.
\begin{align}
\boldsymbol D_i(\mbf m^\pms,  k_{i-1}^\pms,l_{i})&\triangleq \left(B_{i-1,i}(\mbf m^\pms,  k_{i-1}^\pms), C_{i}(\mbf m^\pms,  k_{i-1}^\pms,l_{i})\right),\\
\mbf Y_i&\triangleq (X_{i-1},X_i),\\
\ell_i &\triangleq ( k_{i-1}^\pms,l_{i}),\\
\mathsf{N}_i&\triangleq 2^{n(
\kappa_{i-1}^\ps+\kappa_{i-1}^\ms+\lambda_{i})}.
\end{align}

Now, consider the following arguments.
%\begin{figure*}[th!]
\begin{align}
&\sum\limits_{\mbf m^\pms} \,\frac{ \Exp\left[\dkl\left(\widehat{Q}^{({i},\mbf m^\pms)}_{\hat{\mbf Y}_i^n|\boldsymbol{A}^n} \parallel  {Q}^{\otimes n}_{\mbf Y_i|\boldsymbol A}(\cdot |\mbf A^n (\boldsymbol m^\pms))\right)\right]}{2^{n(\mu_{1,2}^\ps+\mu_{1,2}^\ms+\cdots+\mu_{\h-1,\h}^\ps+\mu_{\h-1,\h}^\ms)}}\stackrel{(a)}{=} \Exp\left[\dkl\left(\widehat{Q}^{({i},\mbf{\underline 1})}_{\hat{\mbf Y}_i^n|\boldsymbol{A}^n} \parallel  {Q}^{\otimes n}_{{\mbf Y}_i | \boldsymbol{A}} (\cdot| \mbf A^n(\boldsymbol{\underline{1})}\right)\right]\\
{}
&\stackrel{(b)}{=}\Exp\left[ \sum_{\boldsymbol{y}^n}{\textstyle \Bigg( \frac{\sum\limits_{\ell_i'} Q^{\otimes n}_{\boldsymbol{Y}_i| {\boldsymbol A}\boldsymbol D_i}(\boldsymbol{y}^n|{\boldsymbol A}^n(\mbf{\underline 1}) \mbf D_i^n (\mbf{\underline 1}, \ell_i'))}{{\displaystyle\mathsf N_i}}\Bigg)\log \left(\frac{\sum\limits_{\ell_i''} Q^{\otimes n}_{\boldsymbol{Y}_i| {\boldsymbol A\mbf D_i}}(\boldsymbol{y}^n|{\boldsymbol A}^n(\mbf{\underline 1}) \mbf D_i^n (\mbf{\underline 1}, \ell_i''))}{\mathsf N_i\mathsf Q^{\otimes n}_{\boldsymbol Y_i|\mbf{A}}(\boldsymbol y^n| \mbf{A}^n(\mbf {\underline 1}))}\right)}\right]\label{eqn-Anlys6}\\
{}
&\stackrel{(c)}{=}\sum_{\boldsymbol{y}^n,\ell_i'}\Exp\left[{\frac{{\scriptstyle Q^{\otimes n}_{\mbf Y_i |\mbf{AD}_i}(\boldsymbol{y}^n|{\boldsymbol A}^n(\mbf{\underline 1}) \mbf D_i^n (\mbf{\underline 1}, \ell_i'))}}{\mathsf N_i}}\Exp\left[\log {\textstyle\frac{\sum\limits_{\ell_i''} Q^{\otimes n}_{\mbf Y_i | \mbf{AD}_i}(\boldsymbol{y}^n|{\boldsymbol A}^n(\mbf{\underline 1}) \mbf D_i^n (\mbf{\underline 1}, \ell_i''))}{\mathsf N_i\mathsf Q^{\otimes n}_{\boldsymbol Y_i|\mbf{A}}(\boldsymbol y^n| \mbf{A}^n(\mbf {\underline 1}))}}\,\Bigg|\,\substack{ \boldsymbol{A}^n(\mbf{\underline 1})\\  \\\mbf D_i^n (\mbf{\underline 1}, \ell_i'))} \right]\right]\\
{}
&\stackrel{(d)}{\leq}\sum_{\boldsymbol{y}^n,\ell_i'}\Exp\left[{\frac{{\scriptstyle Q^{\otimes n}_{\mbf Y_i |\mbf{AD}_i}(\boldsymbol{y}^n|{\boldsymbol A}^n(\mbf{\underline 1}) \mbf D_i^n (\mbf{\underline 1}, \ell_i'))}}{\mathsf N_i}}\log \Exp\left[ {\textstyle\frac{\sum\limits_{\ell_i''} Q^{\otimes n}_{\mbf Y_i | \mbf{AD}_i}(\boldsymbol{y}^n|{\boldsymbol A}^n(\mbf{\underline 1}) \mbf D_i^n (\mbf{\underline 1}, \ell_i''))}{\mathsf N_i\mathsf Q^{\otimes n}_{\boldsymbol Y_i|\mbf{A}}(\boldsymbol y^n| \mbf{A}^n(\mbf {\underline 1}))}}\,\Bigg|\,\substack{ \boldsymbol{A}^n(\mbf{\underline 1})\\  \\\mbf D_i^n (\mbf{\underline 1}, \ell_i'))} \right]\right]\\\
{}
&\stackrel{(e)}{\leq}\hspace{-0.5mm} \sum_{\substack{\boldsymbol{y}^n, \boldsymbol{a}^n\\b^n,{c}^n}} \hspace{-0.4mm} Q^{\otimes n}_{\boldsymbol{Y}_i\boldsymbol{AD}_i}(\boldsymbol{y}^n\hspace{-0.3mm}, \boldsymbol{a}^n\hspace{-0.3mm},b^n\hspace{-0.3mm},{{c}}^n)\log\hspace{-0.5mm}\left[1\hspace{-0.6mm}+\hspace{-0.6mm} \frac{Q^{\otimes n}_{\boldsymbol{Y}_i|\boldsymbol{AD}_i}(\boldsymbol{y}^n| \boldsymbol{a}^n\hspace{-0.2mm}, b^n\hspace{-0.2mm},{{c}}^n)}{\mathsf N_i Q^{\otimes n}_{\mbf{Y}_i|\mbf A} (\boldsymbol y^n|\mbf{a}^n)}\hspace{-0.5mm}+\hspace{-0.5mm} \frac{Q^{\otimes n}_{\boldsymbol{Y}_i|\boldsymbol{A} B_{i-1,i}}(\boldsymbol{y}^n| \boldsymbol{a}^n,b^n)}{ \frac{\mathsf N_i}{2^{n\mathsf r_i}} Q^{\otimes n}_{\mbf{Y}_i|\mbf A} (\boldsymbol y^n|\mbf{a}^n)}  \right].\label{eqn-Anlys7}
\end{align}
%\hrule
%\vspace{-5.75mm}
%\end{figure*}

Now, to find the conditions on the rates, we proceed in a fashion similar to  \cite{MB-JK-Line2} and \cite{MB-JK-Line1}. The notation and arguments for the manipulations  in  \eqref{eqn-Anlys6}-\eqref{eqn-Anlys7} are as follows:
\begin{itemize}
\item In $(a)$, we let $\underline{\mbf 1}$ to be the all-one vector of length $|\mc F| = \binom{\h}{2}$. Note that $(a)$ follows because the codebooks for $\mbf A$, $B_{i-1,i}$, and $C_{i}$ are generated in an i.i.d. fashion.
\item In $(b)$, the expectation is over the codebooks for $B_{i-1,i}$, and $C_{i}$ and the realization of $\mbf A^n(\underline{\mbf 1})$.
\item  $(c)$ follows by the use of the law of iterated expectations, where the inner conditional expectation denotes the expectation over all random codeword constructions $\{\mbf{D}_i(\mbf{\underline 1}, \ell_i''): \ell_i'' \neq \ell_i'\}$ conditioned on the codeword $\mbf{D}_i(\mbf{\underline 1}, \ell_i')$;
\item $(d)$ follows from Jensen's inequality and by dropping the subscripts for the pmfs for the sake of simplicity; and
\item Similar to \eqref{eqn-Anlys4},  $(e)$ follows by splitting the inner summation according to the components where $\ell_i'\triangleq ({k_{i-1}^\pms}',l_{i}')$ and $\ell_i''\triangleq ({k_{i-1}^\pms}'',l_{i}'')$ differ. Unity is an upper bound when the expectation is evaluated for terms corresponding to ${k_{i-1}^\pms}'\neq {k_{i-1}^\pms}''$, the second term is the result when the expectation is evaluated for $\ell_i'=\ell_i'' $, and lastly, the third is the result from terms for which ${k_{i-1}^\pms}'={k_{i-1}^\pms}''$ and $l_i'\neq l_i''$.
\end{itemize}

 Finally,  the rate conditions can be extracted from \eqref{eqn-Anlys7} by splitting the outer sum depending on whether $(\boldsymbol{y}^n, \boldsymbol{a}^n, b^n,{c}^n)\in T^n_\ve[Q_{\mbf{Y}_i\mbf{AD}_i}]$ or not. The sum for non-typical realizations in \eqref {eqn-Anlys7} is no more than
\begin{align}
\Pr\big[(\boldsymbol{Y}^n, \boldsymbol{A}^n, \mbf{D}_i^n)\notin T^n_\ve[Q_{\mbf{Y}_i\mbf{AD}_i}]\big]\cdot \log \left(1+2\eta_{\mbf Y_i}^{-n}\right), \label{eqn-nontypbndprob2}
\end{align}
where $\eta_{\mbf{Y}_i}= \min\limits_{\mbf{y}\,\in\,\mathsf{supp}(X_{i-1},X_i)} \mathsf Q_{\mbf Y_i}(\mbf y).$ This term in \eqref{eqn-nontypbndprob2} goes to zero as $n\rightarrow \infty$. The contribution from typical realizations can be observed to vanish asymptotically, provided \eqref{eqn-Prob2-h+1cond1}-\eqref{eqn-Prob2-h+1cond4} hold.

Now, consider \eqref{eqn-subprobcond2-1}. Let $\mathsf N'_i \triangleq 2^{n(\kappa_{i-1}^\ps+\lambda_{i})}$. Then,
\begin{align}
& \frac{\sum\limits_{\mbf{m}^\pms, k_{i-1}^\ms}\Exp\left[\dkl\left(\widehat{Q}^{({i}, \mbf m^\pms)}_{\hat{X}_{i-1}^n|\boldsymbol{A}^n, K_{i-1}^\ms} (\cdot| k_{i-1}^\ms) \parallel  {Q}_{{X}_{i-1}|\mbf A }^{\otimes n}\big(\cdot | \boldsymbol A^n (\boldsymbol m^\pms)\big)\right)\right]}{2^{n((\mu_{1,2}^\ps+\mu_{1,2}^\ms)+\cdots+(\mu_{\h-1,\h}^\ps+\mu_{\h-1,\h}^\ms)+ \kappa_{i-1}^\ms)}}\notag\\
&\stackrel{(a)}{=} \Exp\left[\dkl\left(\widehat{Q}^{({i}, \underline{\mbf 1})}_{\hat{X}_{i-1}^n|\boldsymbol{A}^n, K_{i-1}^\ms} (\cdot| 1) \bigll  {Q}_{{X}_{i-1}|\mbf A }^{\otimes n}\big(\cdot | \boldsymbol A^n (\underline{\mbf 1})\big)\right)\right]\label{eqn-Anlys7-1}\\
&\stackrel{(b)}{=}\Exp\left[ \sum_{\boldsymbol{y}^n}{ \Bigg( \frac{\sum\limits_{\hat\ell_i':{\hat k_{i-1}^\ms}=1} Q^{\otimes n}_{{X}_{i-1}| {\boldsymbol A}\boldsymbol D_i}({x}_{i-1}^n|{\boldsymbol A}^n(\mbf{\underline 1}) \mbf D_i^n (\mbf{\underline 1}, \hat\ell_i))}{{\displaystyle\mathsf N'_i}}\Bigg)\log \left(\frac{\sum\limits_{\tilde \ell_i:\tilde k_{i-1}^{\ms}=1} Q^{\otimes n}_{{X}_{i-1}| {\boldsymbol A\mbf D_i}}({x}_{i-1}^n|{\boldsymbol A}^n(\mbf{\underline 1}) \mbf D_i^n (\mbf{\underline 1}, \tilde\ell_i))}{\mathsf N'_i\mathsf Q^{\otimes n}_{X_{i-1}|\mbf{A}}(x_{i-1}^n| \mbf{A}^n(\mbf {\underline 1}))}\right)}\right]\notag\\
&\stackrel{(c)}{\leq}\hspace{-0.5mm} \sum_{\substack{\boldsymbol{x}_{i-1}^n, \boldsymbol{a}^n\\b^n,{c}^n}} \hspace{-0.4mm} Q^{\otimes n}_{{X}_{i-1}\boldsymbol{AD}_i}(\boldsymbol{x}_{i-1}^n\hspace{-0.3mm}, \boldsymbol{a}^n\hspace{-0.3mm},b^n\hspace{-0.3mm},{{c}}^n)\log\hspace{-0.5mm}\left[1\hspace{-0.6mm}+\hspace{-0.6mm} \frac{ Q^{\otimes n}_{X_{i-1}|\boldsymbol{AD}_i}({x}_{i-1}^n| \boldsymbol{a}^n\hspace{-0.2mm}, b^n\hspace{-0.2mm},{{c}}^n)}{\mathsf N'_i Q^{\otimes n}_{{X}_{i-1}|\mbf A} (x_{i-1}^n|\mbf{a}^n)}\hspace{-0.5mm}+\hspace{-0.5mm} \frac{Q^{\otimes n}_{{X}_{i-1}|\boldsymbol{A} B_{i-1,i}}({x}_{i-1}^n| \boldsymbol{a}^n,b^n)}{ \frac{\mathsf N'_i}{2^{n\mathsf r_i}} Q^{\otimes n}_{{X}_{i-1}|\mbf A} (x_{i-1}^n|\mbf{a}^n)}  \right]\label{eqn-Anlys7-3}.
\end{align}
The notation and arguments for the manipulations  in  the above equations are as follows:
\begin{itemize}
\item In $(a)$, we let $\underline{\mbf 1}$ to be the all-one vector of length $|\mc F| = \binom{\h}{2}$. As before, $(a)$ follows because the codebooks for $\mbf A$, $B_{i-1,i}$, and $C_{i}$ are generated in an i.i.d. fashion.
\item In $(b)$, the expectation is over the section of codebooks for $B_{i-1,i}$, and $C_{i}$ corresponding to $k_{i-1}^\ms = 1$ and realization of $\mbf A^n(\underline{\mbf 1})$.
\item  $(c)$ follows from arguments similar to those in \eqref{eqn-Anlys6}-\eqref{eqn-Anlys7}.
\end{itemize}
Lastly, as before, by separating the contributions of typical and non-typical sequences, we see that the expression in \eqref{eqn-Anlys7-3} can be made to vanish if:
\begin{align}
\kappa_{i-1}^\ps+ \lambda_i &> I(X_{i-1}; B_{i-1,i}, C_i| \mbf {A}) \stackrel{\eqref{eqn-AuxRVs1}}{=} I(X_{i-1}; B_{i-1,i}| \mbf {A}), \\
\kappa_{i-1}^\ps &> I(X_{i-1}; B_{i-1,i}| \mbf {A}).
\end{align}
Note that the former constraint is redundant, thereby completing this proof. 

\subsection{Proof of Lemma~\ref{lem-1}}\label{App-3}
Fix $\ve>0$. Let us denote $\mathsf N \triangleq 2^{n((\mu_{1,2}^\ps+\mu_{1,2}^\ms)\cdots + (\mu_{\h-1,\h}^\ps+\mu_{\h-1,\h}^\ms))}$. Recall from Remarks~\ref{rem-Pins1} and \ref{rem-Pins2} that 
\begin{align}
\lim_{n\rightarrow \infty} \Exp\left[\parallel\hspace{-0.75mm} \widehat{Q}^{({1})}_{\hat{X}_1^n\cdots\hat{X}_\h^n}- \mathsf Q^{\otimes n}_{X_1\cdots X_\h} \hspace{-0.75mm}\parallel_1\hspace{-0.25mm} +\sum\limits_{i=2}^\h \sum\limits_{\mbf m^\pms}\frac{\parallel\hspace{-0.75mm}  {\widehat{Q}^{({i}, \mbf m^\pms)}_{\hat{X}_{i-1}^n\hat{X}_i^n|\mathbf{A}^n}}- Q^{\otimes n}_{X_{i-1}X_i | \mbf A}(\cdot| \mbf A^n(\mbf m^\pms))\hspace{-0.75mm}\parallel_1\hspace{-0.25mm} }{\mathsf N} \right] = 0.\label{eqn-Rems1and2}
\end{align}
Let for $j=1,\ldots,\h-1$,
\begin{align}
\delta_i\triangleq \lim_{n\rightarrow \infty}\left(\frac{1}{\mathsf N} \sum_{\mbf m^\pms} \Exp\bigll Q^{\otimes n}_{X_{i}|\mbf A}(\cdot| \mbf A^n (\mbf m^\pms)) \prod_{j=i+1}^{\h} \hat Q^{({j},\mbf m^\pms)}_{\hat X_{j}^n | \mbf A^n , \hat{X}^n_{j-1}} - Q^{\otimes n}_{X_{i}\cdots X_{\h} |  \mbf A}(\cdot|\mbf A^n(\mbf m^\pms))  \bigll_1\right),
\end{align}
where we denote
\begin{align}
 \hat Q^{({j},\mbf m^\pms)}_{\hat X_{j}^n | \mbf A^n , \hat{X}^n_{j-1}} \triangleq \frac{\hat Q^{({j},\mbf m^\pms)}_{\hat X_{j-1}^n\hat X_j^n | \mbf A^n}}{ \hat Q^{({j},\mbf m^\pms)}_{\hat X_{j-1}^n | \mbf A^n}}.
\end{align}
 First, consider $\delta_{\h-1}$.
\begin{align}
\delta_{\h-1} &\triangleq  \lim_{n\rightarrow \infty}\left(\frac{1}{\mathsf N} \sum_{\mbf m^\pms} \Exp\bigll Q^{\otimes n}_{X_{\h-1}|\mbf A}(\cdot| \mbf A^n (\mbf m^\pms))  \hat Q^{(\h,\mbf m^\pms)}_{\hat{X}^n_{\h}|  \mbf A^n, \hat{X}^n_{\h-1} } - Q^{\otimes n}_{X_{\h-1},X_{\h} |  \mbf A}(\cdot|\mbf A^n(\mbf m^\pms))\bigll_1\right) \\
{}
&\leq  \lim_{n\rightarrow \infty}\left[\begin{array}{l}  \frac{1}{\mathsf N} \sum\limits_{\mbf m^\pms}\Exp\bigll Q^{\otimes n}_{X_{\h-1}|\mbf A}(\cdot| \mbf A^n (\mbf m^\pms))  \hat Q^{(\h,\mbf m^\pms)}_{\hat{X}^n_{\h}|  \mbf A^n , \hat{X}^n_{\h-1}} -  \hat Q^{(\h,\mbf m^\pms)}_{\hat{X}^n_{\h-1}\hat{X}^n_\h |  \mbf A^n} \bigll_1 \\
\qquad +\quad  \frac{1}{\mathsf N} \sum\limits_{\mbf m^\pms}\Exp\bigll  \hat Q^{(\h,\mbf m^\pms)}_{\hat{X}^n_{\h-1} \hat{X}^n_\h |  \mbf A^n } -  Q^{\otimes n}_{X_{\h-1},X_{\h} |  \mbf A}(\cdot|\mbf A^n(\mbf m^\pms))\bigll_1\end{array} \right]\\
{}
&=  \lim_{n\rightarrow \infty}\left[\begin{array}{l}  \frac{1}{\mathsf N} \sum\limits_{\mbf m^\pms}\Exp\bigll Q^{\otimes n}_{X_{\h-1} |  \mbf A}(\cdot|\mbf A^n(\mbf m^\pms)) -  \hat Q^{(\h,\mbf m^\pms)}_{\hat{X}^n_{\h-1} |  \mbf A^n} \bigll_1 \\
\qquad +\quad  \frac{1}{\mathsf N} \sum\limits_{\mbf m^\pms}\Exp\bigll  \hat Q^{(\h,\mbf m^\pms)}_{\hat{X}^n_{\h-1} \hat{X}^n_\h |  \mbf A^n} - Q^{\otimes n}_{X_{\h-1},X_{\h} |  \mbf A}(\cdot|\mbf A^n(\mbf m^\pms)) \bigll_1\end{array} \right] \stackrel{ \eqref{eqn-Rems1and2}}{=} 0.\label{eqn-Indbasecase}
\end{align}
Now, for $j>1$, the following applies. 
\begin{align}
\delta_{h-j} &\triangleq \lim_{n\rightarrow \infty}\frac{1}{\mathsf N}\left(\sum_{\mbf m^\pms} \bigll Q^{\otimes n}_{X_{\h-j}|\mbf A}(\cdot| \mbf A^n (\mbf m^\pms))  \prod_{s = \h-j+1}^\h \hat Q^{({s},\mbf m^\pms)}_{\hat X_{s}^n | \mbf A^n,\hat{X}^n_{s-1}} -Q_{X_{\h-j}\cdots X_\h |  \mbf A}^{\otimes n}(\cdot | \mbf A^n(\mbf m^\pms) )  \bigll_1\right)\\
{}
&\leq \lim_{n\rightarrow \infty}\frac{\left[\begin{array}{l} \sum\limits_{\mbf m^\pms}\bigll \Big[Q^{\otimes n}_{X_{\h-j}|\mbf A}(\cdot| \mbf A^n (\mbf m^\pms))- \hat Q^{(\h-j+1,\mbf m^\pms)}_{\hat X_{\h-j}^n|\mbf A^n }\Big]\prod\limits_{s = \h-j+1}^\h \hat Q^{({s},\mbf m^\pms)}_{\hat X_{s}^n | \mbf A^n,\hat{X}^n_{s-1}} \bigll_1\\
+ \sum\limits_{\mbf m^\pms}{\bigll\Big[\hat Q^{(\h-j+1,\mbf m^\pms)}_{\hat X_{\h-j}^n \hat X_{\h-j+1}^n|\mbf A^n } - Q_{X_{\h-j} X_{\h-j+1}|\mbf A}^{\otimes n} (\cdot| \mbf A^n(\mbf m^\pms)) \Big]\prod\limits_{ s = \h-j+2}^\h \hat Q^{({s},\mbf m^\pms)}_{\hat X_{s}^n | \mbf A^n,\hat{X}^n_{s-1}}\bigll_1}\\
+   \sum\limits_{\mbf m^\pms}{\bigll Q_{X_{\h-j} X_{\h-j+1}|\mbf A}^{\otimes n}(\cdot| \mbf A^n(\mbf m^\pms))\prod\limits_{ s = \h-j+2}^\h Q^{({s},\mbf m^\pms)}_{\hat X_{s}^n | \mbf A^n  ,\hat{X}^n_{s-1}}-Q_{X_{\h-j}\cdots X_\h |  \mbf A}^{\otimes n}(\cdot | \mbf A^n(\mbf m^\pms) ) \bigll_1}\end{array}\right]}{\mathsf N}\notag\\
&\stackrel{(a)}{\leq} \lim_{n\rightarrow \infty}\frac{\left[\begin{array}{l} \sum\limits_{\mbf m^\pms}\bigll Q^{\otimes n}_{X_{\h-j}|\mbf A}(\cdot| \mbf A^n (\mbf m^\pms))- \hat Q^{(\h-j+1,\mbf m^\pms)}_{\hat X_{\h-j}^n|\mbf A^n } \bigll_1\\
+ \sum\limits_{\mbf m^\pms}{\bigll\hat Q^{(\h-j+1,\mbf m^\pms)}_{\hat X_{\h-j}^n \hat X_{\h-j+1}^n|\mbf A^n } - Q_{X_{\h-j} X_{\h-j+1}|\mbf A}^{\otimes n} (\cdot| \mbf A^n(\mbf m^\pms)) \bigll_1}\\
+   \sum\limits_{\mbf m^\pms}{\bigll Q_{ X_{\h-j+1}|\mbf A}^{\otimes n}(\cdot| \mbf A^n(\mbf m^\pms))\prod\limits_{ s = \h-j+2}^\h Q^{({s},\mbf m^\pms)}_{\hat X_{s}^n | \mbf A^n  ,\hat{X}^n_{s-1}}-Q_{X_{\h-j+1}\cdots X_\h |  \mbf A}^{\otimes n}(\cdot | \mbf A^n(\mbf m^\pms) ) \bigll_1}\end{array}\right]}{\mathsf N}\notag\\
&\stackrel{\eqref{eqn-Rems1and2}}{=} \delta_{h-j+1}. 
\end{align}
Note that in $(a)$, we have used the fact that \eqref{eqn-AuxRVs1} implies $X_{i} \mkv (X_{i+1}, \mbf A) \mkv (X_{i+2},\ldots, X_\h)$, which then allows us to eliminate $X_{h-j}$ within the third variational distance term. By induction, we then conclude that $\delta_1=0$. Finally, by transferring the summation inside the norm, we see that that the following holds.
\begin{align}
\lim_{n\rightarrow \infty}   \frac{\Exp\bigll  \sum\limits _{\mbf m^\pms} \Big(Q^{\otimes n}_{X_{i}|\mbf A}(\cdot| \mbf A^n (\mbf m^\pms)) \prod\limits_{j=i+1}^{\h} \hat Q^{({j},\mbf m^\pms)}_{\hat X_{j}^n | \mbf A^n , \hat{X}^n_{j-1}}\Big)-\sum\limits _{\mbf m^\pms}Q^{\otimes n}_{X_{i}\cdots X_{\h} |  \mbf A}(\cdot|\mbf A^n(\mbf m^\pms))\bigll_1}{\mathsf N}  =0.
\end{align}

The proof is then complete by invoking the triangle inequality to combine the above result with \eqref{eqn-Rem1-1}. 

\subsection{Message Selection at Node 1}\label{app-revchan1}
\begin{figure}[!h]
\centering
 \vspace{-1mm}
 \includegraphics[width=4in]{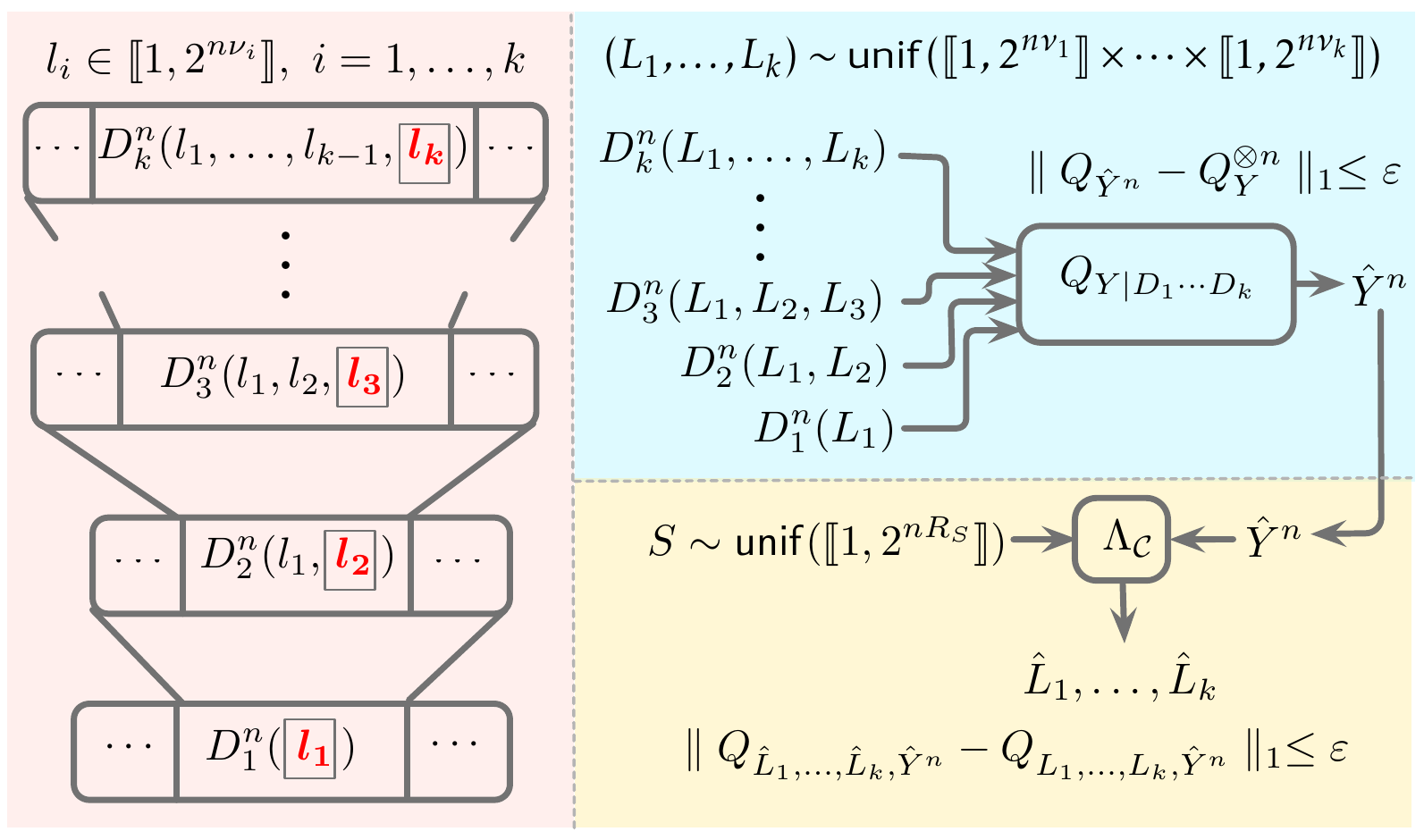}
  \vspace{-5mm}
\caption{A nested codebook structure for channel resolvability and the codeword selection problem.}
 \label{Fig-7}
 \vspace{-1mm}
\end{figure}

Let pmf $Q_{D_1,D_2Y}$ be given. Consider the a nested channel resolvability code for generating $Y\sim Q_{Y}$ via the channel $Q_{Y|D_1,\ldots D_2}$. Let the codebook structure be as given in Fig.~\ref{Fig-7}, where the codebook for $D_i$ is constructed randomly using $Q_{D_i|D_{i-1}\cdots D_1}$. Suppose that the rates of the codebooks satisfy
\begin{align}
\nu_1+\cdots +\nu_i> I (Y; D_1,\ldots, D_i), \quad i=1,\ldots k. \label{eqn-revchanresconstraint1}
\end{align}
For this choice of rates, it can be shown that for any $\ve>0$, one can find $n_0\in \mathbb{N}$ such that for $n>n_0$,
\begin{align}
\Exp\left[\parallel Q_{\hat Y^n }- Q_{Y}^{\otimes n}\parallel_1 \right] \leq \ve,
\end{align}
where
\begin{align}
Q_{\hat Y^n}\triangleq \frac{1}{2^{n(\nu_1+\cdots+\nu_k)}}\sum_{l_1,\ldots l_k} Q_{Y|D_1\cdots D_l}^{\otimes n} (\cdot | D_1^n (l_1), D_2^n(l_1,l_2),\ldots, D_k^n (l_1,\ldots, l_k)).
\end{align}

%---------------------------------------------------------------------------------------------------------------------
%---------------------------------------------------------------------------------------------------------------------
Now suppose that we aim to realize a random selection of indices $(\hat L_1,\ldots, \hat L_k)$ using a function $\Lambda_\mathcal{C}$ that depends on the codebooks $\mc C$ and takes as inputs, an independent and uniform random seed $S$ and the output of the channel $\hat Y^n$. We require that the conditional pmf of the selected indices given the channel output match the \emph{a posteriori} probability of the indices $(L_1,\ldots, L_k)$ given $\hat{Y}^n$, i.e., 
\begin{align}
\Exp\left[\parallel Q_{\hat L_{1},\ldots,\hat{L}_k, \hat{Y}^n }-Q_{L_1,\ldots, L_k, \hat{Y}^n}\parallel_1\right] \leq \varepsilon.
\end{align}
The following result characterizes the rate of random seed required to realize such a random selection.

\begin{theorem}\label{thm-reversechannelres}
Fix $n\in\mathbb N$. Consider the random codebook structure given in Fig.~\ref{Fig-7} with rates $\nu_1,\ldots, \nu_k$ that satisfy \eqref{eqn-revchanresconstraint1}. Let $\hat Y^n$ denote the output of the channel when the input is a codeword  tuple that is uniformly selected from  channel resolvability codebook. Let $S\sim \mathsf{unif} (\llbracket 1,2^{nR_S}\rrbracket)$, where $$R_S> \nu_1 + \cdots + \nu_{k} - I (Y; D_{1},\ldots, D_k).$$

 Then, there exists a function $\Lambda_{\mathcal{C}}: \Y^n\times \llbracket 1,2^{nR_S}\rrbracket \rightarrow \llbracket 1,2^{n\nu_{1}}\rrbracket\times \cdots \times  \llbracket 1,2^{n\nu_k}\rrbracket $ (that depends on the instance of the realized codebooks) such that $(\hat L_{1},\ldots,\hat{L}_k) \triangleq \Lambda_{\mathcal{C}} (\hat{Y}^n, S)$ satisfies:
\begin{align}
\lim_{n\rightarrow\infty} \Exp\left[\parallel Q_{\hat L_{1},\ldots,\hat{L}_k, \hat{Y}^n }-Q_{L_1,\ldots, L_k, \hat{Y}^n}\parallel_1\right] = 0, \label{eqn-explimit}
\end{align}
where the expectation is over all random codebooks.
\end{theorem}
\begin{IEEEproof}
Let $\delta, \ve >0$ be chosen such that 
\begin{align}
R_S -(\nu_1+\cdots+\nu_k)+ I(D_1,\ldots, D_k; Y) - 4\delta \log_2 (|\mc D_1||\mc D_2|\cdots|\mc D_k||\Y|) > \ve. \label{eqn-infitineschoice}
\end{align}

Let $(L_1,\ldots, L_k)\sim\mathsf{unif}(\llbracket 1,2^{n\nu_1}\rrbracket\times\cdots\times \llbracket 1,2^{n\nu_k}\rrbracket)$,  ${L}^i\triangleq (L_1,\ldots, L_i)$ and  $\hat{L}^i\triangleq (\hat L_1,\ldots, \hat L_i)$ for $1\leq i \leq k$. By the random codebook construction, it follows that $(D_1^n(L_1), D_2^n(L^2),\ldots, D_{k}^n(L^k),\hat{Y}^n)$ is indistinguishable from the output of a DMS $Q_{D_1\cdots D_k Y}$. Hence, by \cite[Theorem 1.1]{Kramer-MUIF}, it follows that
\begin{align}
\Pr\left[ (D_1^n(L_1), D_2^n(L^2),\ldots, D_{k}^n(L^k),\hat{Y}^n) \notin T_{\delta}^n[Q_{D_1\cdots D_k Y}]\right] \leq 2K e^{-n \delta^2 \eta},\label{eqn-doubleensembleave1}
\end{align}
where $\eta \triangleq \min\limits_{Q_{D_1\cdots D_k Y}\in\,\mathsf{supp}(Q_{D_1\cdots D_k Y})} Q_{D_1\cdots D_k Y}(d_1,\ldots, d_k,y)$, and
\begin{align}
K&\triangleq |\mc D_1||\mc D_2|\cdots|\mc D_k||\Y|.\label{eqn-constdefn}
 \end{align}
Now, for a codebook $\mc C_\circ \triangleq \{(d_1^n(l_1),\ldots,d_l^n(l^k))\}_{l^k\in \llbracket 1,2^{n\nu_1}\rrbracket\times\cdots\times \llbracket 1,2^{n\nu_k}\rrbracket}$, and channel output $y^n\in\Y^n$, let
 \begin{align}
 \mathcal N_\mathcal{C_\circ}(y^n) \triangleq \big\{ l^k: (d_1^n(l_1),\ldots, d_k^n(l^k),y^n)\in  T_{\delta}^n[Q_{D_1\cdots D_k Y}] \big\}.\label{eqn-listsize}
  \end{align}
  Then, by Lemma \ref{lem-listdecbnd} below, the following holds for sufficiently large $n$.
  \begin{align}
\Exp\left[ | \mathcal N_{\mathcal{C}}(\hat Y^n)|\right] \leq(k+1) 2^{n\left(\nu_1+\cdots+\nu_k - I(D_{1}\ldots,D_{k}; Y) +2\delta \log_2 K\right)}.
  \end{align}
 
 Let $\mc G$ be the set of codebooks $\mc C_\circ \triangleq \{(d_1^n(l_1),\ldots,d_l^n(l^k)):l^k\in \llbracket 1,2^{n\nu_1}\rrbracket\times\cdots\times \llbracket 1,2^{n\nu_k}\rrbracket\}$ such that:
 \begin{align}
\hspace{-2mm}\Pr\left[ (d_1^n(L_1),\ldots, d_{k}^n(L^k),\hat{Y}^n) \notin T_{\delta}^n[Q_{D_1\cdots D_k Y}]\,\big|\,\mc C =\mc C_\circ\right] &\leq \sqrt{2K e^{-n \delta^2 \eta}}\label{eqn-singleensembleave1}\\
(k+1)^{-1}\Exp\left[ | \mathcal N_{\mathcal{C}}(\hat Y^n)|\,\big|\, \mc C =\mc C_\circ \right] &\leq2^{n\left(\nu_1+\cdots +\nu_k - I(D_{1}\cdots D_{k}; Y)+3\delta \log_2 K\right)}. \label{eqn-singleensembleave2}
 \end{align}
 By Markov's inequality, we then have
 \begin{align}
 \Pr[ \mc C \notin \mc G] \leq \sqrt{2K e^{-n \delta^2 \eta}}+2^{-n\delta \log_2 K}. \label{eqn-mcFbnd}
 \end{align}
 Now, pick $\mc C^* \triangleq \big\{({d_1^*}^n(l_1),\ldots,{d_l^*}^n(l^k))\big\}_{l^k\in \llbracket 1,2^{n\nu_1}\rrbracket\times\cdots\times \llbracket 1,2^{n\nu_k}\rrbracket} \in \mc G$ and define $\mc G_{\mc C^*}$ by
 \begin{align}
 \mc G_{\mc C^*}=\left\{ y^n: \begin{array}{ll} \Pr\left[ ({d_1^*}^n(L_1),\ldots, {d_k^*}^n(L^k), \hat Y^n )\notin T_{\delta}^n[Q_{D_1\cdots D_k Y}] \Big|\, \substack{\hat{Y}^n \,=\, y^n\\ \,\,\,\,\,\mc C\,= \,\mc C^*}\right]  \leq \sqrt[4]{2K e^{-n \delta^2 \eta}} \\  | \mathcal N_{\mathcal{C}^*}(y^n)| \leq (k+1) 2^{n\left(\nu_1+\cdots+\nu_k - I(D_{1}\ldots,D_{k}; Y) +4\delta \log_2 K\right)}  \end{array}\right\}. \label{eqn-GC*defn}
 \end{align}
 Again, by an application of Markov's inequality, it follows that
 \begin{align}
 \Pr[ \hat{Y}^n \notin  \mc G_{\mc C^*} | \mc C=\mc C^* ] \leq \sqrt[4]{2K e^{-n \delta^2 \eta}}+2^{-n\delta \log_2 K}. \label{eqn-mcGC^*bnd}
 \end{align}
 Further, it also follows that for each $y^n\in \mc G_{\mc C^*}$,
 \begin{align}
 \sum_{l^k\notin  \mathcal N_{\mathcal{C}^*}(y^n)} Q_{L_1\cdots L_k \hat{Y}^n} (l^k, y^n) \stackrel{\eqref{eqn-GC*defn}}{\leq} \sqrt[4]{2K e^{-n \delta^2 \eta}},
 \end{align}
 where $Q_{L_1\cdots L_k \hat{Y}^n}$ is the joint pmf between indices and the output induced by $\mc C^*$. Then, from Lemma~\ref{lem-randpmfselec} below, for each $y^n\in \mc G_{\mc C^*}$, we can select a function $f_{y^n}:\llbracket 1, 2^{nR_S} \rrbracket \rightarrow  \llbracket 1,2^{n\nu_1}\rrbracket\times\cdots\times \llbracket 1,2^{n\nu_k}\rrbracket $ such that 
 \begin{align}
 \parallel Q_{f_{y^n}(S)} - Q_{L_1\cdots L_k |\hat{Y}^n=y^n} \parallel_1&\leq \frac{|\mc N_{\mc C^*}(y^n)|}{2^{nR_S}} +  \sqrt[4]{2K e^{-n \delta^2 \eta}}\\
 &\leq (k+1) 2^{n\left(\nu_1+\cdots+\nu_k- I(D_{1}\ldots,D_{k}; Y)+4\delta\log_2 S-R_S\right)}+  \sqrt[4]{2K e^{-n \delta^2 \eta}}\\
 &\stackrel{\eqref{eqn-infitineschoice}}{\leq} (k+1)2^{-n\ve}+  \sqrt[4]{2K e^{-n \delta^2 \eta}},
 \end{align}
 where $S\sim \mathsf{unif} (\llbracket 1, 2^{n R_S} \rrbracket)$. Pick ${l^*}^k \in  \llbracket 1,2^{n\nu_1}\rrbracket\times\cdots\times \llbracket 1,2^{n\nu_k}\rrbracket $. Now, we can piece together these functions to define
 \begin{align}
 \Lambda_{\mc C^*} (y^n, S) \triangleq \left\{ \begin{array}{ll} f_{y^n}(S), & y^n \in  \mc G_{\mc C^*}\\ {l^*}^k,&  y^n \notin  \mc G_{\mc C^*} \end{array}\right..
 \end{align}
 By construction, we now have 
 \begin{align}
 \sum_{y^n\in \mc G_{\mc C^*}} Q_{\hat{Y}^n|\mc C = \mc C^*}( y^n ) \parallel Q_{ \Lambda_{\mc C^*} (y^n, S) } - Q_{L^k |\hat{Y}^n=y^n} \parallel_1 &\leq(k+1) 2^{-n\ve} +  \sqrt[4]{2K e^{-n \delta^2 \eta}}.
 \end{align}
 Combining the above bound with \eqref{eqn-mcGC^*bnd} and the fact that the variation distance between two pms is bounded above by 2, we obtain
 \begin{align}
\sum_{y^n} Q_{\hat{Y}^n|\mc C = \mc C^*}( y^n ) \parallel Q_{ \Lambda_{\mc C^*} (y^n, S) } - Q_{L^k |\hat{Y}^n=y^n} \parallel_1 &\stackrel{}{\leq} (k+1) 2^{-n\ve}+ 3 \sqrt[4]{2K e^{-n \delta^2 \eta}}+ 2\cdot2^{-n\delta \log_2 K}. 
 \end{align}
 
Since the RHS does not depend on the choice of $\mc C^*$ in $\mc G$, it follows that 
 \begin{align}
 \Exp \left[ \parallel Q_{ \Lambda_{\mc C} (y^n, S) } - Q_{L^k |\hat{Y}^n=y^n} \parallel_1 \big|\, \mc C \in \mc G\right] &\leq (k+1)2^{-n\ve}+ 3 \sqrt[4]{2K e^{-n \delta^2 \eta}}+ 2\cdot2^{n\delta \log_2 K}.
 \end{align}
Next, using the fact that the variational distance between two pmfs is no more than 2, we also have
 \begin{align}
 \Exp\left[\parallel Q_{ \Lambda_{\mc C} (y^n, S) } - Q_{L^k |\hat{Y}^n=y^n} \parallel_1 \mid \mc C \notin \mc G \right]\ \leq  2.\end{align}
Finally, combining the above two equations with \eqref{eqn-mcFbnd} completes the claim.
\end{IEEEproof}

\begin{lemma}\label{lem-listdecbnd}
Consider the codebook construction of Fig.~\ref{Fig-7} with codebook sizes satisfying \eqref{eqn-revchanresconstraint1}. Let $\mc{N}_{\mc C}(\cdot)$ be as defined in \eqref{eqn-listsize}. Then, for $n$ sufficiently large, 
  \begin{align}
\Exp\left[ | \mathcal N_{\mathcal{C}}(\hat Y^n)|\right] \leq (k+1) 2^{n\left(\nu_1+\cdots+\nu_k - I(D_{1}\ldots,D_{k}; Y) -2\delta \log_2 \left(|\mc D_1||\mc D_2|\cdots|\mc D_k||\Y|\right)\right)}.
  \end{align}
\end{lemma}
\begin{IEEEproof}
Due to the random construction of the codebooks, we see that
\begin{align}
\Exp\left[ | \mathcal N_{\mathcal{C}}(\hat Y^n)|\right] 
&= \sum_{\mathring l^k} \Exp \left[ \mathds{1}\{\mathring l^k \in \mc N_\mathcal{C}(\hat Y^n)\} \big| L^k = (1,\ldots, 1)\right].
\end{align}
To evaluate the conditional expectation, we partition the space $ \llbracket 1,2^{n\nu_1}\rrbracket\times\cdots\times \llbracket 1,2^{n\nu_k}\rrbracket$ as follows.
\begin{align}
\mathcal{S}_i = \left\{ l^k: \begin{array}{ll} l_j=1,& j<i\\ l_{j+1}\neq 1,& j=i \end{array}\right\}, \quad i=0,\ldots k.
\end{align}
Note that $\bigcup\limits_{j=1}^k \mc S_i  =  \llbracket 1,2^{n\nu_1}\rrbracket\times\cdots\times \llbracket 1,2^{n\nu_k}\rrbracket$. By the random nature of codebook construction, we have
\begin{align}
 \Pr \left[ (l_1,\ldots, l_k) \in \mc N_\mathcal{C}(\hat Y^n) \,\big| L^k = (1,\ldots, 1)\right] =  \Pr \left[(l_1',\ldots, l_k') \in \mc N_\mathcal{C}(\hat Y^n) \,\big| L^k = (1,\ldots, 1)\right]. \label{eqn-Jjprop}
\end{align}
for any pair of tuples $(l_1,\ldots, l_k), (l_1',\ldots, l_k') \in \mc S_j$, $j=0,\ldots, k$.  Let for $j=0,1,\ldots, k$, $\ell^k(j)$ be chosen such that $\ell^k(j)\in\mc S _j$, and thus, due to \eqref{eqn-Jjprop}, we have
\begin{align}
\Exp\left[ | \mathcal N_{\mathcal{C}}(\hat Y^n)|\right] &= \sum_{j=0}^k |\mc S_j| \,\Pr \left[ \ell^k(j) \in \mc N_\mathcal{C}(\hat Y^n)\,\big | L^k = (1,\ldots, 1)\right]\leq\sum_{j=0}^k  \bigg[\prod_{\iota>j} 2^{n\nu_\iota}\bigg] \eta_j, \label{eqn-Lsetbnd}
\end{align}
where we let $\eta_j\triangleq \Pr \left[\ell^k(j) \in \mc N_\mathcal{C}(\hat Y^n)\big| L^k = (1,\ldots, 1)\right]$, $j=0,\ldots, k$. Clearly, $\eta_k \leq 1$, and $\eta_0$ is exactly the probability that  realizations  $(D_1^n, D_2^n,\ldots, D_k^n)\sim Q_{D_1\ldots D_k}^{\otimes n}$ and $Y^n\sim Q_{Y}^{\otimes n}$ selected independent of one another are jointly $\delta$-letter typical. Thus, by \cite[Theorem 1.1]{Kramer-MUIF}, it follows that 
\begin{align}
\eta_0 &=\sum_{(d_1^n,\ldots, d_k^n,y^n)\in T_{\delta}^n[Q_{D_1\cdots D_k Y}]} Q_{D_1\cdots D_k}(d_1^n,\ldots, d_k^n) Q_Y(y^n) \\
&\leq 2^{-n I(D_1,\ldots, D_k; Y) + n\delta (H(D_1,\ldots, D_k,Y) + H(D_1,\ldots, D_k)+H(Y))} \leq 2^{-n(I(D_1,\ldots, D_k; Y) - 2\delta \log_2 K)},
\end{align}
where $K$ is defined in \eqref{eqn-constdefn}.
Now, when $0<j<k$,  we observe that $\eta_j$ is the probability that $(D_1^n, D_2^n,\ldots, D_k^n)\sim Q_{D_1\ldots D_k}^{\otimes n}$ and $Y^n$ (i.e., the output when $(D_1,\ldots, D_j)$ is sent through the channel $Q_{Y|D_1,\ldots, D_j}$) are jointly $\delta$-letter typical. Therefore, by use of \cite[Theorems 1.1 and 1.2]{Kramer-MUIF}, we see that
\begin{align}
\eta_j &\leq \sum_{(d_1^n,\ldots, d_k^n,y^n)\in T_{\delta}^n[Q_{D_1\cdots D_k Y}]} Q_{D_{j+1}\cdots D_k|D_1\cdots D_j}(d_{j+1}^n,\ldots, d_k^n| d_1^n,\ldots, d_{j}^n) Q_{D_1\cdots D_j Y}(d_1^n,\ldots, d_{j}^n,y^n) \\
&\leq 2^{-n (I(D_{j+1},\ldots, D_k; Y|D_1,\ldots, D_j) -2\delta \log_2 K)}.
\end{align}
Finally, combining the bounds for $\eta_0,\ldots, \eta_k$ and \eqref{eqn-Lsetbnd}, we see that
\begin{align}
\Exp\left[ | \mathcal N_{\mathcal{C}}(\hat Y^n)|\right] &\leq  1+ \sum_{j=0}^{k-1} 2^{n\left(\nu_{j+1}+\cdots+\nu_k - I(D_{j+1}\ldots,D_{k}; Y| D_{1} \ldots, D_{j}) +2\delta \log_2 K\right)}.
\end{align}
Finally, the claim follows since \eqref{eqn-revchanresconstraint1} ensures that
\begin{align*}
\max_{j=0,\ldots, k-1} \bigg[ \nu_{j+1}+\cdots+\nu_k - I(D_{j+1}\ldots,D_{k}; Y| D_{1} \ldots, D_{j})\bigg]  = \nu_1+\cdots+\nu_k - I(D_1,\ldots, D_k; Y).
\end{align*}
\end{IEEEproof}
\begin{lemma} \label{lem-randpmfselec}
Let $Q$ be a pmf on a finite set $\A$ such that there exists $\B\subseteq \A$ with $|\B|=M$ and $\sum_{b\in\B} Q(b) \geq 1-\ve$ for  $0<\ve<1$. Now, suppose that $L\sim \mathsf{unif} (\llbracket 1, \ell \rrbracket)$. Then, there exists $f: \llbracket 1, \ell \rrbracket \rightarrow \A$ such that $Q_{f(L)}$, the pmf of $f(L)$, satisfies $\parallel Q_{f(L)} - Q \parallel_1 \leq \ve+ \frac{M}{\ell}$.
\end{lemma}
\begin{IEEEproof}
Let $b_1 \preceq b_2 \preceq \cdots \preceq b_M$ be an ordering of $\B$. Let $p_0 = 0$, and for $1\leq i\leq M$, let $p_i \triangleq \sum_{j=1}^i Q(b_j)$ denote the cumulative mass function. Now, let $N_i \triangleq \left\lfloor {p_i}\ell \right\rfloor$, $i=0,\ldots, M$, and let $f:\left\llbracket 1, N_M \right\rrbracket \rightarrow \B$ be defined by the pre-images via $f^{-1}(b_i) = \{ N_{i-1}+1,\ldots, N_i\}$, $i=1,\ldots, M$.
Fig.~\ref{Fig-8} provides an illustration of these operations. Now, by construction, we have
\begin{align}
0\leq p_i - \Pr\left[f(L) \in \{b_1,\ldots, b_i\} \right] \leq\ell^{-1}, \quad i=1,\ldots, M.
\end{align}

%---------------------------------------------------------------------------------------------------------------------
\begin{figure}[!h]
\centering
 \vspace{-4mm}
 \includegraphics[width=4in]{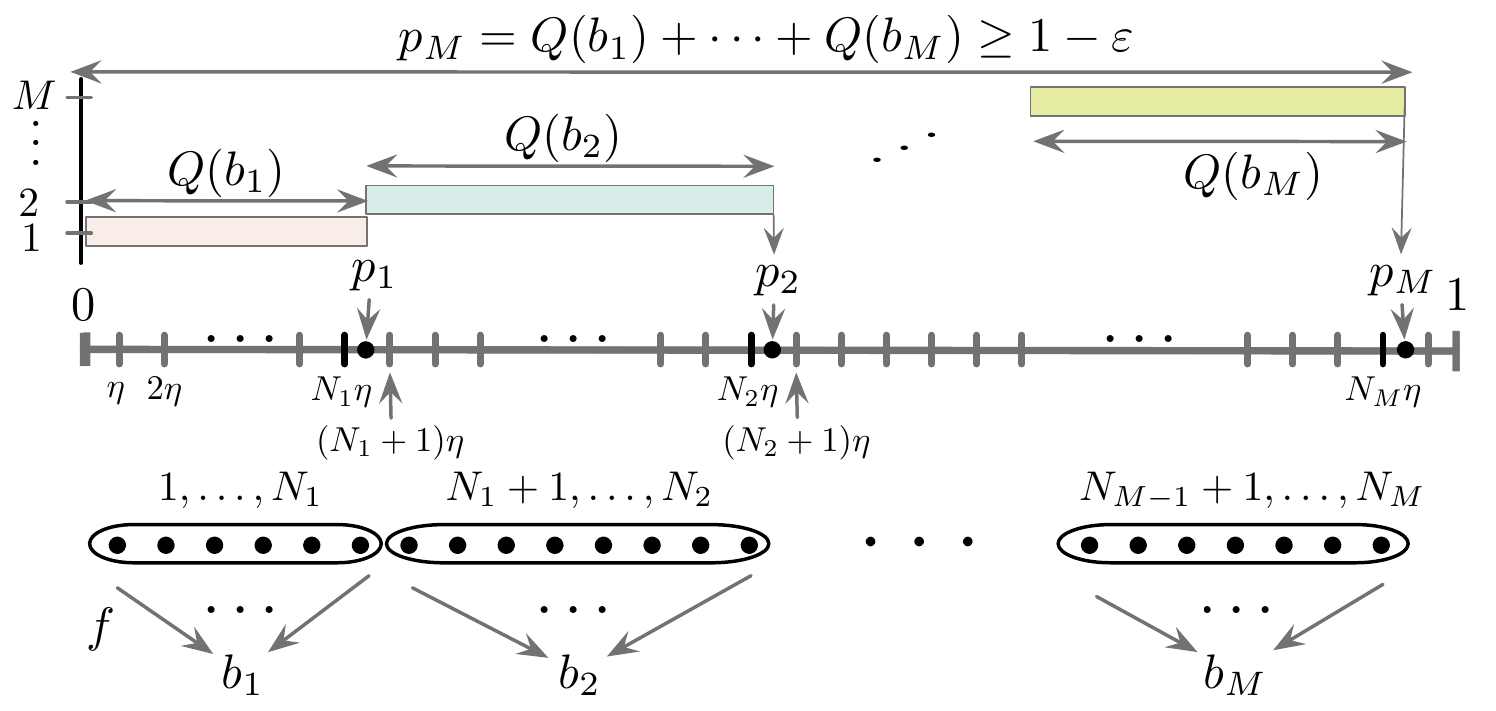}
  \vspace{-5mm}
\caption{An illustration of approximating a pmf using a function of a uniform RV.}
 \label{Fig-8}
 \vspace{-3mm}
\end{figure}
%---------------------------------------------------------------------------------------------------------------------
Consequently, we also have for any $i=1,\ldots, M$,
\begin{align}
-\ell^{-1} \leq p_{i} -p_{i-1} - Q_{f(L)}(b_{i}) = Q(b_{i}) -Q_{f(L)}(b_{i}) \leq \ell^{-1}. \label{eqn-pmfapproxbnd}
\end{align}
Hence, we see that
\begin{align}
\sum_{a\in \A} | Q(a)- Q_{f(L)}(a) | &= \sum_{i=1}^M | Q(b_i)- Q_{f(L)}(b_i) |  + \Pr[ A\notin \B]\stackrel{\eqref{eqn-pmfapproxbnd}}{\leq}  \frac{M}{\ell} +\ve.
\end{align}
\end{IEEEproof}
\subsection{Message Selection at Nodes $2,\ldots, \h-1$}\label{app-revchan2}
Let pmf $p_{D_1,D_2,Y}$ and $n\in\mathbb N$ be given. Suppose that $2^{n\nu_1}$ codewords $\{D_1^n(l_1): l_1=1,\ldots, 2^{n \nu_{1}}\}$ be selected such that $D_1^n(l_1)\sim Q_{D_1}^{\otimes n}$ for each $l_1=1,\ldots, 2^{n\nu_1}$. Note that the selection of the codewords may not be independent of one another. Now, for each $l_1\in\llbracket 1, 2^{n\nu_1}\rrbracket$, generate a codebook for $D_2$ such that codewords $\{D_2^n(i_1,i_2, i_2'): i_2\in\llbracket1, 2^{n\nu_2}\rrbracket,  i_2'\in \llbracket1,2^{n\nu_2'}\rrbracket\}$ are selected independently with each codeword selected using $\prod_{k=1}^n Q_{D_2|D_{1,k}(i_1)}$. Suppose that 
\begin{align}
\nu_2> I (Y; D_2| D_1) \label{eqn-revchanresconstraint2}.
\end{align}
For this choice of rates, it can be shown that for any $\ve>0$, one can find $n_0\in \mathbb{N}$ such that for $n>n_0$,
\begin{align}
\Exp\left[\parallel \hat Q^{(L_1,L_2')}_{\hat Y^n|D_1^n(L_1)}- Q^{\otimes n}_{Y|D_1}(\cdot \mid D_1^n(L_1))\parallel_1\right]=\sum_{l_1,l_2}\frac{\Exp\parallel \hat Q^{(l_1,l_2')}_{\hat Y^n|D_1^n(l_1)}- Q^{\otimes n}_{Y|D_1}(\cdot \mid D_1^n(l_1))\parallel_1}{2^{n(\nu_1+\nu_2')}}  \leq \ve,
\end{align}
where
\begin{align}
\hat Q^{(l_1,l_2')}_{\hat Y^n|D_1^n(l_1)}\triangleq \sum_{l_2=1}^{2^{n\nu_2}} \frac{Q_{Y| D_1, D_2}^{\otimes n} (\cdot | D_1^n(l_1), D_2^n(l_1,l_2,l_2'))}{2^{n\nu_2}}.
\end{align}
\begin{remark}
Since the codewords $D_1^n(\cdot)$ are identically distributed (but not necessarily independent), 
\begin{align}
\Exp\left[\parallel \hat Q^{(L_1,L_2')}_{\hat Y^n|D_1^n(L_1)}- Q^{\otimes n}_{Y|D_1}(\cdot \mid D_1^n(L_1))\parallel_1\right] = \parallel \hat Q^{(1,1)}_{\hat Y^n|D_1^n(1)}- Q^{\otimes n}_{Y|D_1}(\cdot | D_1^n(1))\parallel_1.
\end{align}$\hfill${\tiny$\blacksquare$}
\end{remark}
%---------------------------------------------------------------------------------------------------------------------
\begin{figure}[!h]
\centering
 \vspace{-1mm}
 \includegraphics[width=4in]{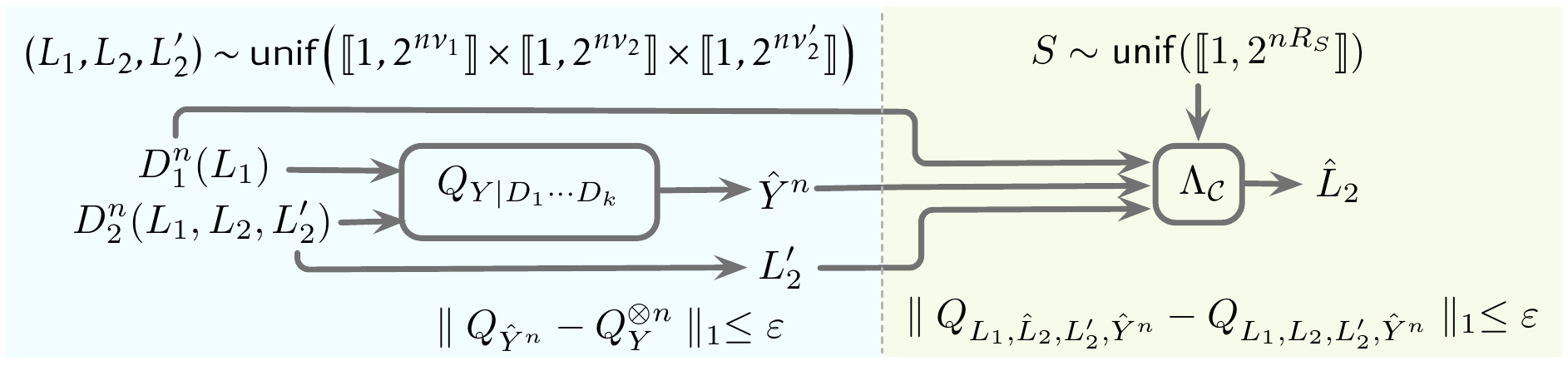}
  \vspace{-5mm}
\caption{A nested codebook structure for channel resolvability and the codeword selection problem.}
 \label{Fig-9}
 \vspace{-3mm}
\end{figure}
%---------------------------------------------------------------------------------------------------------------------

Now suppose that as in Appendix~\ref{app-revchan1}, one would like to characterize the amount of randomness required to generate randomly $\hat L_2$ using a function $\Lambda_\mathcal{C}$ (that depends on the codebooks $\mc C$) that takes as inputs a uniform random seed $S$, the output of the channel $\hat Y^n$, and the actual $(L_1, L_2')$ that was used to generate the channel output. As given in Fig.~\ref{Fig-8}, we want $\hat L_2$ to mimic $L_2$ and that the joint correlation of the RV $\hat L_2$ with $L_1, L_2'$ and the realized $\hat Y^n$ is arbitrarily close to $Q_{L_1,L_2, L_2',\hat{Y}^n}$, i.e., 
\begin{align}
\Exp\left[\parallel Q_{L_1, \hat L_{2},L_2', \hat{Y}^n }-Q_{L_1,L_2, L_2',\hat{Y}^n}\parallel_1\right] \leq \varepsilon.
\end{align}
The following result characterizes the rate of randomness required to realize this random index selection. 
\begin{theorem}\label{thm-reversechannelres1}
Consider the codebook structure described above with $\nu_1$ satisfying \eqref{eqn-revchanresconstraint2}. Let $\hat Y^n$ denote the channel output when the input is a codeword  pair selected uniformly from  the codebooks. Let \begin{align}R_S> \nu_2 - I (Y; D_{2}|D_1).\end{align}

 Then, there exists a function $\Lambda_{\mathcal{C}}: \Y^n \times \llbracket 1, 2^{n\nu_1}\rrbracket \times\llbracket 1, 2^{n\nu_2'}\rrbracket \times\llbracket 1, 2^{nR_S}\rrbracket\rightarrow \llbracket 1,2^{n\nu_{2}}\rrbracket$ (that depends on the instance of the realized codebooks) such that $\hat L_2 \triangleq \Lambda_{\mathcal{C}} (\hat{Y}^n, L_1,L_2', S)$ satisfies:
\begin{align}
\lim_{n\rightarrow\infty} \Exp\left[\parallel Q_{L_1, \hat L_{2}, L_2', \hat{Y}^n }-Q_{L_1,L_2,  L_2',\hat{Y}^n}\parallel_1\right] =0, \label{eqn-explimit}
\end{align}
where the expectation is over all random codebook realizations.
\end{theorem}
\begin{IEEEproof}
The proof mirrors exactly those of Theorem~\ref{thm-reversechannelres} and its associated lemmas, and is omitted. 
\end{IEEEproof}


\end{document}